\documentclass[aps,prd,showpacs,notitlepage,nofootinbib,preprintnumbers,amsmath,amssymb]{revtex4-1}

\usepackage{graphics,graphicx}
\usepackage{dcolumn}
\usepackage{bm}
\usepackage{epsfig}
\usepackage[usenames]{color}
\usepackage{hyperref} 
\usepackage{mathbbol}
\usepackage{epstopdf}
\usepackage{simplewick}
\usepackage[utf8]{inputenc} 
\usepackage{slashed}
\usepackage{stackrel}

\def\eq#1{{Eq.~(\ref{#1})}}
\def\fig#1{{Fig.~\ref{#1}}}
\newcommand{\ben}{\begin{eqnarray*}}
\newcommand{\een}{\end{eqnarray*}}
\newcommand{\un}[1]{\underline{#1}}

\newcommand{\tr}{\mbox{tr}}
\newcommand{\thalf}{\tfrac{1}{2}}
\newcommand{\llangle}{\Big\langle \!\! \Big\langle}
\newcommand{\rrangle}{\Big\rangle \!\! \Big\rangle}

\newcommand{\stackeven}[2]{{{}_{\displaystyle{#1}}\atop\displaystyle{#2}}}

\newcommand{\gsim}{\stackeven{>}{\sim}}
\newcommand{\as}{\alpha_s}

\newcommand{\dhd}{{\textstyle d}
\lower.03ex\hbox{\kern-0.38em$^{\scriptstyle-}$}\kern-0.05em{}}
\newcommand{\dbar}{{\textstyle \delta}
\lower.03ex\hbox{\kern-0.38em$^{\scriptstyle-}$}\kern-0.05em{}}
\newcommand{\half}{{1\over 2}}

\newcommand{\bra}[1]{\left\langle #1 \right|}
\newcommand{\ket}[1]{\left| #1 \right\rangle}

\newcommand{\ul}[1]{\underline{#1}}

\newcommand{\cc}{\mbox{c.c.}}
\newcommand{\ord}[1]{\mathcal{O}\left( #1 \right)}

\begin{document}

\title{Small-$x$ Asymptotics of the Gluon Helicity Distribution}

\author{Yuri V. Kovchegov} 
         \email[Email: ]{kovchegov.1@osu.edu}
         \affiliation{Department of Physics, The Ohio State
           University, Columbus, OH 43210, USA}
\author{Daniel Pitonyak}
	\email[Email: ]{dap67@psu.edu}
	\affiliation{Division of Science, Penn State University-Berks, Reading, PA 19610, USA}
\author{Matthew D. Sievert}
  \email[Email: ]{sievertmd@lanl.gov}
	\affiliation{Theoretical Division, Los Alamos National Laboratory, Los Alamos, NM 87545, USA}

\begin{abstract}
  We determine the small-$x$ asymptotics of the gluon helicity
  distribution in a proton at leading order in perturbative QCD at
  large $N_c$. To achieve this, we begin by evaluating the dipole
  gluon helicity TMD at small $x$. In the process we obtain an
  interesting new result: in contrast to the unpolarized dipole gluon
  TMD case, the operator governing the small-$x$ behavior of the
  dipole gluon helicity TMD is different from the operator
  corresponding to the polarized dipole scattering amplitude (used in
  our previous work to determine the small-$x$ asymptotics of the
  quark helicity distribution). We then construct and solve novel
  small-$x$ large-$N_c$ evolution equations for the operator related
  to the dipole gluon helicity TMD. Our main result is the
  small-$x$ asymptotics for the gluon helicity distribution: $\Delta G
  \sim \left( \tfrac{1}{x} \right)^{\alpha_h^G}$ with $\alpha_h^G =
  \tfrac{13}{4 \sqrt{3}} \, \sqrt{\tfrac{\as \, N_c}{2 \pi}} \approx
  1.88 \, \sqrt{\tfrac{\as \, N_c}{2 \pi}}$. We note that the power
  $\alpha_h^G$ is approximately 20$\%$ lower than the corresponding
  power $\alpha_h^q$ for the small-$x$ asymptotics of the quark
  helicity distribution defined by $\Delta q \sim \left( \tfrac{1}{x}
  \right)^{\alpha_h^q}$ with $\alpha_h^q = \tfrac{4}{\sqrt{3}} \,
  \sqrt{\tfrac{\as \, N_c}{2 \pi}} \approx 2.31 \, \sqrt{\tfrac{\as \,
      N_c}{2 \pi}}$ found in our earlier work.
\end{abstract}

\pacs{12.38.-t, 12.38.Bx, 12.38.Cy}

\maketitle


%
\section{Introduction} \label{sec:Intro}
%

A solid theoretical understanding of the small-$x$ asymptotics of the
quark and gluon helicity distributions $\Delta q (x, Q^2)$ and $\Delta
G (x, Q^2)$ is crucially important for the resolution of the proton
spin puzzle. The quark and gluon components of the proton spin,
\begin{align}
  \label{eq:net_spin}
  S_q (Q^2) = \frac{1}{2} \, \int\limits_0^1 \!dx \, \Delta \Sigma (x,
  Q^2) \, \ \ \ \mbox{and} \ \ \ S_G (Q^2) = \int\limits_0^1 \!dx \,
  \Delta G (x, Q^2)\,,
\end{align}
may receive significant contributions from the small-$x$ region. Given
that the current experimental values (see
\cite{Accardi:2012qut,Aschenauer:2013woa,Aschenauer:2015eha,Aschenauer:2016our}
for reviews), $S_q (Q^2 = 10\, \mbox{GeV}^2) \approx 0.15 \div 0.20$
(integrated over $0.001 < x <1$) and $S_G (Q^2 = 10 \, \mbox{GeV}^2)
\approx 0.13 \div 0.26$ (integrated over $0.05 < x <1$), still do not
add up to the proton spin of $1/2$, the small-$x$ region may turn out
to be important for satisfying helicity sum
rules~\cite{Jaffe:1989jz,Ji:1996ek,Ji:2012sj} (see
\cite{Leader:2013jra} for a review), such as the Jaffe--Manohar sum
rule~\cite{Jaffe:1989jz}
\begin{align}
  \label{eq:sum_rule}
  S_q + L_q + S_G + L_G = \frac{1}{2}\,, 
\end{align}
where $L_q$ and $L_G$ denote the quark and gluon orbital angular
momentum (OAM), respectively.

Moreover, the experimental measurement of the relevant double-longitudinal
spin asymmetry $A_{LL}$ is always
limited to the $x \in [x_{min}, 1]$ subset of the $x \in [0,1]$ range
employed in the integrals of \eq{eq:net_spin}, with $x_{min}$ given by
the experimental coverage of the specific machine and detector. No
matter how high-energy an experiment may be, there will always be some
$x_{min}$ below which it will not be able to measure $A_{LL}$. Therefore,
below that $x_{min}$ one does not have data from which 
to extract $\Delta q (x, Q^2)$ and $\Delta G (x, Q^2)$. 
To be certain that the experimentally
excluded region of $x \in [0, x_{min}]$ does not contribute much to
$S_q$ and $S_G$, or to obtain an accurate estimate of how much spin
resides at $x \in [0, x_{min}]$, one has to develop a quantitative
theoretical understanding of $\Delta q (x, Q^2)$ and $\Delta G (x,
Q^2)$ at small $x$. Then one could hope for the following possible
scenario at future polarized-scattering experiments, such as the ones
to be carried out at the proposed Electron-Ion Collider (EIC) in the
US \cite{Accardi:2012qut}: one may obtain solid agreement between
 theory predictions and experiment for the $x$-dependence of
$A_{LL}$ above $x_{min}$ (but still
at small $x$), that is for $x \gsim x_{min}$, which would allow one to
confidently extrapolate $\Delta q (x, Q^2)$ and $\Delta G (x, Q^2)$ to
the $x < x_{min}$ region. This extrapolation, in turn, would allow one
to make a good estimate of the amount of the proton's spin carried by
the quarks and gluons at $x < x_{min}$. The extrapolation would need
to be further tested by later experiments probing polarization at
smaller values of $x$: if agreement is found again, one may be able
claim that the procedure is converging and that the spin at small $x$
is approaching full theoretical control.

To address the important question of the small-$x$ asymptotics of
$\Delta q (x, Q^2)$ in the flavor-singlet channel, we derived
small-$x$ helicity evolution equations in
\cite{Kovchegov:2015pbl}. The evolution equations were written down
for the polarized dipole amplitude, which can be defined as the part
of the forward scattering amplitude for a $q\bar q$ dipole, with a
longitudinally polarized quark or antiquark in it, on a longitudinally
polarized target proton that depends on the product of the target and
projectile polarizations. The polarized dipole amplitude is related to
the quark helicity transverse-momentum-dependent (TMD) parton
distribution function: knowing the former gives us the latter
\cite{Kovchegov:2015zha}. The evolution equations for the polarized
dipole amplitude are both similar to and different from the
unpolarized Balitsky--Kovchegov (BK)
\cite{Balitsky:1995ub,Balitsky:1998ya,Kovchegov:1999yj,Kovchegov:1999ua}
and Jalilian-Marian--Iancu--McLerran--Weigert--Leonidov--Kovner
(JIMWLK)
\cite{Jalilian-Marian:1997dw,Jalilian-Marian:1997gr,Iancu:2001ad,Iancu:2000hn}
evolution equations. The similarity is in the fact that both the
helicity evolution and BK/JIMWLK evolution involve Wilson
lines. Moreover, just like in the Balitsky hierarchy
\cite{Balitsky:1995ub,Balitsky:1998ya}, the helicity evolution
equations do not close in general, and the large-$N_c$ limit has to be
invoked to produce a closed equation
\cite{Balitsky:1995ub,Balitsky:1998ya,Kovchegov:1999yj,Kovchegov:1999ua}. There
are also important differences: helicity evolution is sub-eikonal, and
involves the so-called ``polarized Wilson line" operator, which is
related to the helicity-dependent part of a high-energy
polarized-quark propagator through a longitudinally polarized target
\cite{Kovchegov:2015pbl}. The helicity evolution equations also become
a closed system of equations in the large-$N_c \& N_f$ limit, in
addition to the large-$N_c$ limit.

Perhaps most importantly, the helicity evolution equations resum
{\it{double logarithms}} of energy, that is, powers of $\as \, \ln^2
\tfrac{1}{x}$ with $\as$ the strong coupling constant. This is in
contrast to the leading-logarithmic resummation of the powers of $\as
\, \ln \tfrac{1}{x}$ in the unpolarized
Balitsky--Fadin--Kuraev--Lipatov (BFKL)
\cite{Kuraev:1977fs,Balitsky:1978ic} along with the BK/JIMWLK
equations. The double-logarithmic approximation (DLA) resulting from
resumming the powers of $\as \, \ln^2 \tfrac{1}{x}$ was considered
before for the $t$-channel quark exchange amplitudes
\cite{Kirschner:1983di, Kirschner:1985cb,
  Kirschner:1994vc,Kirschner:1994rq,Griffiths:1999dj,Itakura:2003jp,Bartels:2003dj}. For
helicity evolution it was first applied by Bartels, Ermolaev and
Ryskin (BER) in \cite{Bartels:1995iu,Bartels:1996wc} (see also
\cite{Ermolaev:1999jx,Ermolaev:2000sg,Ermolaev:2003zx,Ermolaev:2009cq}). (The
DLA parameter $\as \, \ln^2 \tfrac{1}{x}$ does not exist in the
unpolarized BFKL/BK/JIMWLK evolution, and so far has been established
either in $t$-channel quark exchanges \cite{Kirschner:1983di,
  Kirschner:1985cb,
  Kirschner:1994vc,Kirschner:1994rq,Griffiths:1999dj,Itakura:2003jp,Bartels:2003dj}
or for the $t$-channel longitudinal spin transfer
\cite{Bartels:1995iu,Bartels:1996wc}.) To accomplish the DLA
resummation in the $s$-channel small-$x$ formalism we had to introduce
an auxiliary ``neighbor" polarized dipole amplitude
\cite{Kovchegov:2015pbl}, which was never required in the
leading-logarithmic unpolarized dipole evolution
\cite{Mueller:1993rr,Mueller:1994jq,Mueller:1994gb,Balitsky:1995ub,Balitsky:1998ya,Kovchegov:1999yj,Kovchegov:1999ua}.

The derivation of the flavor-singlet helicity evolution equations from
\cite{Kovchegov:2015pbl} was further clarified in
\cite{Kovchegov:2016zex}, where we also derived and solved the
evolution equation for the quark helicity TMD in the flavor
non-singlet case. The resulting small-$x$ (large-$N_c$) asymptotics of
the flavor non-singlet quark helicity distribution were in complete
agreement with that derived previously by BER \cite{Bartels:1995iu}.

The flavor-singlet large-$N_c$ helicity evolution equations from
\cite{Kovchegov:2015pbl} were first solved numerically in
\cite{Kovchegov:2016weo} and then analytically in
\cite{Kovchegov:2017jxc}. The resulting small-$x$ asymptotics of the
quark helicity parton distribution function (PDF) were found to be
\begin{align}\label{q_asym}
  \Delta q (x, Q^2) \sim \left( \frac{1}{x} \right)^{\alpha_h^q} \ \ \
  \mbox{with} \ \ \ \alpha_h^q = \frac{4}{\sqrt{3}} \, \sqrt{\frac{\as
      \, N_c}{2 \pi}} \approx 2.31 \, \sqrt{\frac{\as \, N_c}{2 \pi}}.
\end{align}
The flavor-singlet quark helicity intercept $\alpha_h^q$ at large
$N_c$ was about 30$\%$ smaller than that found by BER in
\cite{Bartels:1996wc}. We discussed the possible origin of our
differences in \cite{Kovchegov:2016zex}; in Appendix~B of that paper
we presented some of the DLA diagram contributions we believe BER did
not include in their analysis.

Having established the small-$x$ asymptotics for the quark helicity
distribution \eqref{q_asym}, we now turn our attention to the gluon
helicity distribution, which is the main topic of this paper. We begin
in Sec.~\ref{sec:recap} by reviewing the central results from the
quark helicity case and by constructing a definition for the
``polarized Wilson line" operator employed previously in
\cite{Kovchegov:2015pbl,Kovchegov:2016zex} without presenting an
explicit form. The polarized Wilson line operator provides us with
the operatorial form of the polarized dipole scattering amplitude from
\cite{Kovchegov:2015pbl,Kovchegov:2016zex}. We proceed in
Sec.~\ref{sec:operators} by evaluating the gluon helicity TMDs at
small $x$. We consider both the dipole and Weizs\"{a}cker-Williams
(WW) gluon helicity TMDs according to the standard prescription
\cite{Dominguez:2011wm}. Starting with their definitions, we express
each of those gluon helicity TMDs in terms of light-cone Wilson lines
and an insertion of the sub-eikonal longitudinal spin-dependent gluon
field of the target. In \cite{Dominguez:2011wm} a similar procedure
expressed the unpolarized dipole gluon TMD in terms of the forward
scattering amplitude for a (fundamental) $q \bar q$ dipole scattering
on the target, hence giving rise to the name for the dipole TMD. (Also see
Ref.~\cite{Metz:2011wb}
for related work on the distribution of linearly polarized gluons.) This
amplitude can be found by solving the BK evolution
equation. Surprisingly, and unlike the unpolarized case, the dipole
gluon helicity TMD turns out not to be directly related to the
polarized dipole scattering amplitude. Instead it is related to a
somewhat different operator as shown in Sec.~\ref{sec:operators}. (The
same applies for the WW gluon helicity TMD: it is not directly related
to the polarized dipole amplitude. However, this is not unlike the
unpolarized case, in which the unpolarized WW gluon TMD was found to
be related to the color-quadrupole amplitude \cite{Dominguez:2011gc}
and not to the dipole one.)

The small-$x$ evolution for the dipole gluon helicity TMD is
constructed in Sec.~\ref{sec:evolution}. There we begin by
reconstructing the DLA evolution equations for the polarized dipole
amplitude from \cite{Kovchegov:2015pbl}; since now we have an operator
expression for the polarized dipole amplitude, we use the operator
language, similar to that developed by Balitsky in
\cite{Balitsky:1995ub,Balitsky:1998ya}. This is a cross-check of both
our equations in \cite{Kovchegov:2015pbl} as well as the operator
definition and approach. We proceed by applying the operator method to
evaluate the operator related to the dipole gluon helicity TMD. The
result, in the large-$N_c$ limit, is the evolution equations
\eqref{M:evol4} which mix this ``gluon helicity operator" with the
``quark helicity operator" given by the polarized dipole
amplitude. These equations are solved in Sec.~\ref{sec:solution}, both
analytically and numerically. The end result is the following
small-$x$ asymptotics of the gluon helicity distribution:
\begin{align} 
\label{G_asym} 
\Delta G (x, Q^2) \sim \left( \frac{1}{x} \right)^{\alpha_h^G} \ \ \
\mbox{with} \ \ \ \alpha_h^G = \frac{13}{4 \sqrt{3}} \,
\sqrt{\frac{\as \, N_c}{2 \pi}} \approx 1.88 \, \sqrt{\frac{\as \,
    N_c}{2 \pi}}.
\end{align}
Equations \eqref{q_asym} and \eqref{G_asym} give us the
leading-in-$\as$ small-$x$ asymptotics of both the quark and gluon
helicity distributions. It is interesting to note that
$\alpha_h^G < \alpha_h^q$; we explore the phenomenological consequences of this in Sec.~\ref{sec:pheno} and Sec.~\ref{sec:concl}.

In Sec.~\ref{sec:pheno} we estimate the amount of the proton's spin carried by small-$x$ gluons
using a simple phenomenological approach.
As depicted in \fig{f:run_spinG}, we observe a
$5\div 10 \%$ increase in the amount of gluon spin if we use our
intercept \eqref{G_asym} to augment the existing
DSSV14~\cite{deFlorian:2014yva} PDF parameterization.  We also discuss the importance of
incorporating our work into future fits of helicity PDFs.

We conclude in Sec.~\ref{sec:concl} by summarizing our main results
and by outlining further steps which need to be made in order to
perform a detailed comparison with the experimental data.

%
\section{The Quark Helicity TMD and the Polarized Dipole Amplitude}
\label{sec:recap}
%

%
\subsection{Review}
%

In \cite{Kovchegov:2015pbl}, we derived the polarized small-$x$
evolution equations for the TMD quark
helicity distribution \cite{Mulders:1995dh},
\begin{align} 
  \label{e:qTMD} 
  g_{1L}^q (x, k_T^2) = \frac{1}{(2\pi)^3} \, \half\sum_{S_L} S_L \,
  \int d^2 r \, dr^- \, e^{i x P^+ r^-} \, e^{- i \ul{k} \cdot \ul{r}}
  \bra{P, S_L} \bar{\psi}(0) \: \mathcal{U}[0,r] \: \frac{\gamma^+
    \gamma^5}{2} \: \psi(r) \ket{P, S_L}_{r^+ = 0},
\end{align}
by relating it to a ``polarized dipole amplitude'' $G(x_{10}^2 , z
s)$, giving
\begin{align} 
  \label{e:qTMD2}
  g_{1L}^{q , S} (x, k_T^2) = \frac{8 N_c}{(2\pi)^6} \sum_f
  \int\limits^1_{\Lambda^2/s} \frac{dz}{z} \int d^2 x_{01} \, d^2
  x_{0'1} \, e^{- i \ul{k} \cdot (\ul{x}_{01} - \ul{x}_{0'1})} \:
  \frac{\ul{x}_{01} \cdot \ul{x}_{0'1}}{x_{01}^2 \, x_{0'1}^2} \:
  G(x_{10}^2 , z s = \tfrac{z}{x} Q^2)
\end{align}
in the flavor-singlet case \cite{Kovchegov:2016zex}.  In the above and
throughout this paper, we use light-front coordinates $x^\pm \equiv
\frac{1}{\sqrt{2}} (x^0 \pm x^3)$, denote transverse vectors
$(x_\bot^1 , x_\bot^2)$ by $\ul{x}$ and their magnitudes by $x_T
\equiv | \ul{x} |$, and indicate differences in transverse coordinates
by the abbreviated notation $\ul{x}_{10} \equiv \ul{x}_1 -
\ul{x}_0$. The center-of-mass energy squared for the scattering process is
$s$, the infrared (IR) transverse momentum cutoff is $\Lambda$, and
$z$ is the fraction of the light-cone momentum of the dipole carried
by the polarized (anti-)quark.  As is well-known, the TMD
\eqref{e:qTMD} contains a process-dependent gauge link
$\mathcal{U}[0,r]$.  For specificity, in \cite{Kovchegov:2015pbl} we
considered semi-inclusive deep inelastic scattering (SIDIS), although
the resulting small-$x$ evolution equations also apply to the
collinear quark helicity distribution, which is process independent.

The impact-parameter integrated polarized dipole amplitude is
\begin{align}
  \label{eq:Gint}
  G(x_{10}^2 , z s) = \int d^2 b_{10} \, G_{10} (z s)
\end{align}
with $\un{b}_{10} = (\un{x}_1 + \un{x}_0)/2$.  The polarized dipole
scattering amplitude $G_{10} (z s)$ was defined as the polarized
generalization of the forward dipole S-matrix in terms of Wilson lines
\cite{Kovchegov:2015pbl}:
\begin{align} 
  \label{eq:Gdef} 
  G_{10} (z s) &\equiv \frac{1}{2 N_c} \,
  \llangle \tr \left[ V_{\ul 0} V_{\ul 1}^{pol \, \dagger} \right] +
  \tr \left[V_{\ul 1}^{pol} V_{\ul 0}^\dagger \right] \rrangle (z s)
\notag \\ &\equiv
\frac{z s}{2 N_c} \, \left\langle \tr \left[ V_{\ul 0} V_{\ul 1}^{pol
      \, \dagger} \right] + \tr \left[V_{\ul 1}^{pol} V_{\ul
      0}^\dagger \right] \right\rangle (z s) ,
\end{align}
where the double-angle brackets are defined to scale out the
center-of-mass energy $z s$ between the polarized (anti)quark and the
target.  While the unpolarized Wilson lines in \eq{eq:Gdef} are the standard
eikonal gauge links (in the fundamental representation),
\begin{align} \label{e:Wlineunp} V_{\ul 0} \equiv V_{\ul{x}_0}
  [+\infty, -\infty] \equiv \mathcal{P} \exp\left[ i g
    \int\limits_{-\infty}^{+\infty} dx^- A^+ (x^+ =0, x^- , \ul{x}_0)
  \right] ,
\end{align}
the polarized Wilson lines $V_{\ul 1}^{pol}$ are more complex
operators.  Wilson lines in general correspond to the eikonal
propagators of partons in the background field of the target, with the
eikonal gauge link \eqref{e:Wlineunp} being manifestly
spin-independent.  The polarized Wilson line $V_{\ul 1}^{pol}$
represents the spin-dependent propagator of a quark in the background
field of the target, which in the high-energy limit is suppressed by
one factor of the center-of-mass energy, motivating the rescaling
performed in \eq{eq:Gdef}.  Spin dependence is introduced into the
polarized Wilson line by the insertion of exactly one sub-eikonal
interaction which is sensitive to the spins of the parton and the
target.  As discussed in \cite{Kovchegov:2015pbl}, the spin-dependent
interaction may correspond either to the $t-$channel exchange of two
quarks or of the transverse component of the gluon field.  Because each
such sub-eikonal interaction leads to a suppression of the Wilson line
by a factor of the energy, additional spin-dependent exchanges can be
neglected as power suppressed.  While we leave the determination of
the quark-exchange part of the polarized Wilson line operator for
future work, we will show by explicit calculation below that the
gluon-exchange component takes the form
\begin{align} \label{M:Vpol1}
(V_{\ul x}^{pol})^g = \int\limits_{-\infty}^{+\infty} dx^- \: V_{\ul{x}}[+\infty, x^-] \: 
\hat{\mathcal{O}}_{pol}^g (0^+ , x^- , \ul{x}) \: V_{\ul{x}} [x^- , -\infty]
\end{align}
with the effective vertex $\hat{\mathcal{O}}_{pol}^{g}$ computed in
\eq{eq:noneik2} (see also \eq{M:VpolF12}). Here we have defined an
abbreviated notation for the light-cone Wilson line in the fundamental
representation,
\begin{align}
  V_{\un{x}} [b^-, a^-] = \mathcal{P} \exp \left[ i g
    \int\limits_{a^-}^{b^-} d x^- \, A^+ (x^+=0, x^-, {\un x})
  \right].
\end{align}

The small-$x$ limit of the quark helicity distribution \eqref{e:qTMD2}
corresponds to the large-$s$ limit of the polarized dipole amplitude
$G(x_{10}^2 , z s)$.  The evolution equations for the latter, derived
in \cite{Kovchegov:2015pbl}, resum double logarithms of the energy,
$\alpha_s \ln^2\tfrac{s}{\Lambda^2} \sim \alpha_s \ln^2\tfrac{1}{x}
\sim 1$.  Interestingly, in addition to the ``soft logarithm'' coming
from the longitudinal momentum integral which is also generated by the
unpolarized BFKL/BK/JIMWLK evolution, the polarized dipole amplitude
is especially sensitive to short-distance fluctuations about the
polarized Wilson line, generating an additional logarithm of energy
coming from the transverse momentum integration.  Preserving these
transverse logarithms of energy in the double-logarithmic
approximation (DLA) requires imposing a lifetime ordering constraint
on the successive steps of evolution, similar to the ``kinematical
improvements'' which become important in the unpolarized evolution at
NLO (see, for example, \cite{Beuf:2014uia}).  Like in the unpolarized
case, the small-$x$ evolution equations for the polarized dipole
amplitude lead to an infinite operator hierarchy, but simplify to a
closed set of equations in the large-$N_c$ limit, where $N_c$ is the
number of colors.  In the large-$N_c$ limit, with DLA accuracy, the
polarized evolution equations are
\cite{Kovchegov:2015pbl,Kovchegov:2016zex}
\begin{subequations} \label{e:oldevol1}
\begin{align}
  G(x_{10}^2 , z s) &= G^{(0)} (x_{10}^2 , z s) + \frac{\alpha_s
    N_c}{2\pi} \int\limits_{\frac{1}{x_{10}^2 s}}^{z} \frac{dz'}{z'}
  \int\limits_{\frac{1}{z' s}}^{x_{10}^2}\frac{dx_{21}^2}{x_{21}^2}
  \left[ \Gamma(x_{10}^2 , x_{21}^2 , z' s) + 3 G(x_{21}^2 , z' s)
  \right] , \label{Geq111}
\\
\Gamma(x_{10}^2 , x_{21}^2 , z' s) &= G^{(0)} (x_{10}^2 , z' s) +
\frac{\alpha_s N_c}{2\pi} \int\limits_{\frac{1}{x_{10}^2 s}}^{z'}
\frac{dz''}{z''} \int\limits_{\frac{1}{z'' s}}^{\min\left[ x_{10}^2 \,
    , \, x_{21}^2 \frac{z'}{z''} \right]} \frac{dx_{32}^2}{x_{32}^2}
\left[ \Gamma(x_{10}^2 , x_{32}^2 , z'' s) + 3 G(x_{32}^2 , z'' s)
\right], \label{Gameq111}
\end{align}
\end{subequations}
where $G^{(0)}$ are the initial conditions.  Because of the lifetime
ordering condition necessary to preserve the double-logarithmic
structure, the polarized dipole $G$ depends upon an auxiliary function
$\Gamma$, termed the ``neighbor dipole amplitude'', in which further
evolution is constrained by the lifetime of an adjacent dipole.  We
also note that, although nonlinear saturation corrections can be
incorporated straightforwardly, even at leading order they resum only
leading logarithms $\as \, \ln \tfrac{1}{x}$ and are beyond DLA
accuracy.  As such, the evolution equations \eqref{e:oldevol1} are the
quark helicity analog of the linear BFKL equation.

Equations~\eqref{e:oldevol1} were solved numerically in
\cite{Kovchegov:2016weo} and analytically in \cite{Kovchegov:2017jxc}
for the high-energy asymptotics yielding
\begin{subequations} \label{M:asymsol}
\begin{align}
  G(x_{10}^2 , z s) &= \frac{1}{3} G_0 \: (z s \, x_{10}^2)^{\alpha_h^q} \\
  \Gamma(x_{10}^2 , x_{21}^2 , z s) &= \frac{1}{3} G_0 \: (z s \,
  x_{21}^2)^{\alpha_h^q} \left[ 4
    \left(\frac{x_{10}^2}{x_{21}^2}\right)^{\frac{\alpha_h^q}{4}} - 3
  \right] ,
\end{align}
\end{subequations}
where the exponent of the energy, known as the ``quark helicity
intercept'' in analogy to the Pomeron intercept, is given by
\begin{align} 
  \label{M:ahel}
  \alpha_h^q = \frac{4}{\sqrt{3}} \sqrt{\frac{\alpha_s N_c}{2\pi}}
  \approx 2.31 \sqrt{\frac{\alpha_s N_c}{2\pi}} .
\end{align}
The numerical solution of \eqref{e:oldevol1} found in
\cite{Kovchegov:2016weo} possesses two features which are not
immediately obvious from the evolution equations \eqref{e:oldevol1}: a
negligible dependence on the initial conditions $G^{(0)}$ and an
emergent scaling behavior. The scaling behavior is an observation that
for
\begin{align} 
\label{e:scaleonset} 
z s > \frac{1}{x_{10}^2} \, e^{\zeta_0} , \hspace{1cm}
\zeta_0 \approx (1 \div 2) \sqrt{\frac{2\pi}{\alpha_s N_c}} ,
\end{align}
the polarized dipole and neighbor dipole become functions only of the
product of the energy and transverse distances, $G(x_{10}^2 , z s) = G
(z s x_{10}^2)$ and $\Gamma(x_{10}^2 , x_{21}^2 , z s) = \Gamma (z s
x_{10}^2 , z s x_{21}^2)$, rather than being dependent on each
variable (made dimensionless with the help of the IR cutoff $\Lambda$)
individually.  The coefficient $G_0$ in \eq{M:asymsol} is then the
``scaling initial condition'' for when this behavior sets in, or, more
precisely, the effective value of the inhomogeneous term $G^{(0)}$ at
the onset of scaling.  In \cite{Kovchegov:2017jxc}, $G_0$ was set to
$1$ as irrelevant for the determination of the intercept, but it is
useful to keep here for power-counting purposes.

The main purpose of this paper is to extend the analysis summarized
above for the quark helicity distribution to the gluon helicity
distribution.  We will proceed to derive a relation analogous to
\eqref{e:qTMD2} between the gluon helicity distribution and a
polarized dipole operator, derive its large-$N_c$ evolution equations
similar to \eqref{e:oldevol1} which employ the solution
\eqref{M:asymsol}, and obtain the gluon helicity intercept analogous
to \eqref{M:ahel}.

%
\subsection{The Gluonic Contribution to the Polarized Wilson Line
  Operator}
%

\begin{figure}[ht]
\begin{center}
\includegraphics[width= 0.6 \textwidth]{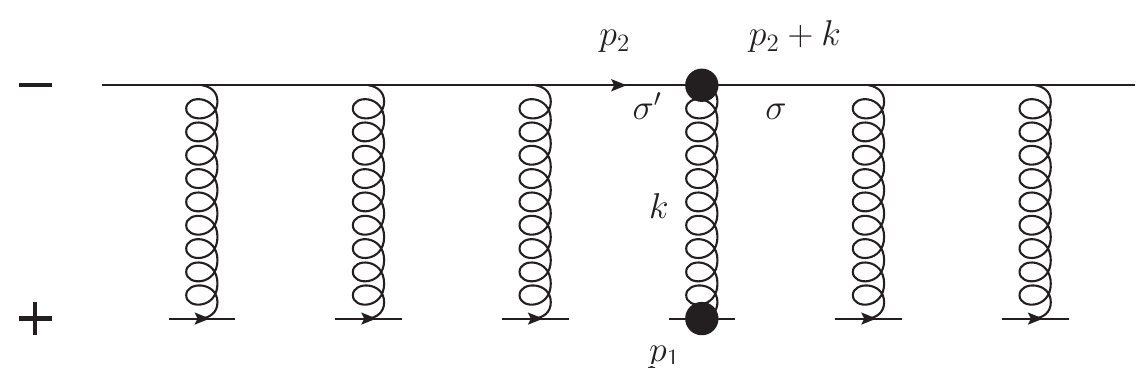} 
\caption{The polarized Wilson line \eqref{M:Vpol1} in the
  quasi-classical approximation in $A^-=0$ gauge. The filled circles
  denote the spin-dependent sub-eikonal scattering.}
\label{vpol}
\end{center}
\end{figure}

Before proceeding to the gluon helicity distribution, it is a useful
exercise to construct the operator $\hat{\mathcal{O}}_{pol}^{g}$
corresponding to $t-$channel gluon exchange in the polarized Wilson
line.  This will provide a valuable cross-check of the quark helicity
evolution equations \eqref{e:oldevol1} at the operator level later on.
We will evaluate \eq{M:Vpol1} directly by computing the
polarization-dependent propagator of a quark in the quasi-classical
background field of a heavy nucleus.  For consistency with
\eq{M:Vpol1}, we choose a frame in which the quark is moving in the
light-cone minus direction and the target is moving in the plus
direction, and we will work in the $A^- = 0$ gauge.  The sub-eikonal
vertex $\hat{\mathcal{O}}_{pol}^{g}$ carries polarization information,
while all other interactions are eikonal, as illustrated in
\fig{vpol}.  As usual, the Fourier transform to the longitudinal
coordinate $x^-$ puts the intermediate quark lines between scatterings
on mass shell \cite{Mueller:1989st,Kovchegov:2012mbw}.

%

With the momenta labeled as in \fig{vpol}, the non-eikonal vertex is
straightforward to compute:
\begin{align} 
  \label{eq:noneik}
  \sigma \, \delta_{\sigma , \sigma'} \: \hat{\mathcal{O}}_{pol}^{g}
  (k) \equiv - \frac{1}{2 p_2^-} \, {\bar u}_\sigma (p_2+k)
  \gamma^i_\perp u_{\sigma'} (p_2) \, i g \, A^i_\perp (k) = - \frac{i
    \, \sigma}{2 p_2^-} \, \delta_{\sigma \sigma'} \, \un{k} \times
  \un{A} (k) \, i g,
\end{align}
where we only keep the spin-dependent terms proportional to $\sigma$
and $A_\mu$ denotes the color matrix $A^{a}_\mu t^a$ with $t^a$ the
fundamental generators of SU($N_c$).  Fourier transforming to
coordinate space gives
\begin{align} \label{eq:noneik2} \hat{\mathcal{O}}_{pol}^{g} (x^- ,
  \ul{x}) &\equiv \int \frac{dk^+}{2\pi} \frac{d^2 k}{(2\pi)^2} \,
  e^{- i k^+ x^-} \, e^{i \ul{k} \cdot \ul{x}} \: \left[ \frac{g}{2
      p_2^-} \un{k} \times \un{A} (k) \right]
\notag \\ & =
\frac{1}{s} \, (- i g p_1^+) \: \epsilon_T^{i j} \:
\frac{\partial}{\partial x_\bot^i} \, A_\bot^j (x^-, \un{x}) \equiv
\frac{1}{s} \, (- i g p_1^+) \: \un{\nabla} \times \un{A} (x^-,
\un{x}),
\end{align}
where $s = 2 p_1^+ p_2^-$ is the center-of-mass energy of the
polarization-dependent interaction and $p_1^+$ is the momentum of the
polarized nucleon. We have defined $\un{\nabla} \equiv
(\partial/\partial x^1, \partial/\partial x^2)$ and the cross-product
$\un{u} \times \un{v} = \epsilon_T^{i j} u^i v^j = u^1 \, v^2 - u^2 \,
v^1$. Here $\epsilon_T^{12} = 1 = - \epsilon_T^{21}$ and
$\epsilon_T^{11} = \epsilon_T^{22} =0$ and Latin indices denote
transverse components of 4-vectors, $i,j = 1,2$.
We observe that $\ul{\nabla} \times \ul{A} = - (\partial_\bot^1
A_\bot^2 - \partial_\bot^2 A_\bot^1)$ is the negative of the Abelian
part of the field-strength tensor $F^{12}$.  In the $A^- = 0$ gauge we
are working in, the non-Abelian contribution $\sim [ A_\bot^1 ,
A_\bot^2 ]$ is further suppressed by an extra $1/s$, but will appear
in other gauges. We therefore conclude\footnote{We thank Ian Balitsky
  for this insight.} that the non-eikonal vertex $\hat{O}_{pol}^g$
when expressed in the most-general gauge-covariant form is
proportional to the gluon field-strength tensor $F^{1 2}$; we write
\begin{align} 
  \label{M:VpolF12}
  (V_{\ul x}^{pol})^g = \frac{i \, g \, p_1^+}{s}
  \int\limits_{-\infty}^{+\infty} dx^- \: V_{\ul{x}}[+\infty, x^-] \:
  F^{12} (x^+ =0 , x^- , \ul{x}) \: V_{\ul{x}} [x^- , -\infty].
\end{align}

It is also instructive to calculate the spin-dependent field $A_\bot^j
(x^- , \ul{x})$ explicitly for a quark target with helicity $S_L$ and
momentum $p_1^+$, giving
\begin{align} 
  \label{eq:field}
  A^{a \, i} (x^-, \un{x}) = \frac{g}{2 \pi} \, (t^a) \, S_L \,
  \delta_{S_L S_L^\prime} \, \frac{1}{2 p_1^+} \, \delta (x^-) \,
  \epsilon^{ij} \frac{x_\perp^j}{x_\perp^2}
\end{align}
for the transverse field entering \eq{eq:noneik2}.  The exact form of
\eqref{eq:field} is specific to the quark target model, but the
$1/p^+$ suppression is a general feature of the sub-eikonal
spin-dependent exchange. Hence we may write
\begin{align}
  \label{eq:scale_out}
  \un{A} (x^-, \un{x}) = \frac{S_L}{2 p_1^+} \, \un{\bar A} (x^-,
  \un{x})
\end{align}
to scale out the sub-eikonal suppression of the emission vertex in the
target and equivalently write the polarized Wilson line as
\begin{align} 
  \label{eq:Wpol2} 
  (V^{pol}_{\un x})^g & = \frac{1}{2 \, s} \,
  \int\limits_{-\infty}^\infty d x^- \, V_{\ul x} [+\infty, x^-] \:
  \left( - i g \, \epsilon_T^{i j} \: \frac{\partial}{\partial
      x_\bot^i} \, \bar{A}_\bot^j (x^-, \un{x}) \right) \: V_{\ul x}
  [x^- , -\infty] \\ & = \frac{i g}{2 \, s} \,
  \int\limits_{-\infty}^\infty d x^- \, V_{\ul x} [+\infty, x^-] \:
  \bar{F}^{12} (x^-, \un{x}) \: V_{\ul x} [x^- , -\infty], \notag
\end{align}
where $\bar{F}^{12} = (2 p_1^+/S_L) \, F^{12}$ and all of the energy
suppression is contained in the prefactor $1/s$ which is then scaled
out in the definition of the polarized dipole amplitude in
\eq{eq:Gdef}. We have also put $S_L = +1$, which will be our standard
assumption about the helicity of the target from now on, unless
specified otherwise by notation.

Employing \eq{eq:Wpol2} in \eq{eq:Gdef} we can finally write down an
explicit operator expression for the polarized dipole scattering
amplitude (in $A^- =0$ gauge):
\begin{align} 
\label{eq:Gdef22} 
G_{10} (z s) \equiv \frac{p_1^+}{2 N_c} \,
\int\limits_{-\infty}^\infty d x_1^- \, \left\langle \tr \left[ V_{\ul
      0} \:\: V_{\ul 1} [-\infty, x_1^-] \, \left( i g \,
      \epsilon_T^{i j} \: \frac{\partial}{\partial (x_1)_\bot^i}
      A_\bot^j (x_1^- , \ul{x}_1) \right) \, V_{\ul 1} [x_1^-, \infty]
  \right] + \cc \right\rangle (z s) .
\end{align}

%
\section{The Gluon Helicity TMDs and New Polarized Dipole
  Amplitude(s)}
\label{sec:operators}
%
%

The gluon helicity TMD is defined
\footnote{Note the differing normalizations and conventions, e.g. Refs.~
\cite{Mulders:2000sh, Dominguez:2011wm, Collins:1981uw, Ji:2005nu, Bomhof:2006dp}.}
similarly to \eqref{e:qTMD} as~\cite{Bomhof:2006dp}
\begin{align} 
\label{eq:TMDdef} 
g_{1L}^{G} (x, k_T^2) = \frac{-2 i}{x \, P^+} \, \half\sum_{S_L} S_L
\, \int \frac{d \xi^- \, d^2\xi}{(2 \pi)^3} \, e^{i x P^+ \, \xi^- - i
  \un{k} \cdot \un{\xi}} \, \bra{P, S_L} \epsilon_T^{ij} \, \tr \left[
  F^{+i} (0) \: {\cal U}[0,\xi] \: F^{+j} (\xi) \: {\cal
    U}^\prime[\xi, 0] \right] \ket{P, S_L}_{\xi^+ = 0}.
\end{align}
For gluon TMD distributions, the field strength operators are
connected by two fundamental gauge links $\mathcal{U} \, , \,
\mathcal{U}^\prime$ which may separately be either future-pointing
$([+])$ or past-pointing $([-])$, with
\begin{subequations}\label{U+-}
\begin{align}
\mathcal{U}^{[+]} [y, x] &\equiv 
\mathcal{P} \exp \! \left[ i g \!\! \int\limits_{+\infty}^{y^-} \!\! 
dz^- A^+ (0^+, z^-, \ul{y}) \right] \!
\mathcal{P} \exp \! \left[ - i g \! \int\limits_{\ul{x}}^{\ul{y}} \! 
d \ul{z} \cdot \ul{A} (0^+, +\infty^-, \ul{z}) \right] \!
\mathcal{P} \exp \! \left[ i g \!\! \int\limits_{x^-}^{+\infty} \!\! 
dz^- \, A^+ (0^+, z^-, \ul{x}) \right]
\\
\mathcal{U}^{[-]} [y, x] &\equiv \mathcal{P} \exp \! \left[ i g \!\!
  \int\limits_{-\infty}^{y^-} \!\!  dz^- A^+ (0^+, z^-, \ul{y})
\right] \!  \mathcal{P} \exp \! \left[ - i g \!
  \int\limits_{\ul{x}}^{\ul{y}} \!  d \ul{z} \cdot \ul{A} (0^+,
  -\infty^-, \ul{z}) \right] \!  \mathcal{P} \exp \! \left[ i g \!\!
  \int\limits_{x^-}^{-\infty} \!\!  dz^- \, A^+ (0^+, z^-, \ul{x})
\right] .
\end{align}
\end{subequations}
(The minus sign in the middle exponent in both equations \eqref{U+-}
is due to the metric.) Of particular interest are the ``dipole
distribution'' $g_{1L}^{G \, dip}$ for which one is future pointing
and the other is past pointing, $\mathcal{U} = \mathcal{U}^{[+]} \, ,
\, \mathcal{U}^\prime = \mathcal{U}^{[-]}$, and the
``Weizs\"{a}cker-Williams distribution'' $g_{1L}^{G \, WW}$ for which
both are future pointing, $\mathcal{U} = \mathcal{U}^{[+]} \, , \,
\mathcal{U}^\prime = \mathcal{U}^{[+]}$.

%
\subsection{Dipole Gluon Helicity TMD}
%

In this paper we will focus primarily on the ``dipole-type'' gluon
helicity distribution.  Starting with \eq{eq:TMDdef} with the
appropriate gauge links, we multiply and divide by a volume factor
$V^- = \int d^2 x \, dx^-$ and shift the operators in the matrix
element to write
\begin{align}
  g_{1L}^{G \, dip} (x, k_T^2) &= \frac{-2 i}{x \, P^+ V^-}
  \frac{1}{(2\pi)^3} \, \half\sum_{S_L} S_L \, \int d \xi^- \, d^2\xi
  \: d\zeta^- \, d^2\zeta \:\: e^{i x P^+ \, (\xi^- - \zeta^-)} \: e^{
    - i \un{k} \cdot (\un{\xi} - \ul{\zeta})}
\notag \\ & \hspace{1cm} \times
\bra{P, S_L} \epsilon_T^{ij} \, \tr \left[ F^{+i} (\zeta) \: {\cal
    U}^{[+]}[\zeta,\xi] \: F^{+j} (\xi) \: {\cal U}^{[-]} [\xi, \zeta]
\right] \ket{P, S_L}_{\zeta^+ = \xi^+ = 0}.
\end{align}
We next convert from the matrix element of a momentum-space eigenstate
to a wave packet which is localized in both impact parameter and
momentum space:
\begin{align} 
\label{M:CGCavg} 
\frac{1}{2 P^+ V^-} \bra{P, S_L} \cdots \ket{P, S_L} = \int d^2b \,
db^- \, \rho(\ul{b}, b^-) \: \bra{p, b, S_L} \cdots \ket{p, b, S_L}
\equiv \langle \cdots \rangle_{P, S_L}.
\end{align}
This procedure is standard in the color-glass-condensate framework and
is used to match the ``unintegrated gluon distribution'' and the gluon
TMD $f_1^g$ in the unpolarized sector \cite{Dominguez:2011wm,
  Sievert:2014psa}; it is also similar to the calculation of the TMDs
of a heavy nucleus in the quasi-classical approximation
\cite{Kovchegov:2015zha}.  Applying this to the dipole gluon helicity
TMD gives
\begin{align}   
\label{eq:TMDdef2}
g_{1L}^{G \, dip} (x, k_T^2) &= \frac{-4 i}{x} \frac{1}{(2\pi)^3} \,
\int d \xi^- \, d^2\xi \: d\zeta^- \, d^2\zeta \:\: e^{i x P^+ \,
  (\xi^- - \zeta^-)} \: e^{ - i \un{k} \cdot (\un{\xi} - \ul{\zeta})}
\notag \\ & \hspace{1cm} \times
\left\langle \epsilon_T^{ij} \, \tr \left[ F^{+i} (\zeta) \: {\cal
      U}^{[+]}[\zeta,\xi] \: F^{+j} (\xi) \: {\cal U}^{[-]} [\xi,
    \zeta] \right] \right\rangle ,
\end{align}
where we have again put $S_L = +1$ for simplicity and dropped the $P,S_L$ subscript off the angle brackets for brevity.

To go further, we need to specify a gauge; we will work in the $A^- =
0$ light-cone gauge, which is equivalent to the covariant gauge in the
quasi-classical approximation and is also convenient for including
logarithmic small-$x$ evolution.  In this gauge, the target field is localized in $x^-$ such that the transverse
segments of the staple-shaped gauge links $\mathcal{U}^{[\pm]}$ at
$x^- = \pm \infty$ do not contribute, leaving
\begin{align} 
\label{eq:TMDdef3} 
g_{1L}^{G \, dip} (x, k_T^2) &= \frac{-4 i}{x} \frac{1}{(2\pi)^3} \,
 \int d \xi^- \, d^2\xi \: d\zeta^- \, d^2\zeta
\:\: e^{i x P^+ \, (\xi^- - \zeta^-)} \: e^{ - i \un{k} \cdot
  (\un{\xi} - \ul{\zeta})}
\notag \\ & \hspace{1cm} \times
\epsilon_T^{ij} \, \left\langle \tr \left[ V_{\ul \zeta} [-\infty,
    \zeta^-] \: F^{+i} (\zeta) \: V_{\ul \zeta} [\zeta^-, +\infty]
    \:\: V_{\ul \xi} [+\infty, \xi^-] \: F^{+j} (\xi) \: V_{\ul \xi}
    [\xi^-, -\infty] \right] \right\rangle,
\end{align}
where we have used the cyclicity of the color trace.  For the
unpolarized gluon distribution, it is sufficient to replace the
field-strength tensors by their eikonal approximations, $F^{+ i }
\approx - \partial_\bot^i A^+$, but since the gluon helicity
distribution contains a sub-eikonal contribution, we must expand the
product of field-strength tensors to the first non-vanishing
sub-eikonal order:
\begin{align}
  & F^{+ i} (\zeta) \cdots F^{+ j} (\xi) = \! \Big( \partial^+ A_\bot^i (\zeta) - \partial^i A^+ (\zeta) - i g \, [A^+ (\zeta) \, , \, A_\bot^i (\zeta) ] \Big) \! \cdots \! \Big( \partial^+ A_\bot^j (\xi) - \partial^j A^+ (\xi) - i g \, [A^+ (\xi) \, , \, A_\bot^j (\xi) ] \Big) \\
  &\approx \left( \frac{\partial}{\partial \zeta^-} A_\bot^i (\zeta) -
    i g \, [A^+ (\zeta) \, , \, A_\bot^i (\zeta) ]\right) \cdots
  \left( \frac{\partial}{\partial \xi_\bot^j} A^+ (\xi) \right)
%
%
  + \left( \frac{\partial}{\partial \zeta_\bot^i} A^+ (\zeta) \right)
  \cdots \left( \frac{\partial}{\partial \xi^-} A_\bot^j (\xi) - i g
    \, [A^+ (\xi) \, , \, A_\bot^j (\xi) ]\right). \notag
\end{align}
We next convert the sub-eikonal part of the field-strength tensor
$F^{+ i} (\zeta)$ into a total derivative,
\begin{align}
  V_{\ul \zeta} [-\infty, \zeta^-] \left( \frac{\partial}{\partial
      \zeta^-} A_\bot^i (\zeta) - i g [ A^+ (\zeta) \, , \, A_\bot^i
    (\zeta) ] \right) V_{\ul \zeta} [\zeta^- , +\infty] =
  \frac{\partial}{\partial \zeta^-} \left( V_{\ul \zeta} [-\infty,
    \zeta^-] \: A_\bot^i (\zeta) \: V_{\ul \zeta} [\zeta^- , +\infty]
  \right) ,
\end{align}
which can then be integrated by parts to act on the Fourier factor and
generate a net factor of $+ i x P^+$.  In the same way, the
sub-eikonal part of the $F^{+ j} (\xi)$ field-strength tensor can be
converted into a net factor of $- i x P^+$ and the operator $A_\bot^j
(\xi)$.  After taking these derivatives, we can safely set $e^{i x P^+
  (\xi^- - \zeta^-)} \approx 1$ (thus neglecting higher powers of
$x \ll 1$), giving
\begin{align}   
  g_{1L}^{G \, dip} (x, k_T^2) &= 4 \, P^+ \frac{1}{(2\pi)^3} \,
   \int d \xi^- \, d^2\xi \: d\zeta^- \,
  d^2\zeta \:\: e^{ - i \un{k} \cdot (\un{\xi} - \ul{\zeta})} \:
  \epsilon_T^{ij}
\notag \\ & \hspace{1cm} \times
\Bigg\{ \left\langle \tr \left[ V_{\ul \zeta} [-\infty, \zeta^-] \:
    A^i (\zeta) \: V_{\ul \zeta} [\zeta^-, +\infty] \:\: V_{\ul \xi}
    [+\infty, \xi^-] \: \left( \frac{\partial}{\partial \xi_\bot^j} A^+ (\xi) \right)
    \: V_{\ul \xi} [\xi^-, -\infty] \right] \right\rangle
\notag \\ & \hspace{1.5cm} -
\left\langle \tr \left[ V_{\ul \zeta} [-\infty, \zeta^-] \:
    \left( \frac{\partial}{\partial \zeta_\bot^i} A^+ (\zeta) \right) \: V_{\ul
      \zeta} [\zeta^-, +\infty] \:\: V_{\ul \xi} [+\infty, \xi^-] \:
    A^j (\xi) \: V_{\ul \xi} [\xi^-, -\infty] \right] \right\rangle
\Bigg\} .
\end{align}
We can now similarly convert the eikonal parts of the field-strength
tensors into total derivatives,
\begin{align}
  \int\limits_{-\infty}^\infty d\zeta^- \, V_{\ul \zeta} [-\infty,
  \zeta^-] \, \left( \frac{\partial}{\partial \zeta_\bot^i} A^+ (\zeta) \right) \,
  V_{\ul \zeta} [\zeta^-, +\infty] = \frac{i}{g}
  \frac{\partial}{\partial \zeta_\bot^i} \, V_{\ul \zeta} [-\infty,
  +\infty] ,
\end{align}
which absorbs the $d\zeta^-$ integral from the TMD and can be
integrated by parts to generate a net factor of $\frac{1}{g}
k_\bot^i$:
\begin{align}   
  g_{1L}^{G \, dip} (x, k_T^2) &= \frac{-4}{g (2\pi)^3} \, P^+ \,
   \int d^2\xi \, d^2\zeta \:\: e^{ - i \un{k}
    \cdot (\un{\xi} - \ul{\zeta})} \: k_\bot^i \epsilon_T^{i j}
\notag \\ & \hspace{1cm} \times
\Bigg\{ \left\langle \tr \left[ \left( \int d\zeta^- \, V_{\ul \zeta}
      [-\infty, \zeta^-] \: A^j (\zeta) \: V_{\ul \zeta} [\zeta^-,
      +\infty] \right) \:\: V_{\ul \xi} [+\infty, -\infty] \right]
\right\rangle
\notag \\ & \hspace{1.5cm} +
\left\langle \tr \left[ V_{\ul \zeta} [-\infty, +\infty] \:\: \left(
      \int d\xi^- \, V_{\ul \xi} [+\infty, \xi^-] \: A^j (\xi) \:
      V_{\ul \xi} [\xi^-, -\infty] \right) \right] \right\rangle
\Bigg\} ,
\end{align}
where we also swapped $i \leftrightarrow j$ in the first term.

We observe that the sub-eikonal gluon vertex enters in a form similar
to \eq{M:Vpol1}, but with an explicit transverse index.  Defining the
analogous polarized Wilson line (one may call it the polarized Wilson
line of the second kind to distinguish it from \eq{eq:Wpol2})
\begin{align} \label{M:Vpol2} (V_{\ul x}^{pol})_\bot^i &\equiv
  \int\limits_{-\infty}^{+\infty} dx^- \, V_{\ul x} [+\infty, x^-] \:
  \left( i g \, P^+ \, A_\bot^i (x) \right) \: V_{\ul x} [x^- ,
  -\infty]
  \notag \\ &= \half \int\limits_{-\infty}^{+\infty} dx^- \, V_{\ul x}
  [+\infty, x^-] \: \left( i g \, \bar{A}_\bot^i (x) \right) \:
  V_{\ul x} [x^- , -\infty]
\end{align}
allows us to write the dipole gluon helicity TMD in a more compact
form (compare this with a very similar Eq.~(47) in
\cite{Hatta:2016aoc})
\begin{align}   
  g_{1L}^{G \, dip} (x, k_T^2) &= \frac{-4 i}{g^2 (2\pi)^3} \,
   \int d^2\xi \, d^2\zeta \:\: e^{ - i \un{k}
    \cdot (\un{\xi} - \ul{\zeta})} \: k_\bot^i \epsilon_T^{i j} \:
  \Bigg\{ \left\langle \tr \left[ V_{\ul \xi} \, (V_{\ul \zeta}^{pol
        \, \dagger})_\bot^j \right] \right\rangle - \left\langle \tr
    \left[ (V_{\ul \xi}^{pol})_\bot^j \, V_{\ul \zeta}^\dagger \right]
  \right\rangle \Bigg\} ,
\end{align}
where, for brevity, we have also dropped the explicit integration
limits from the infinite unpolarized Wilson lines.
Swapping $\ul{\zeta} \leftrightarrow \ul{\xi}$ in the last term
generates a minus sign and makes the two terms in braces complex
conjugates of one another.  Relabeling the dummy integration variables
$\ul{\zeta}$ and $\ul{\xi}$ as $\ul{x}_1$ and $\ul{x}_0$,
respectively, and changing variables to $d^2 x_0 \, d^2 x_1 = d^2
x_{10} \, d^2 b_{10}$ with $\ul{b}_{10} \equiv \thalf (\ul{x}_1 +
\ul{x}_0)$ the impact parameter, we can write
\begin{align}   
\label{M:dip1}
g_{1L}^{G \, dip} (x, k_T^2) &= \frac{- 4 i}{g^2 (2\pi)^3} \,
 \int d^2 x_{10} \, d^2 b_{10} \:\: e^{+ i
  \un{k} \cdot \un{x}_{10}} \: k_\bot^i \epsilon_T^{i j} \: \Bigg\{
\left\langle \tr \left[ V_{\ul 0} \, (V_{\ul 1}^{pol \,
      \dagger})_\bot^j \right] \right\rangle + \cc \Bigg\} .
\end{align}
Defining another dipole-like polarized operator
\begin{align}   \label{eq:Gidef}
G^i_{10} (z s) \equiv \frac{1}{2 N_c} \,  
\left\langle \tr \left[ V_{\ul 0} (V_{\ul 1}^{pol \, \dagger})_\bot^i
\right] + \cc \right\rangle (z s)
\end{align}
we rewrite the dipole gluon helicity TMD as
\begin{align}   
  g_{1L}^{G \, dip} (x, k_T^2) &= \frac{- 8 i \, N_c}{g^2 (2\pi)^3} \,
  \int d^2 x_{10} \, e^{i \un{k} \cdot \un{x}_{10}} \: k_\bot^i
  \epsilon_T^{i j} \: \left[ \int d^2 b_{10} \, G_{10}^j (z s =
    \tfrac{Q^2}{x}) \right] .
\end{align}
The dipole gluon helicity TMD is related to an operator which is,
surprisingly, different from the polarized dipole amplitude in
\eq{eq:Gdef22}. This is very different from the situation with the
unpolarized gluon TMDs for which the dipole gluon TMD was related to
the (unpolarized adjoint) dipole scattering amplitude on the target
proton or nucleus \cite{Dominguez:2011wm}. This relation gave rise to
the ``dipole'' designation of this TMD. Here we see that this relation
is not universal and is not valid for the dipole gluon helicity TMD,
therefore putting the designation in question as well.

After the integration over all impact parameters, the new polarized
dipole amplitude is a vector-valued function of $\ul{x}_{10}$ alone,
allowing us to write the decomposition
\begin{align}\label{decomp}
  \int d^2 b_{10} \, G^{i}_{10} (z s) = (x_{10})_\bot^i \, G_1
  (x_{10}^2, z s) + \epsilon_T^{ij} \, (x_{10})_\bot^j \, G_2
  (x_{10}^2, z s) .
\end{align} 
By further writing $(x_{10})_\bot^i$ as a derivative $- i
\frac{\partial}{\partial k_\bot^i}$ on the Fourier factor, we see that
the scalar function $G_1$ does not contribute to the dipole gluon
helicity TMD, leaving only
\begin{align} 
  \label{M:dipdef}
  g_{1L}^{G \, dip} (x, k_T^2) &= \frac{8 i \, N_c}{g^2 (2\pi)^3}
  \:\: \int d^2 x_{10} \, e^{i \un{k} \cdot \un{x}_{10}} \, \ul{k}
  \cdot \ul{x}_{10} \: G_2 (x_{10}^2, z s = \tfrac{Q^2}{x})
\notag \\ &=
\frac{N_c}{2 \pi^4 \alpha_s} \: k_T^2 \frac{\partial}{\partial k_T^2}
\, \left[ \int d^2 x_{10} \, e^{i \un{k} \cdot \un{x}_{10}} \: G_2
  (x_{10}^2, z s = \tfrac{Q^2}{x}) \right].
\end{align}
For future purposes, it is also useful to convert the derivatives back
into coordinate space, writing
\begin{align} 
  \label{M:Gdipcoord}
  g_{1L}^{G \, dip} (x, k_T^2) 
 &= \frac{1}{\alpha_s \, 8\pi^4} \,
   \int d^2 x_0 \, d^2 x_1 \, e^{i \ul{k} \cdot
    \ul{x}_{10}} \: \epsilon_T^{i j} \: \left\langle \tr \left[
      (V_{\ul 1}^{pol})_\bot^i \, \left(\frac{\partial}{\partial
          (x_0)_\bot^j} V_{\ul 0}^\dagger \right) \right] + \cc
  \right\rangle
 \notag \\ &=
  \frac{- N_c}{\alpha_s \, 2\pi^4} \int d^2 x_{10} \, e^{i \ul{k} \cdot
    \ul{x}_{10}} \, \left[ 1 + x_{10}^2 \frac{\partial}{\partial
      x_{10}^2} \right] G_2 (x_{10}^2 , z s = \tfrac{Q^2}{x}).
\end{align}

We have thus expressed the dipole gluon helicity TMD in terms of a
polarized dipole operator; Eqs.~\eqref{M:dipdef} and
\eqref{M:Gdipcoord} should be compared with \eq{e:qTMD2} from the
quark helicity TMD.  Unexpectedly, however, the polarized dipole
operator \eqref{eq:Gidef} which determines the dipole gluon helicity
TMD is different from the polarized dipole amplitude \eqref{eq:Gdef22}
which determines the quark helicity TMD.  Comparing the underlying
polarized Wilson lines, we see that the quark case \eqref{eq:Wpol2} is
sensitive to a {\textit{local}} derivative $\ul{\nabla} \times \ul{A}
(x^-)$ reflecting spin-dependent coupling at some point in the
propagation through the target.  On the other hand, the gluon case
\eqref{M:dip1} is sensitive to a {\textit{total}} derivative $\ul{k}
\times \ul{V}^{pol} \rightarrow \ul{\nabla} \times \ul{V}^{pol}$
reflecting an overall circular polarization which remains after the
entire interaction with the target.  In principle, it would seem that
quark helicity and gluon helicity are very different quantities, with
the gluon helicity requiring not only that a spin-dependent scattering
take place but also that the circular-polarized structure survive the
rest of the rescattering.  We will thus need to derive new evolution
equations analogous to \eq{e:oldevol1} for the new polarized dipole
amplitude $G_2$ in order to determine the small-$x$ asymptotics of the
dipole gluon helicity distribution.

%
\subsection{Weizs\"{a}cker-Williams Gluon Helicity TMD}
%

For completeness and further comparison, we will also evaluate the
``Weizs\"{a}cker-Williams (WW) gluon helicity TMD'':
\begin{align} \label{eq:TMDWW1} g_{1L}^{G \, WW} (x, k_T^2) &=
  \frac{-4 i}{x} \frac{1}{(2\pi)^3} \,  \int d
  \xi^- \, d^2\xi \: d\zeta^- \, d^2\zeta \:\: e^{i x P^+ \, (\xi^- -
    \zeta^-)} \: e^{ - i \un{k} \cdot (\un{\xi} - \ul{\zeta})}
\notag \\ & \hspace{1cm} \times
\left\langle \epsilon_T^{ij} \, \tr \left[ F^{+i} (\zeta) \: {\cal
      U}^{[+]}[\zeta,\xi] \: F^{+j} (\xi) \: {\cal U}^{[+]} [\xi,
    \zeta] \right] \right\rangle .
\end{align}
Because both gauge links are now future-pointing, it is possible to
choose a gauge in which the WW gluon distributions possess a simple
partonic interpretation; specifically, we choose the $A^+ =0$
light-cone (LC) gauge with the $\un{\nabla} \cdot \un{A} (x^- = +
\infty) =0$ sub-gauge condition (see \cite{Chirilli:2015fza} for a
discussion of the LC gauge and its sub-gauges).\footnote{Throughout this subsection, 
we denote the fields in the $A^+ = 0 \: ,
\: \un{\nabla} \cdot \un{A} (x^- = + \infty) =0$ gauge with the
explicit subscript ``$LC$''; fields without explicit subscripts
correspond to the $A^- = 0$ gauge used elsewhere in this paper.}  With this choice, the
gauge links are unity on both the light-like segments and on the
transverse segments at $x^- = +\infty$ (with the physical content of
the gauge links having been encoded in the boundary at $x^- =
-\infty$), and we also have $F^{+ i} = \partial^+ A^i_{LC}$.
Integrating the derivatives by parts in the usual way gives
\begin{align}   
\label{eq:TMDWW2}
g_{1L}^{G \, WW} (x, k_T^2) &= \frac{-4 i }{(2\pi)^3} \, x (P^+)^2 \,
 \int d \xi^- \, d^2\xi \: d\zeta^- \, d^2\zeta
\:\: e^{i x P^+ \, (\xi^- - \zeta^-)} \: e^{ - i \un{k} \cdot
  (\un{\xi} - \ul{\zeta})} \: \left\langle \epsilon_T^{ij} \, \tr
  \left[ A^i_{LC} (\zeta) \: A^j_{LC} (\xi) \right] \right\rangle .
\end{align}
From here, the rest of the calculation is similar to the standard
textbook treatment of the unpolarized WW gluon distribution
\cite{Kovchegov:2012mbw}.  We first determine the explicit gauge
transformation which achieves the form of \eq{eq:TMDWW2} in terms of
the fields in the $A^- = 0$ or covariant gauge we have used elsewhere.
The desired gauge condition
\begin{align}
  0 = A^+_{LC} = S A^+ S^{-1} - \frac{i}{g} (\partial^+ S) \, S^{-1},
\end{align}
and sub-gauge condition $\un{\nabla} \cdot \un{A}_{LC} (x^- = +
\infty) =0$ \cite{Kovchegov:1996ty,Kovchegov:1997pc} are easily seen
to be satisfied by the gauge transformation
\begin{align}\label{eq:Solution}
  S (x) = \mathcal{P} \exp \left\{ i g \int\limits_{x^-}^{+\infty} d
    x^- \, A^+ (x^-, {\un x}) \right\} = V_{\un x} [+\infty, x^-].
\end{align}

The transverse components $A_{LC}^i$ we need are given by
\begin{align}\label{Ai}
A^i_{LC} = S A_\bot^i S^{-1} - \frac{i}{g} (\partial^i S) \, S^{-1},
\end{align}
where in the eikonal approximation we would normally neglect the first
term compared to the second term on the right-hand side.  But for the gluon helicity, we must
keep the first sub-eikonal polarization-dependent correction to the
product of the two fields, which enters \eq{Ai} through $A_\bot^i$:
\begin{align}
  A_{LC}^i (\zeta) \, A_{LC}^j (\xi) &\approx \frac{i}{g} \: \left(
    V_{\ul \zeta} [+\infty, \zeta^-] \, A_\bot^i (\zeta) \, V_{\ul
      \zeta} [\zeta^- , +\infty] \right) \left(
    \frac{\partial}{\partial \xi_\bot^j} V_{\ul \xi} [+\infty, \xi^-]
  \right) V_{\ul \xi} [\xi^- , +\infty]
\notag \\ & \hspace{1cm} + \frac{i}{g} \left( \frac{\partial}{\partial
    \zeta_\bot^i} V_{\ul \zeta} [+\infty, \zeta^-] \right) V_{\ul
  \zeta} [\zeta^- , +\infty] \left( V_{\ul \xi} [+\infty, \xi^-] \,
  A_\bot^j (\xi) \, V_{\ul \xi} [\xi^- , +\infty] \right) .
\end{align}
In the small-$x$ limit, the longitudinal coordinate integrals are
\begin{align}
  \int\limits_{-\infty}^\infty d\zeta^- \, e^{i x P^+ \zeta^-} \,
  V_{\ul \zeta} [+\infty, \zeta^-] \, A_\bot^i (\zeta) \, V_{\ul
    \zeta} [\zeta^- , +\infty] &\approx \int\limits_{-\infty}^\infty
  d\zeta^- \, V_{\ul \zeta} [+\infty, \zeta^-] \, A_\bot^i (\zeta) \,
  V_{\ul \zeta} [\zeta^- , +\infty]
\notag \\ &=
\frac{-i}{g P^+} (V_{\ul \zeta}^{pol})_\bot^i \: V_{\ul \zeta}^\dagger =
\frac{i}{g P^+} V_{\ul \zeta}  \: (V_{\ul \zeta}^{pol \, \dagger})_\bot^i
\end{align}
and
\begin{align}
  \int\limits_{-\infty}^\infty d\xi^- \, e^{i x P^+ \xi^-} \, & \left(
    \frac{\partial}{\partial \xi_\bot^j} V_{\ul \xi} [+\infty, \xi^-]
  \right) V_{\ul \xi} [\xi^- , +\infty] =
\notag \\ & \hspace{1cm} =
\int\limits_{-\infty}^{+\infty} d\xi^- \, e^{i x P^+ \xi^-} \,
\int\limits_{\xi^-}^{+\infty} dz^- \, V_{\ul \xi} [+\infty, z^-]
\left( i g \, \frac{\partial}{\partial \xi_\bot^j} A^+ (0^+, z^- ,
  \ul{\xi}) \right) V_{\ul \xi} [z^- , +\infty]
\notag \\ & \hspace{1cm} =
\int\limits_{-\infty}^{+\infty} dz^- \, \left[
  \int\limits_{-\infty}^{z^-} d\xi^- \, e^{i x P^+ \xi^-} \right] \,
V_{\ul \xi} [+\infty, z^-] \left( i g \, \frac{\partial}{\partial
    \xi_\bot^j} A^+ (0^+, z^- , \ul{\xi}) \right) V_{\ul \xi} [z^- ,
+\infty]
\notag \\ & \hspace{1cm} \approx
\frac{-i}{x P^+} \int\limits_{-\infty}^{+\infty} dz^- \, V_{\ul \xi}
[+\infty, z^-] \left( i g \, \frac{\partial}{\partial \xi_\bot^j} A^+
  (0^+, z^- , \ul{\xi}) \right) V_{\ul \xi} [z^- , +\infty]
\notag \\ & \hspace{1cm} =
\frac{-i}{x P^+} \left( \frac{\partial}{\partial \xi_\bot^j} V_{\ul
    \xi} \right) \, V_{\ul \xi}^\dagger = \frac{+i}{x P^+} V_{\ul \xi}
\left( \frac{\partial}{\partial \xi_\bot^j} V_{\ul \xi}^\dagger
\right),
\end{align}
where we have expanded the exponent to the first non-vanishing term.
Inserting all of these expressions into \eq{eq:TMDWW2} gives
\begin{align}
  g_{1L}^{G \, WW} (x, k_T^2) &= \frac{4}{g^2 (2\pi)^3} \,
   \int d^2\xi \, d^2\zeta \: e^{ - i \un{k}
    \cdot (\un{\xi} - \ul{\zeta})} \: \epsilon_T^{ij}
\notag \\ & \hspace{1cm} \times
\left\langle \tr \left[ (V_{\ul \zeta}^{pol})_\bot^i \, V_{\ul
      \zeta}^\dagger \:\: V_{\ul \xi} \left( \frac{\partial}{\partial
        \xi_\bot^j} V_{\ul \xi}^\dagger \right) \right] - \tr \left[
    \left( \frac{\partial}{\partial \zeta_\bot^i} V_{\ul \zeta}
    \right) \, V_{\ul \zeta}^\dagger \: \: V_{\ul \xi} \, (V_{\ul
      \xi}^{pol \, \dagger})_\bot^j \right] \right\rangle.
\end{align}
Swapping $\ul{\zeta} \leftrightarrow \ul{\xi}$ and $i \leftrightarrow
j$ in the second term makes it the complex conjugate of the first
term.  Relabeling the dummy integration variables $\ul \zeta$ and $\ul
\xi$ as $\ul{x}_1$ and $\ul{x}_0$, respectively, and changing
variables to $d^2 x_0 \, d^2 x_1 = d^2 x_{10} \, d^2 b_{10}$ with
$\ul{b}_{10} = \thalf (\ul{x}_1 + \ul{x}_0)$ the impact parameter, we
can write
\begin{align} 
  \label{M:GWWcoord} 
  g_{1L}^{G \, WW} (x, k_T^2) &= \frac{4}{g^2 (2\pi)^3} \,
   \int d^2 x_{10} \, d^2 b_{10} \: e^{i \un{k}
    \cdot \ul{x}_{10}} \: \epsilon_T^{ij} \: \left\langle \tr \left[
      (V_{\ul 1}^{pol})_\bot^i \, V_{\ul 1}^\dagger \:\: V_{\ul 0}
      \left( \frac{\partial}{\partial (x_0)_\bot^j} V_{\ul 0}^\dagger
      \right) \right] + \cc \right\rangle.
\end{align}

It seems that the WW gluon helicity TMD is determined by yet another
polarized dipole-like operator
\begin{align} 
\label{eq:Gijdef}
G_{10}^{j i} (z s) &\equiv \frac{-1}{2 N_c} \, \left\langle \tr \left[
    (V_{\ul 1}^{pol})_\bot^i \, V_{\ul 1}^\dagger \:\: V_{\ul 0}
    \left( \frac{\partial}{\partial (x_0)_\bot^j} V_{\ul 0}^\dagger
    \right) \right] + \cc \right\rangle (z s)
\end{align}
which is a rank-2 tensor in the transverse plane.  After integration
over impact parameters, we can correspondingly define a scalar
function
\begin{align}
  G_3 (x_{10}^2 , z s) &\equiv \int d^2 b_{10} \, \epsilon_T^{i j} \,
  G_{10}^{j i} (z s)
\notag \\ &=
\frac{-1}{2 N_c} \, \int d^2 b_{10} \, \epsilon_T^{i j} \,
\left\langle \tr \left[ (V_{\ul 1}^{pol})_\bot^i \, V_{\ul 1}^\dagger
    \:\: V_{\ul 0} \left( \frac{\partial}{\partial (x_0)_\bot^j}
      V_{\ul 0}^\dagger \right) \right] + \cc \right\rangle (z s)
\end{align}
in terms of which the WW gluon helicity TMD is written
\begin{align} 
\label{M:WWdef}
g_{1L}^{G \, WW} (x, k_T^2) &= \frac{- N_c}{4 \pi^4 \alpha_s} \int d^2
x_{10} \, e^{i \un{k} \cdot \ul{x}_{10}} \: G_3 (x_{10}^2 , z s =
\tfrac{Q^2}{x}).
\end{align}

We have now expressed the Weizs\"{a}cker-Williams gluon helicity TMD
as well in terms of a yet another new polarized dipole operator;
\eq{M:WWdef} for the WW gluon helicity distribution is directly
comparable to \eq{M:dipdef} for the dipole gluon helicity distribution
and \eq{e:qTMD2} for the quark helicity distribution.  The polarized
dipole operator \eqref{eq:Gijdef} for the WW gluon helicity
distribution is different still from both the operator
\eqref{eq:Gidef} for the dipole gluon helicity distribution and the
amplitude \eqref{eq:Gdef22} for the quark helicity distribution.
Although the WW gluon helicity distribution is built from the same
polarized Wilson line \eqref{M:Vpol2} as the dipole gluon helicity
distribution, it is incorporated into a more complicated operator due
to the future-pointing structure of the WW gauge links: this feature
is similar to the unpolarized WW gluon TMD, which is related to the
color quadrupole operator instead of a dipole
\cite{Dominguez:2011wm,Dominguez:2011gc}.

%
\section{Operator Evolution Equations At Small {\bf \it  x}}
\label{sec:evolution}
%

Having constructed the appropriate polarized dipole amplitudes for the
dipole gluon helicity distribution \eqref{eq:Gidef} and
Weizs\"{a}cker-Williams gluon helicity distribution \eqref{eq:Gijdef},
we will now proceed to derive small-$x$ evolution equations, focusing
on the dipole distribution.  We will do this at the operator level
using a procedure which is similar in spirit (although different in
gauge) to the background field method employed in
\cite{Balitsky:1995ub}.

Beginning with the operator definitions of the polarized Wilson lines
and dipole amplitudes, we will separate the gauge fields $A^\mu$ of
the target into ``classical'' fields $A^\mu_{cl}$ and ``quantum''
fields $a^{\mu}$:
\begin{align} \label{e:Bkgd1}
A^\mu (x) = A^{\mu}_{cl} (x) + a^\mu (x).
\end{align}
This separation can be done using a rapidity regulator $\eta$, such
that the ``fast'' quantum fields have rapidities greater than $\eta$,
while the ``slow'' classical fields have rapidities less than $\eta$
and are effectively frozen from the point of view of the quantum
fluctuations. (Here ``greater'' and ``smaller'' rapidities depend on
the choice of a coordinate system, and may be interchanged.)  This is
essentially the rapidity factorization approach used in
\cite{Balitsky:2015qba}, and the evolution equations we will derive
can be understood as renormalization group equations in the rapidity
cutoff $\eta$.  The classical fields of the target, being enhanced by
the target density, will be resummed to all orders.  These classical
fields (in the $A^- = 0$ light-cone gauge) are localized in $x^-$ to a
parametrically small window, which we choose to be centered on the
origin: $x^- \in [- R^- , + R^-] \sim [ - \frac{1}{p^+} , +
\frac{1}{p^+} ]$, with $p^+$ the large momentum of the target.
Although the classical fields are Lorentz-contracted to a delta
function at $x^- = 0$, the quantum fields can extend far beyond the
target; we will calculate the first correction due to these quantum
fields in perturbation theory.

As a warm-up exercise and as a cross-check of our previous work
\cite{Kovchegov:2015pbl}, we will first employ this method to rederive
the evolution equations for the polarized dipole amplitude
\eqref{eq:Gdef} (or \eqref{eq:Gdef22}) which governs the quark helicity distribution at small
$x$.  We will then repeat this exercise to derive new evolution
equations for the polarized dipole amplitude \eqref{eq:Gidef} which
governs the dipole gluon helicity distribution.  We leave the
corresponding evolution equations for the Weizs\"{a}cker-Williams
gluon helicity distribution for future work, although we note that the
small-$x$ asymptotics of both gluon helicity distributions must
coincide.

%
\subsection{Evolution of the Polarized Dipole Operator For Quark
  Helicity}
%

%
\begin{figure}[th]
\begin{center}
\includegraphics[width= 0.7 \textwidth]{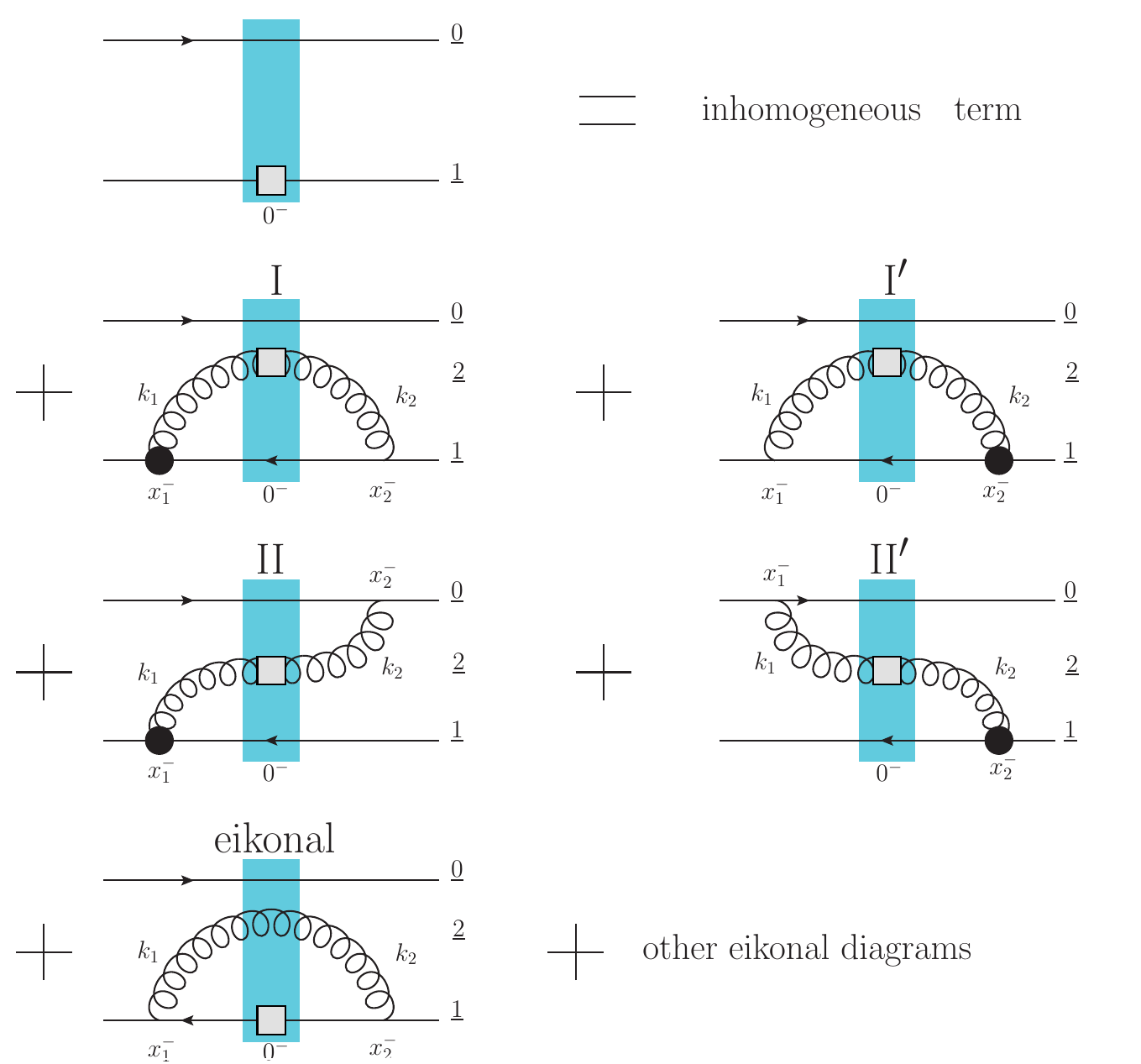} 
\caption{Diagrams illustrating contractions \eqref{e:qcontractions}
  contributing to the evolution of the polarized dipole amplitude
  \eqref{eq:Gdef2} for the quark helicity distribution.  The blue band
  represents the classical fields (shock wave), the black vertex
  represents the sub-eikonal operator insertion \eqref{eq:noneik2},
  and the gray box represents the polarized Wilson line.}
\label{fig:Gevol_v2}
\end{center}
\end{figure}
%
%

We begin with the polarized dipole amplitude for the quark helicity
distribution \eq{eq:Gdef}, using the explicit operator form
\eqref{eq:Wpol2} for the polarized Wilson line (cf. \eq{eq:Gdef22}):
\begin{align} 
\label{eq:Gdef2} 
G_{10} (z s) \equiv \frac{p^+}{2 N_c} \, 
\int\limits_{-\infty}^\infty d x_1^- \, \left\langle \tr \left[ V_{\ul
      0} \:\: V_{\ul 1} [-\infty, x_1^-] \, \left( i g \,
      \epsilon_T^{i j} \: \frac{\partial}{\partial (x_1)_\bot^i}
      A_\bot^j (x_1^- , \ul{x}_1) \right) \, V_{\ul 1} [x_1^-, \infty]
  \right] + \cc \right\rangle (z s) .
\end{align}
Because this operator contains only $t$-channel gluon exchange, it
will not couple directly to soft quarks.  This procedure will
therefore only test the gluon emission sector of the quark helicity
evolution equations, but this is precisely what is needed to verify
the evolution equations in the large-$N_c$ limit.

As in \eq{e:Bkgd1}, we first expand the gauge fields into classical
and quantum components, both in the Wilson lines and in the explicit
operator insertion.  We then keep the first quantum correction to the
classical background by contracting two of the quantum fields to form
a quantum propagator in the background of the classical
fields.\footnote{ One may note a subtlety of this procedure: strictly,
  the fields must be time-ordered in order to apply Wick's theorem and
  form contractions.  The fields entering the operators here are not
  time-ordered but rather all sit at $x^+ = 0$, which plays the role
  of time in light-front quantization.  Time ordering may be achieved
  by inserting a complete set of ``out'' states, as in
  \cite{Dominguez:2011wm}, although the resulting Schwinger-Keldysh
  ordering is still different from the forward-scattering time
  ordering implicit in the background field method.  The equivalence
  between these two time-ordered structures was verified in
  \cite{Mueller:2012bn} up to next-to-leading order, which is more
  than sufficient precision for our purposes here.  }
We may distinguish the following classes of contractions shown
diagrammatically in \fig{fig:Gevol_v2}: ``polarized ladder'' emissions
(I and I${}^\prime$) in which a polarized gluon is emitted and
absorbed by line 1; ``polarized non-ladder'' emissions (II and
II${}^\prime$) in which a polarized gluon is exchanged between lines 1
and 0; and unpolarized gluon emissions (dubbed ``eikonal'' in
\fig{fig:Gevol_v2}).  As visualized in Fig.~\ref{fig:Gevol_v2}, these
contractions are
\begin{subequations} \label{e:qcontractions}
\begin{align}
\mathrm{I}: \hspace{1cm}&
\contraction[2ex]
{\tr\Big[ V_{\ul 0} \:\: V_{\ul 1} [-\infty, x_1^- ] \: \ul{\nabla}\times}
{\ul{a}}
{(x_1^- , \ul{x}_1) \:}
{V_{\ul 1}}
\tr\Big[ V_{\ul 0} \:\: V_{\ul 1} [-\infty, x_1^- ] \: 
\ul{\nabla}\times \ul{a} (x_1^- , \ul{x}_1) \:
V_{\ul 1} [x_1^- , \infty] \Big]
\\
\mathrm{I}': \hspace{1cm}&
\contraction[2ex]
{\tr\Big[ V_{\ul 0} \:\:}
{V_{\ul 1}}
{[-\infty, x_1^- ] \: \ul{\nabla}\times}
{\ul{a}}
\tr\Big[ V_{\ul 0} \:\: V_{\ul 1} [-\infty, x_1^- ] \: 
\ul{\nabla}\times \ul{a} (x_1^- , \ul{x}_1) \:
V_{\ul 1} [x_1^- , \infty] \Big]
\\ \notag \\
\mathrm{II} + \mathrm{II}': \hspace{1cm}&
\contraction[2ex]
{\tr\Big[}
{V_{\ul 0}}
{\:\: V_{\ul 1} [-\infty, x_1^- ] \: \ul{\nabla}\times}
{\ul{a}}
\tr\Big[ V_{\ul 0} \:\: V_{\ul 1} [-\infty, x_1^- ] \: 
\ul{\nabla}\times \ul{a} (x_1^- , \ul{x}_1) \:
V_{\ul 1} [x_1^- , \infty] \Big]
\\ \notag \\
\mathrm{eikonal}: \hspace{1cm}&
\contraction[2ex]
{\tr\Big[}
{V}
{}
{ {}_{\ul 0} }
\tr\Big[ V {}_{\ul 0} \:\: V_{\ul 1} [-\infty, x_1^- ] \: 
\ul{\nabla}\times \hat{\ul{A}}_{cl} (x_1^- , \ul{x}_1) \:
V_{\ul 1} [x_1^- , \infty] \Big]
\notag \\ &+
\contraction[2ex]
{\tr\Big[ }
{V_{\ul 0}}
{\:\:}
{V_{\ul 1}}
\tr\Big[ V_{\ul 0} \:\: V_{\ul 1} [-\infty, x_1^- ] \: 
\ul{\nabla}\times \hat{\ul{A}}_{cl} (x_1^- , \ul{x}_1) \:
V_{\ul 1} [x_1^- , \infty] \Big]
\notag \\ &+
\contraction[2ex] {\tr\Big[ } {V_{\ul 0}} {\:\: V_{\ul 1} [-\infty,
  x_1^- ] \: \ul{\nabla}\times \hat{\ul{A}}_{cl} (x_1^- , \ul{x}_1)
  \:} {V_{\ul 1}}
\tr\Big[ V_{\ul 0} \:\: V_{\ul 1} [-\infty, x_1^- ] \: 
\ul{\nabla}\times \hat{\ul{A}}_{cl} (x_1^- , \ul{x}_1) \:
V_{\ul 1} [x_1^- , \infty] \Big]
\notag \\ &+
\contraction[2ex]
{\tr\Big[ V_{\ul 0} \:\: }
{V}
{}
{{}_{\ul 1}}
\tr\Big[ V_{\ul 0} \:\: V {}_{\ul 1} [-\infty, x_1^- ] \: 
\ul{\nabla}\times \hat{\ul{A}}_{cl} (x_1^- , \ul{x}_1) \:
V_{\ul 1} [x_1^- , \infty] \Big]
\notag \\ &+
\contraction[2ex]
{\tr\Big[ V_{\ul 0} \:\: }
{V_{\ul 1}}
{[-\infty, x_1^- ] \: \ul{\nabla}\times \hat{\ul{A}}_{cl} (x_1^- , \ul{x}_1) \:}
{V_{\ul 1}}
\tr\Big[ V_{\ul 0} \:\: V_{\ul 1} [-\infty, x_1^- ] \: 
\ul{\nabla}\times \hat{\ul{A}}_{cl} (x_1^- , \ul{x}_1) \:
V_{\ul 1} [x_1^- , \infty] \Big] 
\notag \\ &+
\contraction[2ex]
{\tr\Big[ V_{\ul 0} \:\: V_{\ul 1} [-\infty, x_1^- ] \: 
\ul{\nabla}\times \hat{\ul{A}}_{cl} (x_1^- , \ul{x}_1) \:}
{V}
{}
{ {}_{\ul 1} }
\tr\Big[ V_{\ul 0} \:\: V_{\ul 1} [-\infty, x_1^- ] \: 
\ul{\nabla}\times \hat{\ul{A}}_{cl} (x_1^- , \ul{x}_1) \:
V {}_{\ul 1} [x_1^- , \infty] \Big] .
\end{align}
\end{subequations}

Consider first the contraction I.  Expanding the Wilson line $V_{\ul
  1} [x_1^- , \infty]$ to first order in the quantum field, we have
\begin{align} 
\label{e:qIevol1}
(\delta G_{10})_\mathrm{I} = \frac{g^2 \, p^+}{2 N_c}
\, \int\limits_{-\infty}^0 dx_1^-
\int\limits_0^\infty dx_2^- \, \left\langle \tr\left[ V_{\ul 0} t^a
    V_{\ul 1}^\dagger t^b \right] \:\:
\contraction[2ex]
{\Big( \frac{\partial}{\partial (x_1)_\bot^i} \epsilon_T^{i j} \: }
{a_{\bot}^{j \, a}}
{(x_1^- , \ul{x}_1) \Big) \:}
{a^{+ \, b}}
\Big( \frac{\partial}{\partial (x_1)_\bot^i} \epsilon_T^{i j} \: 
a_{\bot}^{j \, a} (x_1^- , \ul{x}_1) \Big) \:
a^{+ \, b} (x_2^- , \ul{x}_1) 
\:\: + \cc \right \rangle .
\end{align}
After forming the contraction of these two quantum fields, we set
$a^\mu = 0$ in the rest of the Wilson lines, such that only the
classical background fields contribute. Since these classical
fields are localized at $x^- = 0$ we replace the remaining
semi-infinite Wilson lines by the fully infinite ones: this is in
accordance with the standard calculation in the shock wave background
\cite{Balitsky:1995ub}.  The contraction between the operator
insertion $a_{\bot}^{j a}$ and the semi-infinite Wilson line $V_{\ul 1}
[x_1^-, \infty]$ explicitly requires $x_2^- > x_1^-$, but in principle
there are contributions from $x_1^- < x_2^- < 0$ and $0 < x_1^- <
x_2^-$ in addition to the $x_1^- < 0 < x_2^-$ written here.  We
neglect these sub-eikonal virtual diagrams, since then the antiquark
would again need to scatter in the classical field in a spin-dependent
way, making them further energy suppressed.  Thus only the $x_1^- < 0
< x_2^-$ ``real'' diagram shown in Fig.~\ref{fig:Gevol_v2}
contributes.  Similarly, only the diagram in which the radiated gluon
scatters in a spin-dependent way is capable of receiving logarithmic
enhancement at small $x$.

The contraction in \eq{e:qIevol1} is the gluon propagator from the
sub-eikonal emission vertex to the Wilson line in the background of
the classical fields.  In general, we can write it as a free
propagator from the emission vertex to the shock wave, a Wilson line
for the interaction with the shock wave, and another free propagator to
the absorption vertex:
\begin{align}
\int\limits_{-\infty}^0 dx_1^- 
\int\limits_0^\infty dx_2^- \,
\contraction[2ex]
{\epsilon_T^{i j} \Big( \frac{\partial}{\partial (x_1)_\bot^i}  \: }
{a_{\bot}^{j \, a}}
{(x_1^- , \ul{x}_1) \Big) \:}
{a^{+ \, b}}
\epsilon_T^{i j} \Big( \frac{\partial}{\partial (x_1)_\bot^i} & \: 
a_{\bot}^{j \, a} (x_1^- , \ul{x}_1) \Big) \:
a^{+ \, b} (x_2^- , \ul{x}_1) =
\notag \\ & =
\int d^2 x_2 \left[ \epsilon_T^{i j} \frac{\partial}{\partial (x_1)_\bot^i} 
\int\limits_{-\infty}^0 dx_1^- \int \frac{d^4 k_1}{(2\pi)^4} e^{i k_1^+ x_1^-}
e^{i \ul{k}_1 \cdot \ul{x}_{21}} \frac{-i}{k_1^2 + i\epsilon} N^{j \mu} (k_1) \right]
\notag \\ & \hspace{1cm} \times
\Bigg[ (U_{\ul 2}^{b a})_{\mu \nu} \:  (2 k_1^-) 2\pi \: \delta(k_1^- - k_2^-) \Bigg]
\notag \\ & \hspace{1cm} \times
\left[ \int\limits_0^\infty dx_2^- \int \frac{d^4 k_2}{(2\pi)^4} e^{-i k_2^+ x_2^-}
e^{-i \ul{k}_2 \cdot \ul{x}_{21}} \frac{-i}{k_2^2 + i\epsilon} N^{\nu +} (k_2) \right].
\end{align}
Here the numerator of the free gluon propagator in the $\eta \cdot A
\equiv A^- = 0$ light-cone gauge is
\begin{align} 
  \label{e:polsum}
  N^{\mu \nu} (k) &= g^{\mu \nu} - \frac{\eta^{\mu} k^\nu + k^\mu
    \eta^{\nu}}{k^-}
  = - \sum_{\lambda = \pm} \, (\epsilon_\lambda^* (k))^\mu \,
  (\epsilon_\lambda (k))^\nu - \frac{k^2}{(k^-)^2} \eta^{\mu}
  \eta^{\nu}.
\end{align}
The contribution from the instantaneous gluon term (last term on the right-hand side of \eq{e:polsum}) is proportional to a delta function in $x^-$ and cannot propagate across the classical shockwave; it therefore does not contribute to real gluon emission.  This allows us to replace the numerators by polarization sums and write the interaction with the shockwave as a polarization matrix:
\begin{align}
  (\epsilon_\lambda (k_1))_\mu (U_{\ul x}^{b a})^{\mu \nu}
  (\epsilon_{\lambda'}^* (k_2))_\nu &= \delta_{\lambda \lambda'}
  (U_{\ul x})^{b a} + \lambda \: \delta_{\lambda \lambda'} \, (U_{\ul
    x}^{pol})^{b a} + \dots \, ,
\end{align}
where the ellipsis represents sub-eikonal terms which do not contribute to helicity evolution.

This gives
\begin{align} \label{e:bkprop1}
\int\limits_{-\infty}^0 dx_1^- 
\int\limits_0^\infty dx_2^- \,
\contraction[2ex]
{\epsilon_T^{i j} \Big( \frac{\partial}{\partial (x_1)_\bot^i}  \: }
{a_{\bot}^{j \, a}}
{(x_1^- , \ul{x}_1) \Big) \:}
{a^{+ \, b}}
\epsilon_T^{i j} \Big( \frac{\partial}{\partial (x_1)_\bot^i} &\: 
a_{\bot}^{j \, a} (x_1^- , \ul{x}_1) \Big) \:
a^{+ \, b} (x_2^- , \ul{x}_1) =
\notag \\ & =
 \sum_{\lambda} \lambda \int d^2 x_2 \left[
  \epsilon_T^{i j} \frac{\partial}{\partial (x_1)_\bot^i}
  \int\limits_{-\infty}^0 dx_1^- \int \frac{d^4 k_1}{(2\pi)^4} e^{i
    k_1^+ x_1^-} e^{i \ul{k}_1 \cdot \ul{x}_{21}} \frac{-i}{k_1^2 +
    i\epsilon} (\epsilon_\lambda^*)_\bot^j \right]
\notag \\ & \hspace{1cm} \times
\Bigg[ (U_{\ul 2}^{pol})^{b a} \: 2\pi (2 k_1^-) \: \delta(k_1^- -
k_2^-) \Bigg]
\notag \\ & \hspace{1cm} \times
\left[ \int\limits_0^\infty dx_2^- \int \frac{d^4 k_2}{(2\pi)^4} e^{-i
    k_2^+ x_2^-} e^{-i \ul{k}_2 \cdot \ul{x}_{21}} \frac{-i}{k_2^2 +
    i\epsilon} [\epsilon_\lambda (k_2)]^+ \right].
\end{align}
Each factor in brackets now has a transparent interpretation as the
emission vertex of a gluon with physical polarization $\lambda$, the
polarized Wilson line for that gluon to scatter in the classical
field, and the absorption vertex.  Performing the spin sum gives
$\sum_\lambda \lambda (\epsilon_\lambda^*)_\bot^j [\epsilon_\lambda
(k_2)]^+ = i \epsilon_T^{j \ell} (k_2)_\bot^\ell / k_2^-$
(equivalently, we could have just kept the appropriate terms in the
numerators \eqref{e:polsum}), such that
\begin{align}
 &
\int\limits_{-\infty}^0 dx_1^- 
\int\limits_0^\infty dx_2^-  \,
\contraction[2ex]
{\epsilon_T^{i j}  \Big( \frac{\partial}{\partial (x_1)_\bot^i} \: }
{a_{\bot}^{j \, a}}
{(x_1^- , \ul{x}_1) \Big) \:}
{a^{+ \, b}}
\epsilon_T^{i j} \Big( \frac{\partial}{\partial (x_1)_\bot^i} \:
a_{\bot}^{j \, a} (x_1^- , \ul{x}_1) \Big) \: a^{+ \, b} (x_2^- ,
\ul{x}_1) =
\notag \\ & =
\frac{i}{\pi} \int\limits_{-\infty}^\infty dk^- \int d^2 x_2 \left[
  \frac{\partial}{\partial (x_1)_\bot^i} \int\limits_{-\infty}^0
  dx_1^- \int \frac{d^2 k_1 \, dk_1^+}{(2\pi)^3} e^{i k_1^+ x_1^-}
  e^{i \ul{k}_1 \cdot \ul{x}_{21}} \frac{1}{k_1^2 + i\epsilon} \right]
\notag \\ & \hspace{1cm} \times
\left[ \int\limits_0^\infty dx_2^- \int \frac{d^2 k_2 \,
    dk_2^+}{(2\pi)^3} e^{-i k_2^+ x_2^-} e^{-i \ul{k}_2 \cdot
    \ul{x}_{21}} \frac{1}{k_2^2 + i\epsilon} (k_2)_\bot^i \right] \:
 (U_{\ul 2}^{pol})^{b a}_{(k^-)} \Bigg|_{k_1^- =
  k_2^- = k^-}.
\end{align}

The integrals in brackets are straightforward to perform:
\begin{subequations} \label{e:bkInts}
\begin{align}
  \int\limits_{-\infty}^0 dx_1^- \int \frac{d^2 k_1 \,
    dk_1^+}{(2\pi)^3} e^{i k_1^+ x_1^-} e^{i \ul{k}_1 \cdot
    \ul{x}_{21}} \frac{1}{k_1^2 + i\epsilon} &= \frac{-1}{2\pi}
  \ln\frac{1}{x_{21} \Lambda} \, \theta (k_1^-)
\\
\frac{\partial}{\partial (x_1)_\bot^i} \left[ \frac{-1}{2\pi}
  \ln\frac{1}{x_{21} \Lambda} \right] &= \frac{-1}{2\pi}
\frac{(x_{21})_\bot^i}{x_{21}^2}
\\
\int\limits_0^\infty dx_2^- \int \frac{d^2 k_2 \, dk_2^+}{(2\pi)^3}
e^{-i k_2^+ x_2^-} e^{-i \ul{k}_2 \cdot \ul{x}_{21}} \frac{1}{k_2^2 +
  i\epsilon} (k_2)_\bot^i &=
\frac{i}{2\pi} \frac{(x_{21})_\bot^i}{x_{21}^2} \, \theta (k_2^-),
\end{align}
\end{subequations}
such that the full propagator for contraction I is
\begin{align} 
\label{e:qIevol2}
\int\limits_{-\infty}^0 dx_1^- 
\int\limits_0^\infty dx_2^- \,
\contraction[2ex]
{\Big( \frac{\partial}{\partial (x_1)_\bot^i} \epsilon_T^{i j} \: }
{a_{\bot}^{j \, a}}
{(x_1^- , \ul{x}_1) \Big) \:}
{a^{+ \, b}}
\Big( \frac{\partial}{\partial (x_1)_\bot^i} &\epsilon_T^{i j} \: 
a_{\bot}^{j \, a} (x_1^- , \ul{x}_1) \Big) \:
a^{+ \, b} (x_2^- , \ul{x}_1) 
=
\frac{1}{4\pi^3} \int\limits_0^\infty dk^- \int \frac{d^2
  x_2}{x_{21}^2} \, (U_{\ul 2}^{pol})^{b
  a}_{(k^-)}.
\end{align}
The propagator \eqref{e:qIevol2} is the backbone of the calculation,
trivially giving for diagram I
\begin{align} 
\label{e:qIevol3} 
(\delta G_{10})_{\mathrm{I}} (z s) &= \frac{g^2 p^+}{8 \pi^3 N_c}
\int\limits_0^\infty dk^- \int \frac{d^2 x_2}{x_{21}^2} \left\langle
  \tr\left[ V_{\ul 0} t^a V_{\ul 1}^\dagger t^b \right] (U_{\ul
    2}^{pol})^{b a} + \cc \right\rangle (z' s = 2 p^+ k^-)
\notag \\ &=
\frac{\alpha_s N_c}{4 \pi^2} \int\limits_{\frac{\Lambda^2}{s}}^{z}
\frac{dz'}{z'} \int \frac{d^2 x_2}{x_{21}^2} \: \llangle
\frac{1}{N_c^2} \tr\left[ V_{\ul 0} t^a V_{\ul 1}^\dagger t^b \right]
(U_{\ul 2}^{pol})^{b a} + \cc \rrangle (z' s) ,
\end{align}
where we have used the double-angle brackets defined in
\eq{eq:Gdef}. In the second line of \eq{e:qIevol3} we have also
modified the limits of $k^-$ integration to make sure that $k^-$ does
not exceed the large $p^-$ momentum of the projectile in the actual
diagrammatic calculation.

It is straightforward to show that the propagators are symmetric, such
that diagrams I and I${}^\prime$ are equal and diagrams II and
II${}^\prime$ are equal.  In the case of diagram II, the only
difference is that the momenta are conjugate to different coordinates
on opposite sides of the shock wave (note that $a^+$ in the contraction
is now evaluated at $\ul{x}_0$):
\begin{align} \label{e:qIIevol1}
\int\limits_{-\infty}^0 dx_1^- 
\int\limits_0^\infty dx_2^- \,
\contraction[2ex]
{\Big( \frac{\partial}{\partial (x_1)_\bot^i} \epsilon_T^{i j} \: }
{a_{\bot}^{j \, a}}
{(x_1^- , \ul{x}_1) \Big) \:}
{a^{+ \, b}}
\Big( \frac{\partial}{\partial (x_1)_\bot^i} &\epsilon_T^{i j} \: 
a_{\bot}^{j \, a} (x_1^- , \ul{x}_1) \Big) \:
a^{+ \, b} (x_2^- , \ul{x}_0) 
=
\frac{1}{4\pi^3} \int\limits_0^\infty dk^- \int d^2 x_2 \:
\frac{\ul{x}_{21} \cdot \ul{x}_{20}}{x_{21}^2 \, x_{20}^2} \, (U_{\ul
  2}^{pol})^{b a}_{(k^-)},
\end{align}
which reduces back to \eqref{e:qIevol2} in the limit $\ul{x}_0
\rightarrow \ul{x}_1$.  This gives for diagram II
\begin{align} \label{e:qIIevol2} (\delta G_{10})_{\mathrm{II}} (z s)
  &\equiv \frac{- g^2 \, p^+}{2 N_c} \int\limits_{-\infty}^0 dx_1^-
  \int\limits_0^\infty dx_2^- \, \left\langle \tr\left[ V_{\ul 0} t^a
      V_{\ul 1}^\dagger t^b \right] \:\:
\contraction[2ex]
{\Big( \frac{\partial}{\partial (x_1)_\bot^i} \epsilon_T^{i j} \: }
{a_{\bot}^{j \, a}}
{(x_1^- , \ul{x}_1) \Big) \:}
{a^{+ \, b}}
\Big( \frac{\partial}{\partial (x_1)_\bot^i} \epsilon_T^{i j} \: 
a_{\bot}^{j \, a} (x_1^- , \ul{x}_1) \Big) \:
a^{+ \, b} (x_2^- , \ul{x}_0) 
\:\: + \cc \right \rangle 
\notag \\ &=
- \frac{\alpha_s N_c}{4 \pi^2} \int\limits_{\frac{\Lambda^2}{s}}^{z}
\frac{dz'}{z'} \int d^2 x_2 \: \frac{\ul{x}_{21} \cdot
  \ul{x}_{20}}{x_{21}^2 \, x_{20}^2} \: \llangle \frac{1}{N_c^2}
\tr\left[ V_{\ul 0} t^a V_{\ul 1}^\dagger t^b \right] (U_{\ul
  2}^{pol})^{b a} + \cc \rrangle (z' s) .
\end{align}
The extra minus sign from diagram II comes from having expanded
$V_{\ul 0}$ rather than $V_{\ul 1}^\dagger$; that is, from the
opposite charge $(-g)$ of the antiquark.

The last ingredient in the evolution is the unpolarized eikonal
contribution, which can simply be read off of the literature; the only
difference is that line $1$ for us is polarized.
\begin{align} 
  \label{e:qBFKL1}
  (\delta G_{10})_{\mathrm{eik}} (z s) &= \frac{\alpha_s
    N_c}{2\pi^2} \int\limits^z_{\frac{\Lambda^2}{s}} \frac{dz'}{z'}
  \int d^2 x_2 \, \frac{x_{10}^2}{x_{21}^2 \, x_{20}^2} \, \llangle
  \frac{1}{N_c^2} \tr\left[ V_{\ul 0} t^a V_{\ul 1}^{pol \, \dagger}
    t^b \right] (U_{\ul 2})^{b a} - \frac{C_F}{N_c^2} \tr\left[ V_{\ul
      0} V_{\ul 1}^{pol \, \dagger} \right] + \cc \rrangle (z' s).
\end{align}
With all three contributions from polarized ladder gluons (I $+$
I${}^\prime$, \eqref{e:qIevol3}), polarized non-ladder gluons (II $+$
II${}^\prime$, \eqref{e:qIIevol2}), and unpolarized gluons (eikonal,
\eqref{e:qBFKL1}), the complete evolution of the polarized dipole
amplitude for the quark helicity distribution is
\begin{align} 
  \label{e:qevol}
  G_{10} (z s) &= G_{10}^{(0)} (z s) + 2 (\delta G_{10})_{\mathrm{I}}
  (z s) + 2 (\delta G_{10})_{\mathrm{II}} (z s) + (\delta
  G_{10})_{\mathrm{eik}} (z s)
\notag \\ &=
G_{10}^{(0)} (z s) + \frac{\alpha_s N_c}{2\pi^2}
\int\limits^z_{\frac{\Lambda^2}{s}} \frac{dz'}{z'} \int d^2 x_2
\Bigg\{ \left[ \frac{1}{x_{21}^2} - \frac{\ul{x}_{21} \cdot
    \ul{x}_{20}}{x_{21}^2 \, x_{20}^2} \right] \llangle
\frac{1}{N_c^2} \tr\left[ V_{\ul 0} t^a V_{\ul 1}^\dagger t^b \right]
(U_{\ul 2}^{pol})^{b a} + \cc \rrangle (z' s)
\notag \\ & \hspace{1cm} +
\frac{x_{10}^2}{x_{21}^2 \, x_{20}^2} \, \llangle \frac{1}{N_c^2}
\tr\left[ V_{\ul 0} t^a V_{\ul 1}^{pol \, \dagger} t^b \right] (U_{\ul
  2})^{b a} - \frac{C_F}{N_c^2} \tr\left[ V_{\ul 0} V_{\ul 1}^{pol \,
    \dagger} \right] + \cc \rrangle (z' s) \Bigg\},
\end{align}
in complete agreement with Eq.~(50) of \cite{Kovchegov:2015pbl}.  We
should note that the limits of the $x_2$ integral in each term are set
by enforcing a lifetime ordering condition: the lifetime of the
quantum fluctuation should be much longer than the subsequent
classical interactions, in accordance with the rapidity factorization
scheme.  The fact that we have successfully re-derived the evolution
equation \eqref{e:qevol} for the polarized dipole amplitude serves as
an independent check of Eq.~(50) in \cite{Kovchegov:2015pbl}.  It also
validates both the operator definition \eqref{eq:Wpol2} of the
polarized Wilson line and our implementation of the operator-level
evolution using the background field / rapidity factorization methods.
We will next repeat this analysis for the new polarized dipole
amplitude \eqref{eq:Gidef} for the dipole gluon helicity distribution.


Before we do that, let us make the connection between \eq{e:qevol} and
Eqs.~\eqref{e:oldevol1}. Reinstating the lifetime ordering condition
on the $x_2$ integration in the first term in the curly brackets of
\eq{e:qevol} multiplies $1/x_{21}^2$ by $\theta (x_{10}^2 \, z -
x_{21}^2 z')$ while multiplying $(\ul{x}_{21} \cdot
\ul{x}_{20})/(x_{21}^2 \, x_{20}^2)$ by $\theta (x_{10}^2 \, z -
\mbox{max} \{ x_{21}^2 , x_{20}^2\} z')$. The DLA limit of the
resulting kernel is obtained by the following substitution:
    \begin{align}\label{subst1}
      \frac{1}{x_{21}^2} \, \theta (x_{10}^2 z - x_{21}^2 z') -
      \frac{\ul{x}_{21} \cdot \ul{x}_{20}}{x_{21}^2 \, x_{20}^2} \,
      \theta (x_{10}^2 z - \mbox{max} \{ x_{21}^2, x_{20}^2 \} z')
      \approx \frac{1}{x_{21}^2} \, \theta (x_{10} - x_{21}).
\end{align}
To simplify the second term in the curly brackets of \eq{e:qevol} we
employ the Fierz identity, which gives
\begin{align}\label{Fierz}
  2 \, \mbox{tr} \left[ V_{\ul{0}} \, t^a \, V_{\ul{1}}^{pol \,
      \dagger} \, t^b \right] \, (U_{\ul 2})^{b a} = \mbox{tr} \left[
    V_{\ul{0}} \, V^{\dagger}_{\ul{2}} \right] \, \mbox{tr} \left[
    V_{\ul{2}} \, V^{pol \, \dagger}_{\ul{1}} \right] - \frac{1}{N_c}
  \, \mbox{tr} \left[ V_{\ul{0}} \, V^{pol \, \dagger}_{\ul{1}}
  \right].
\end{align}
The $x_2$ integral in the second term of \eqref{e:qevol} is
logarithmic only in the $x_{21} \ll x_{10}$ and $x_{20} \ll x_{10}$
regions. In the $x_{20} \ll x_{10}$ region \eq{Fierz} ensures that the
expression in the double angle brackets in the second term inside the
curly brackets of \eq{e:qevol} approaches zero; thus the transverse
logarithm coming from the $x_{20} \ll x_{10}$ region vanishes. This is
in complete analogy with the unpolarized small-$x$ evolution
\cite{Balitsky:1995ub,Balitsky:1998ya,Kovchegov:1999yj,Kovchegov:1999ua,Jalilian-Marian:1997dw,Jalilian-Marian:1997gr,Iancu:2001ad,Iancu:2000hn}. The
physical reason behind this cancellation is that when the emitted
unpolarized gluon is very close to the unpolarized quark (that it is
emitted by) in the transverse plane, the system is identical to the
original unpolarized quark.

In the $x_{21} \ll x_{10}$ region, however, the second term inside the
curly brackets of \eq{e:qevol} does not vanish, as again can be seen
from \eq{Fierz}. The formal reason behind this is that the zero-size
polarized dipole does not have a unit $S$-matrix. In other words,
polarized dipoles do not have the color-transparency property that the
unpolarized dipoles have, since when the polarized quark line overlaps
with the unpolarized anti-quark line in the transverse plane, their
interactions with the target do not cancel. Somewhat more physically,
one can argue that when an unpolarized gluon is emitted by a polarized
quark, the system does not become equivalent to the original polarized
quark even if the gluon is very close to the quark in the transverse
plane.

In order to keep only the logarithmic $x_{21} \ll x_{10}$ region, we
replace
\begin{align}\label{subst2}
  \frac{x_{10}^2}{x_{21}^2 \, x_{20}^2} \to \frac{1}{x_{21}^2} \,
  \theta (x_{10} - x_{21}) 
\end{align}
in the second term in the curly brackets of \eq{e:qevol}. With the
substitutions \eqref{subst1} and \eqref{subst2}, \eq{e:qevol} becomes
\begin{align} 
  \label{e:qevol1}
  G_{10} (z s) &=
  G_{10}^{(0)} (z s) + \frac{\alpha_s N_c}{2\pi^2}
  \int\limits^z_{\frac{\Lambda^2}{s}} \frac{dz'}{z'}
  \int \frac{d^2 x_2}{x_{21}^2} \, 
  \theta(x_{10}^2 - x_{21}^2) \, \theta (x_{21}^2 - \tfrac{1}{z' s}) \,
  \Bigg\{ \llangle \frac{1}{N_c^2} \tr\left[
    V_{\ul 0} t^a V_{\ul 1}^\dagger t^b \right] (U_{\ul 2}^{pol})^{b
    a} + \cc \rrangle (z' s)
\notag \\ & \hspace{1cm} +
\llangle \frac{1}{N_c^2}
\tr\left[ V_{\ul 0} t^a V_{\ul 1}^{pol \, \dagger} t^b \right] (U_{\ul
  2})^{b a} - \frac{C_F}{N_c^2} \tr\left[ V_{\ul 0} V_{\ul 1}^{pol \,
    \dagger} \right] + \cc \rrangle (z' s) \Bigg\}.
\end{align}

To further simplify \eq{e:qevol1}, we invoke the DLA approximation
(and discard all the leading-logarithmic evolution, such as BFKL, BK
or JIMWLK; that is, put all the $S$-matrices for the dipoles without
polarized Wilson lines equal to one). We also employ the large-$N_c$
limit. With these approximations, we replace (see
\cite{Kovchegov:2015pbl} and Appendix~A of \cite{Kovchegov:2016zex})
\begin{align}\label{subst3}
  \left\langle \!\! \left\langle \mbox{tr} \left[ V_{\ul{0}} \, t^a \,
        V_{\ul{1}}^{\dagger} \, t^b \right] \, (U_{\ul 2}^{pol})^{b a}
    \right\rangle \!\! \right\rangle \to \frac{N_c}{2} \, \left\langle
    \!\! \left\langle \mbox{tr} \left[ V_{\ul{0}} \, V^{pol \,
          \dagger}_{\ul{2}} \right] \right\rangle \!\! \right\rangle +
  \frac{N_c}{2} \, \left\langle \!\! \left\langle \mbox{tr} \left[
        V^{pol}_{\ul{2}} \, V^{\dagger}_{\ul{1}} \right] \right\rangle
    \!\! \right\rangle
\end{align}
and obtain
\begin{align} 
  \label{e:qevol2}
  G_{10} (z s) &=
  G_{10}^{(0)} (z s) + \frac{\alpha_s N_c}{2\pi^2}
  \int\limits^z_{\frac{\Lambda^2}{s}} \frac{dz'}{z'}
  \int \frac{d^2 x_2}{x_{21}^2} \, 
  \theta(x_{10}^2 - x_{21}^2) \, \theta (x_{21}^2 - \tfrac{1}{z' s}) \,
  \Bigg\{ \llangle \frac{1}{2 N_c} \, \mbox{tr}
  \left[ V_{\ul{0}} \, V^{pol \, \dagger}_{\ul{2}} \right] 
\notag \\ & \hspace{1cm} +
  \frac{1}{2 N_c} \, \mbox{tr} \left[ V^{pol}_{\ul{2}} \,
    V^{\dagger}_{\ul{1}} \right] + \cc \rrangle (z' s) +
   \llangle \frac{1}{2 N_c} \tr\left[ V_{\ul 2} V_{\ul 1}^{pol \,
    \dagger} \right] - \frac{1}{2 N_c} \tr\left[ V_{\ul 0} V_{\ul
    1}^{pol \, \dagger} \right] + \cc \rrangle (z' s) \Bigg\}.
\end{align}

Equation \eqref{e:qevol2} has been derived for a polarized quark
dipole evolution. The large-$N_c$ limit of helicity evolution, as
considered in \cite{Kovchegov:2015pbl,Kovchegov:2016zex}, involves
only gluons: the corresponding dipole amplitude $G_{10} (z s) $ would
correspond to the interaction of the quark line of one large-$N_c$
gluon and the anti-quark line of another large-$N_c$ gluon with the
target \cite{Mueller:1993rr,Mueller:1994jq,Mueller:1994gb}. Here lies
another important difference between the small-$x$ helicity evolution
at hand and the unpolarized evolution
\cite{Mueller:1993rr,Mueller:1994jq,Mueller:1994gb,Balitsky:1995ub,Balitsky:1998ya,Kovchegov:1999yj,Kovchegov:1999ua,Jalilian-Marian:1997dw,Jalilian-Marian:1997gr,Iancu:2001ad,Iancu:2000hn}:
in the case of helicity evolution, the difference between a polarized
gluon emission by a polarized quark versus by a polarized gluon is not
only in the color factor. For instance, for helicity splitting
functions at small-$z$ and large $N_c$ one has $\Delta P_{GG} (z) = 4
\, \Delta P_{Gq} (z)$. Out of this factor of 4 difference, 2 is due to
the color factors, while another 2 is due to helicity dynamics in the
splitting. This means that, when going from the quark dipole of
\eq{e:qevol2} to the quark part of the gluon dipole, we need to
multiply the polarized gluon emission term (the first term in the
curly brackets) by 2 \cite{Kovchegov:2016zex}. (Ideally we would not
be needing to do this ``ad hoc" operation if we had started with the
polarized gluon dipole operator above.) We thus have
\begin{align} 
  \label{e:qevol3}
  G_{10} (z s) &=
  G_{10}^{(0)} (z s) + \frac{\alpha_s N_c}{2\pi^2}
  \int\limits^z_{\frac{\Lambda^2}{s}} \frac{dz'}{z'}
  \int \frac{d^2 x_2}{x_{21}^2} \, 
  \theta(x_{10}^2 - x_{21}^2) \, \theta (x_{21}^2 - \tfrac{1}{z' s}) \,
  \Bigg\{ \llangle \frac{1}{N_c} \, \mbox{tr}
  \left[ V_{\ul{0}} \, V^{pol \, \dagger}_{\ul{2}} \right] 
\notag \\ & \hspace{1cm} +
   \frac{1}{N_c} \, \mbox{tr} \left[ V^{pol}_{\ul{2}} \,
    V^{\dagger}_{\ul{1}} \right] + \cc \rrangle (z' s) +
   \llangle \frac{1}{2 N_c} \tr\left[ V_{\ul 2} V_{\ul 1}^{pol \,
    \dagger} \right] - \frac{1}{2 N_c} \tr\left[ V_{\ul 0} V_{\ul
    1}^{pol \, \dagger} \right] + \cc \rrangle (z' s) \Bigg\}.
\end{align}

The last step, which does not automatically follow from our formalism,
is to identify whether various $V V^\dagger$ correlators in
\eq{e:qevol3} combine into the amplitude $G_{10} (z s)$ or into the
auxiliary neighbor-dipole amplitude $\Gamma$. This depends on the
lifetime ordering for the subsequent evolution in those dipoles. For
instance, since in \eq{e:qevol3} we have $x_{21} \ll x_{10}$, the
subsequent evolution in the dipole $02$ in the non-eikonal emission
diagrams of \fig{fig:Gevol_v2} ``knows" about the dipole $21$, and
hence $\mbox{tr} \left[ V_{\ul{0}} \, V^{pol \, \dagger}_{\ul{2}}
\right]$ in \eq{e:qevol3} \cite{Kovchegov:2015pbl,Kovchegov:2016zex}
gives us $\Gamma_{02,21} (z' s) \approx \Gamma_{01,21} (z'
s)$. Similarly, one can show that $\tr\left[ V_{\ul 0} V_{\ul 1}^{pol
    \, \dagger} \right]$ in \eq{e:qevol3} contributes $\Gamma_{01,21}
(z' s)$ \cite{Kovchegov:2016zex}. The remaining traces give us
$G$'s. After performing this identification and integrating over impact parameters, we get
\begin{align} 
  \label{e:qevol4}
  G(x_{10}^2 , z s) &=
  G(x_{10}^2, z s) + \frac{\alpha_s N_c}{2\pi}
  \int\limits^z_{\frac{1}{x_{10}^2 s}} \frac{dz'}{z'}
  \int\limits_{\frac{1}{z' \, s}}^{x_{10}^2} \frac{d
    x_{21}^2}{x_{21}^2} \left[ \Gamma (x_{10}^2 , x_{21}^2 , z' s) + 3 \, G (x_{21}^2 , z' s) \right],
\end{align}
in agreement with \eq{Geq111}. (To arrive at \eq{e:qevol4} one also needs to notice
that, due to the bounds of the $x_{21}$ integral, $z' > 1/(x_{21}^2 \,
s) > 1/(x_{10}^2 \, s)$ which is a stronger lower bound on the $z'$
integration than $\Lambda^2/s$ of \eq{e:qevol3}.) \eq{Gameq111} is
obtained by analogy, with a slightly more subtle way of imposing the
lifetime ordering.

%
\subsection{Evolution of the Polarized Dipole Operator For the Dipole
  Gluon Helicity}
%

\begin{figure}[ht]
\begin{center}
\includegraphics[width= 0.7 \textwidth]{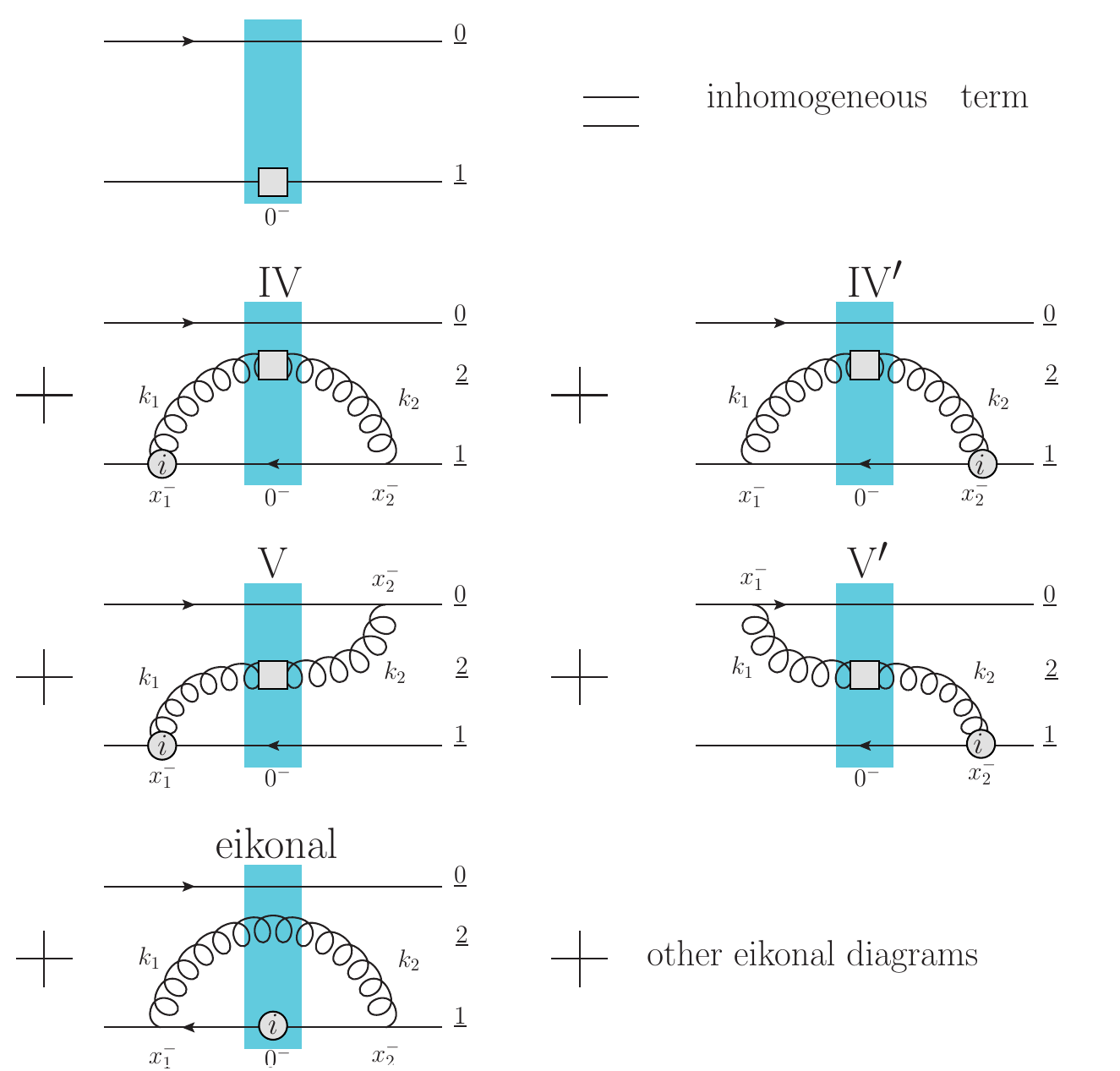} 
\caption{Diagrams illustrating contractions contributing the evolution
  of the polarized dipole amplitude for the dipole gluon helicity
  distribution.  The blue band represents the classical gluon fields
  (shock wave), the vertex $(i)$ denotes the sub-eikonal operator
  insertion, and the gray box represents the polarized Wilson line.}
\label{fig:Gievol_v2}
\end{center}
\end{figure}
%

The dipole gluon helicity distribution is governed by the polarized
dipole amplitude \eqref{eq:Gidef} and the (local) polarized Wilson
line \eqref{M:Vpol2}.  Written explicitly, this operator is
\begin{align}
  G_{10}^i (z s) = \frac{p^+}{2 N_c} \, \int
  \limits_{-\infty}^{\infty} dx_1^- \left\langle \tr\left[ V_{\ul 0}
      \:\: V_{\ul 1} [-\infty, x_1^-] \, (- i g) \, A_\bot^i (x_1^- ,
      \ul{x}_1) \, V_{\ul 1} [x_1^- , \infty] \right] + \cc
  \right\rangle (z s) .
\end{align}
In the same way as before, we expand the fields in terms of classical
and quantum components, contracting the lowest-order contributions in
the quantum fields.  Again, there are three general classes of
contractions / diagrams: ``polarized ladder'' emissions (IV and
IV${}^\prime$), ``polarized non-ladder'' emissions (V and
V${}^\prime$), and unpolarized emissions (``eikonal''), as illustrated
in \fig{fig:Gievol_v2}.  In analogy to \eq{e:qcontractions}, the
specific contractions are
\begin{subequations} \label{e:Gcontractions}
\begin{align}
\mathrm{IV}: \hspace{1cm}&
\contraction[2ex]
{\tr\Big[ V_{\ul 0} \:\: V_{\ul 1} [-\infty, x_1^- ] \: }
{a_{\bot}^i}
{(x_1^- , \ul{x}_1) \:}
{V_{\ul 1}}
\tr\Big[ V_{\ul 0} \:\: V_{\ul 1} [-\infty, x_1^- ] \: 
a_{\bot}^i (x_1^- , \ul{x}_1) \:
V_{\ul 1} [x_1^- , \infty] \Big]
\\
\mathrm{IV}': \hspace{1cm}&
\contraction[2ex]
{\tr\Big[ V_{\ul 0} \:\: }
{V_{\ul 1}}
{[-\infty, x_1^- ] \: }
{a_{\bot}^i}
\tr\Big[ V_{\ul 0} \:\: V_{\ul 1} [-\infty, x_1^- ] \: 
a_{\bot}^i (x_1^- , \ul{x}_1) \:
V_{\ul 1} [x_1^- , \infty] \Big]
\\ \notag \\
\mathrm{V} + \mathrm{V}': \hspace{1cm}&
\contraction[2ex]
{\tr\Big[ }
{V_{\ul 0}}
{\:\: V_{\ul 1} [-\infty, x_1^- ] \: }
{a_{\bot}^i}
\tr\Big[ V_{\ul 0} \:\: V_{\ul 1} [-\infty, x_1^- ] \: 
a_{\bot}^i (x_1^- , \ul{x}_1) \:
V_{\ul 1} [x_1^- , \infty] \Big]
\\ \notag \\
\mathrm{eikonal}: \hspace{1cm}&
\contraction[2ex]
{\tr\Big[ }
{V}
{}
{ {}_{\ul 0} }
\tr\Big[ V {}_{\ul 0} \:\: V_{\ul 1} [-\infty, x_1^- ] \: 
A_{cl \, \bot}^i (x_1^- , \ul{x}_1) \:
V_{\ul 1} [x_1^- , \infty] \Big]
\notag \\ &+
\contraction[2ex]
{\tr\Big[ }
{V_{\ul 0}}
{\:\:}
{V_{\ul 1}}
\tr\Big[ V_{\ul 0} \:\: V_{\ul 1} [-\infty, x_1^- ] \: 
A_{cl \, \bot}^i (x_1^- , \ul{x}_1) \:
V_{\ul 1} [x_1^- , \infty] \Big]
\notag \\ &+
\contraction[2ex]
{\tr\Big[ }
{V_{\ul 0}}
{\:\: V_{\ul 1} [-\infty, x_1^- ] \: A_{cl \, \bot}^i (x_1^- , \ul{x}_1) \:}
{V_{\ul 1}}
\tr\Big[ V_{\ul 0} \:\: V_{\ul 1} [-\infty, x_1^- ] \: 
A_{cl \, \bot}^i (x_1^- , \ul{x}_1) \:
V_{\ul 1} [x_1^- , \infty] \Big]
\notag \\ &+
\contraction[2ex]
{\tr\Big[ V_{\ul 0} \:\: }
{V}
{}
{{}_{\ul 1}}
\tr\Big[ V_{\ul 0} \:\: V {}_{\ul 1} [-\infty, x_1^- ] \: 
A_{cl \, \bot}^i (x_1^- , \ul{x}_1) \:
V_{\ul 1} [x_1^- , \infty] \Big]
\notag \\ &+
\contraction[2ex]
{\tr\Big[ V_{\ul 0} \:\: }
{V_{\ul 1}}
{[-\infty, x_1^- ] \: A_{cl \, \bot}^i (x_1^- , \ul{x}_1) \:}
{V_{\ul 1}}
\tr\Big[ V_{\ul 0} \:\: V_{\ul 1} [-\infty, x_1^- ] \: 
A_{cl \, \bot}^i (x_1^- , \ul{x}_1) \:
V_{\ul 1} [x_1^- , \infty] \Big] 
\notag \\ &+
\contraction[2ex]
{\tr\Big[ V_{\ul 0} \:\: V_{\ul 1} [-\infty, x_1^- ] \: 
A_{cl \, \bot}^i (x_1^- , \ul{x}_1) \:}
{V}
{}
{ {}_{\ul 1} }
\tr\Big[ V_{\ul 0} \:\: V_{\ul 1} [-\infty, x_1^- ] \: 
A_{cl \, \bot}^i (x_1^- , \ul{x}_1) \:
V {}_{\ul 1} [x_1^- , \infty] \Big] .
\end{align}
\end{subequations}

As we saw in Eqs.~\eqref{e:qIevol2} and \eqref{e:qIIevol1}, the
propagator for the ladder diagram I or IV is just a special case of
the propagator for the non-ladder diagram II or V.  We therefore begin
by calculating diagram V, which is the contraction of the operator
insertion with the unpolarized Wilson line in the time ordering $x_1^-
< 0 < x_2^-$.  Expanding the unpolarized Wilson line gives
\begin{align} \label{e:GVevol1}
(\delta G_{10}^i)_{\mathrm{V}} (z s) &= \frac{g^2 \, p^+}{2 N_c} \, 
\int\limits_{-\infty}^0 dx_1^- \int\limits_0^\infty dx_2^- \, \left\langle \tr\Big[ 
\contraction[2ex]
{V_{\ul 0} \,}
{a_{\bot}^i}
{(x_1^- , \ul{x}_1) \, V_{\ul 1}^\dagger \,}
{a^+}
V_{\ul 0} \, a_{\bot}^i (x_1^- , \ul{x}_1) \, V_{\ul 1}^\dagger \, a^+ (x_2^- , x_{\ul 0}) 
\Big] + \cc \right\rangle
\notag \\ &=
\frac{g^2 \, p^+}{2 N_c} \left\langle \tr\left[ V_{\ul 0} t^a V_{\ul
      1}^\dagger t^b \right] \: (\Delta_{cl}^{i +})^{b a}_{pol}
  (\ul{x}_1 , \ul{x}_0) \: + \cc \right\rangle ,
\end{align}
where we have defined the propagator in the classical background field
as
\begin{align} \label{e:GVevol2}
(\Delta_{cl}^{i +})^{b a}_{pol} (\ul{x}_1 , \ul{x}_0) & \equiv
\int\limits_{-\infty}^0 dx_1^- \int\limits_0^\infty dx_2^- \:
\contraction[2ex]
{}
{a_{\bot}^{i a}}
{(x_1^- , \ul{x}_1) \: \:}
{a^{+ b}}
a_{\bot}^{i a} (x_1^- , \ul{x}_1) \: \: a^{+ b} (x_2^- , x_{\ul 0}) .
\end{align}
As before, we will find that the propagator $\Delta_{cl}^{\mu \nu}$ is
symmetric, such that all of the polarized emissions shown in
Fig.~\ref{fig:Gievol_v2} can be written as
\begin{subequations}
\begin{align}
  (\delta G_{10}^i)_{\mathrm{IV}} (z s) = (\delta
  G_{10}^i)_{\mathrm{IV}'} (z s) &= - \frac{g^2 \, p^+}{2 N_c}
  \left\langle \tr\left[ V_{\ul 0} t^a V_{\ul 1}^\dagger t^b \right]
    \: (\Delta_{cl}^{i +})^{b a}_{pol} (\ul{x}_1 , \ul{x}_1) \: + \cc
  \right\rangle
\\
(\delta G_{10}^i)_{\mathrm{V}} (z s) = (\delta G_{10}^i)_{\mathrm{V}'}
(z s) &= \frac{g^2 \, p^+}{2 N_c} \left\langle \tr\left[ V_{\ul 0} t^a
    V_{\ul 1}^\dagger t^b \right] \: (\Delta_{cl}^{i +})^{b a}_{pol}
  (\ul{x}_1 , \ul{x}_0) \: + \cc \right\rangle .
\end{align}
\end{subequations}
The two classes of diagrams differ only in two respects: a sign
difference in the prefactor (due to expanding $V_{\ul 0}$ vs. $V_{\ul
  1}^\dagger$) and the arguments of the propagator (for ladder
vs. non-ladder emissions).

Thus the calculation is reduced to finding the propagator
\eqref{e:GVevol2}.  In analogy to \eq{e:bkprop1}, we write the
propagator as
\begin{align} 
  \label{e:GVevol33}
  (\Delta_{cl}^{i +})^{b a}_{pol} (\ul{x}_1 , \ul{x}_0) &=
  \sum_\lambda \lambda \int d^2 x_2 \left[ \int\limits_{-\infty}^0
    dx_1^- \int \frac{d^4 k_1}{(2\pi)^4} e^{i k_1^+ x_1^-} e^{i
      \ul{k}_1 \cdot \ul{x}_{21}} \frac{-i}{k_1^2 + i\epsilon}
    (\epsilon_\lambda^*)_\bot^i \right]
\notag \\ & \hspace{1cm} \times
\Bigg[ (U_{\ul 2}^{pol})^{b a} \: (2 k_1^-) 2\pi \: \delta(k_1^- -
k_2^-) \Bigg]
\notag \\ & \hspace{1cm} \times
\left[ \int\limits_0^\infty dx_2^- \int \frac{d^4 k_2}{(2\pi)^4} e^{-i
    k_2^+ x_2^-} e^{-i \ul{k}_2 \cdot \ul{x}_{20}} \frac{-i}{k_2^2 +
    i\epsilon} [\epsilon_\lambda (k_2)]^+ \right]
\notag \\ \notag \\ &=
-\frac{i}{\pi} \epsilon_T^{i j} \int dk^- \int d^2 x_2 \left[
  \int\limits_{-\infty}^0 dx_1^- \int \frac{d^2 k_1 \,
    dk_1^+}{(2\pi)^3} e^{i k_1^+ x_1^-} e^{i \ul{k}_1 \cdot
    \ul{x}_{21}} \frac{1}{k_1^2 + i\epsilon} \right]
\notag \\ & \hspace{1cm} \times
\left[ \int\limits_0^\infty dx_2^- \int \frac{d^2 k_2 \,
    dk_2^+}{(2\pi)^3} e^{-i k_2^+ x_2^-} e^{-i \ul{k}_2 \cdot
    \ul{x}_{20}} \frac{1}{k_2^2 + i\epsilon} (k_2)_\bot^j \right] \:
(U_{\ul 2}^{pol})^{b a} .
\end{align}
Employing the integrals in \eqref{e:bkInts} we recast this as
\begin{align}  
  \label{e:GVevol3}
  (\Delta_{cl}^{i +})^{b a}_{pol} (\ul{x}_1 , \ul{x}_0) =
  -\frac{1}{4\pi^3} \int dk^- \int d^2 x_2 \, \ln\frac{1}{x_{21}
    \Lambda} \, \frac{\epsilon_T^{i j} (x_{20})_\bot^j}{x_{20}^2} \:
  (U_{\ul 2}^{pol})^{b a}_{(k^-)} .
\end{align}
With the propagator \eqref{e:GVevol3}, it is straightforward to obtain
the evolution kernels IV $-$ V${}^\prime$:
\begin{subequations}
\begin{align}
  (\delta G_{10}^i)_{\mathrm{IV}} (z s) &= (\delta
  G_{10}^i)_{\mathrm{IV}'} (z s)
\notag \\ &=
\frac{\alpha_s N_c}{4 \pi^2} \int\limits_{\frac{\Lambda^2}{s}}^{z}
\frac{dz'}{z'} \int d^2 x_2 \: \ln\frac{1}{x_{21} \Lambda} \,
\frac{\epsilon_T^{i j} (x_{21})_\bot^j}{x_{21}^2} \llangle
\frac{1}{N_c^2} \tr\left[ V_{\ul 0} t^a V_{\ul 1}^\dagger t^b \right]
\: (U_{\ul 2}^{pol})^{b a} + \cc \rrangle (z' s) ,
\\ \notag \\ 
(\delta G_{10}^i)_{\mathrm{V}} (z s) &= (\delta
G_{10}^i)_{\mathrm{V}'} (z s) 
\notag \\ &=
- \frac{\alpha_s N_c}{4 \pi^2} \int\limits_{\frac{\Lambda^2}{s}}^{z}
\frac{dz'}{z'} \int d^2 x_2 \: \ln\frac{1}{x_{21} \Lambda} \,
\frac{\epsilon_T^{i j} (x_{20})_\bot^j}{x_{20}^2} \llangle
\frac{1}{N_c^2} \tr\left[ V_{\ul 0} t^a V_{\ul 1}^\dagger t^b \right]
\: (U_{\ul 2}^{pol})^{b a} + \cc \rrangle (z' s) .
\end{align}
\end{subequations}
The only other ingredient necessary is the unpolarized eikonal gluon
contribution, which is identical to \eqref{e:qBFKL1} except for the
replacement of the polarized Wilson lines $V_{\ul 1}^{pol \, \dagger}
\rightarrow (V_{\ul 1}^{pol \, \dagger})_\bot^i$:
\begin{align} 
  \label{e:GBFKL1}
  (\delta G_{10}^i)_{\mathrm{eik}} (z s) &= \frac{\alpha_s N_c}{2\pi^2}
  \int\limits^z_{\frac{\Lambda^2}{s}} \frac{dz'}{z'} \int d^2 x_2 \,
  \frac{x_{10}^2}{x_{21}^2 \, x_{20}^2} \, \llangle \frac{1}{N_c^2}
  \tr\left[ V_{\ul 0} t^a (V_{\ul 1}^{pol \, \dagger})_\bot^i t^b
  \right] (U_{\ul 2})^{b a}
\notag \\ & \hspace{2cm} 
- \frac{C_F}{N_c^2} \tr\left[ V_{\ul 0} (V_{\ul 1}^{pol \,
    \dagger})_\bot^i \right] + \cc \rrangle (z' s).
\end{align}

Including all these contributions, we can immediately write down the
evolution equation for the polarized dipole amplitude $G_{10}^i$ as
\begin{align} 
  \label{e:Gevol1}
  G_{10}^i (z s) &= G_{10}^{i \, (0)} (z s) + 2 (\delta
  G_{10}^i)_{\mathrm{IV}} (z s) + 2 (\delta G_{10}^i)_{\mathrm{V}} (z
  s) + (\delta G_{10}^i)_{\mathrm{eik}} (z s)
\notag \\ &=
G_{10}^{i \, (0)} (z s) + \frac{\alpha_s N_c}{2 \pi^2}
\int\limits_{\frac{\Lambda^2}{s}}^{z} \frac{dz'}{z'} \int d^2 x_2 \:
\notag \\ & \hspace{1cm} \times
\Bigg\{ \ln\frac{1}{x_{21} \Lambda} \, \epsilon_T^{i j} \left[ \frac{
    (x_{21})_\bot^j}{x_{21}^2} - \frac{ (x_{20})_\bot^j}{x_{20}^2}
\right] \llangle \frac{1}{N_c^2} \tr\left[ V_{\ul 0} t^a V_{\ul
    1}^\dagger t^b \right] \: (U_{\ul 2}^{pol})^{b a} + \cc \rrangle
(z' s)
\notag \\ & \hspace{1.5cm} +
\frac{x_{10}^2}{x_{21}^2 \, x_{20}^2} \, \llangle \frac{1}{N_c^2}
\tr\left[ V_{\ul 0} t^a (V_{\ul 1}^{pol \, \dagger})_\bot^i t^b
\right] (U_{\ul 2})^{b a} - \frac{C_F}{N_c^2} \tr\left[ V_{\ul 0}
  (V_{\ul 1}^{pol \, \dagger})_\bot^i \right] + \cc \rrangle (z' s)
\Bigg\} .
\end{align}
As expected, this evolution equation represents just the first of an
infinite tower of operator equations; we will remedy this problem in
the usual way by taking the large-$N_c$ limit.  We will also linearize
the evolution equation, keeping the essential polarization-dependent
dipoles and neglecting additional unpolarized rescattering (e.g., the
non-linear saturation corrections); this will be necessary to generate
double logarithms of energy.  With these simplifications, we replace\footnote{Note that \eq{subst3} suggested in Appendix~A of \cite{Kovchegov:2016zex} is missing an overall factor of 2 on its right-hand side \cite{Kovchegov:2018znm}. The correct version of equation \eqref{subst3} would lead directly to \eq{e:qevol3}, bypassing \eq{e:qevol2} above and the discussion that follows \cite{Kovchegov:2018znm}. This factor of 2 is included in obtaining Eqs.~\eqref{traces}.}
\begin{subequations}\label{traces}
\begin{align}
  \frac{1}{N_c^2} \tr\left[ V_{\ul 0} t^a V_{\ul 1}^\dagger t^b
  \right]& \: (U_{\ul 2}^{pol})^{b a} + \cc \rightarrow
\notag \\ &\rightarrow 
\frac{1}{N_c} \tr\left[ V_{\ul 0} V_{\ul 2}^{pol \, \dagger} \right]
+ \frac{1}{N_c} \tr\left[ V_{\ul 2}^{pol} V_{\ul 1}^\dagger \right]
+ \cc
\\ \notag \\
\frac{1}{N_c^2} \tr\left[ V_{\ul 0} t^a (V_{\ul 1}^{pol \,
    \dagger})_\bot^i t^b \right]& (U_{\ul 2})^{b a} -
\frac{C_F}{N_c^2} \tr\left[ V_{\ul 0} (V_{\ul 1}^{pol \,
    \dagger})_\bot^i \right] + \cc \rightarrow
\notag \\ &\rightarrow 
\frac{1}{2 N_c} \tr\left[ V_{\ul 2} (V_{\ul 1}^{pol \,
    \dagger})_\bot^i \right] - \frac{1}{2 N_c} \tr\left[ V_{\ul
    0}^{pol} (V_{\ul 1}^{pol \, \dagger})_\bot^i \right] + \cc ,
\end{align}
\end{subequations}
giving
\begin{align} 
  \label{e:Gevol2}
  G_{10}^i (z s) &= G_{10}^{i \, (0)} (z s) + \frac{\alpha_s N_c}{2
    \pi^2} \int\limits_{\frac{\Lambda^2}{s}}^{z} \frac{dz'}{z'} \int
  d^2 x_2 \:
\notag \\ & \hspace{1cm} \times
\Bigg\{ \ln\frac{1}{x_{21}^2 \Lambda^2} \, \epsilon_T^{i j} \left[ \frac{
    (x_{21})_\bot^j}{x_{21}^2} - \frac{ (x_{20})_\bot^j}{x_{20}^2}
\right] \llangle \frac{1}{2 N_c} \tr\left[ V_{\ul 0} V_{\ul 2}^{pol \,
    \dagger} \right] + \frac{1}{2 N_c} \tr\left[ V_{\ul 2}^{pol}
  V_{\ul 1}^\dagger \right] + \cc \rrangle (z' s)
\notag \\ & \hspace{1.5cm} +
\frac{x_{10}^2}{x_{21}^2 \, x_{20}^2} \, \llangle \frac{1}{2 N_c}
\tr\left[ V_{\ul 2} (V_{\ul 1}^{pol \, \dagger})_\bot^i \right] -
\frac{1}{2 N_c} \tr\left[ V_{\ul 0} (V_{\ul 1}^{pol \,
    \dagger})_\bot^i \right] + \cc \rrangle (z' s) \Bigg\} .
\end{align}
The right-hand side of \eq{e:Gevol2} are now polarized dipole
amplitudes, but we must think carefully before identifying them with
$G$ or $G^i$.  Depending on the precise limits of the $x_2$
integration, these dipoles may instead be ``neighbor dipoles''
$\Gamma$ or $\Gamma^i$.  These limits, in turn, are dictated by the
regions of transverse phase space which generate the greatest
logarithmic enhancement of the evolution.

Consider first the unpolarized eikonal emissions in the last line of
\eq{e:Gevol2}.  Just like in the quark helicity case, we see that the
dipole BFKL kernel $x_{10}^2 / (x_{21}^2 \, x_{20}^2)$ is potentially
DLA in both the $x_{21}^2 \ll x_{10}^2$ limit and in the $x_{20}^2 \ll
x_{10}^2$ limit.  In the latter case, $\ul{x}_2 \rightarrow \ul{x}_0$,
however, the operators multiplying the kernel cancel and destroy the
DLA contribution.  Therefore, similar to the quark case
\cite{Kovchegov:2015pbl}, we conclude that only the $x_{21}^2 \ll
x_{10}^2$ region in that term is DLA and simplify the dipole BFKL
kernel to $\frac{1}{x_{21}^2} \theta (x_{10}^2 - x_{21}^2) \theta
(x_{21}^2 - \tfrac{1}{z' s})$, where the available energy $z' s$ acts
as a UV cutoff.  For each of the associated dipoles $\tr\left[ V_{\ul
    2} (V_{\ul 1}^{pol \, \dagger})_\bot^i \right]$ and $\tr\left[
  V_{\ul 0} (V_{\ul 1}^{pol \, \dagger})_\bot^i \right]$, we must
impose a lifetime ordering condition on their subsequent evolution to
ensure that the ``fast'' quantum fields computed here live longer than
the ``slow'' classical fields.  The first term $\tr\left[ V_{\ul 2}
  (V_{\ul 1}^{pol \, \dagger})_\bot^i \right]$ depends only on the
distance $x_{21}$ associated with the quantum fluctuation and can be
identified as $G_{12}^i (z' s)$.  The second term $\tr\left[ V_{\ul 0}
  (V_{\ul 1}^{pol \, \dagger})_\bot^i \right]$ appears to depend only
on the distance $x_{10}$, but must also respect the lifetime ordering
with respect to the virtual gluon loop of transverse size $x_{21}$
that gave rise to this term in the equation.  This term is therefore a
neighbor dipole $\Gamma_{10 , 21}^i (z' s)$ because it ``remembers''
about the lifetime of the neighboring $x_{21}$ quantum fluctuation
(see \cite{Kovchegov:2016zex} for a detailed calculation explaining
this conclusion).  We therefore simplify the eikonal terms to write
\begin{align} 
  \label{e:Gevol3}
  G_{10}^i (z s) &= G_{10}^{i \, (0)} (z s) + \frac{\alpha_s N_c}{2
    \pi^2} \int\limits_{\frac{\Lambda^2}{s}}^{z} \frac{dz'}{z'} \int
  d^2 x_2 \:
\notag \\ & \hspace{1cm} \times
\Bigg\{ \ln\frac{1}{x_{21}^2 \Lambda^2} \, \epsilon_T^{i j} \left[ \frac{
    (x_{21})_\bot^j}{x_{21}^2} - \frac{ (x_{20})_\bot^j}{x_{20}^2}
\right] \llangle \frac{1}{2 N_c} \tr\left[ V_{\ul 0} V_{\ul 2}^{pol \,
    \dagger} \right] + \frac{1}{2 N_c} \tr\left[ V_{\ul 2}^{pol}
  V_{\ul 1}^\dagger \right] + \cc \rrangle (z' s) \Bigg\}
\notag \\ & \hspace{1.5cm} +
\frac{\alpha_s N_c}{2\pi^2} \int\limits_{\frac{\Lambda^2}{s}}^z
\frac{dz'}{z'} \int \frac{d^2 x_2}{x_{21}^2} \: \theta\Big(x_{10}^2 -
x_{21}^2\Big) \, \theta\Big(x_{21}^2 - \frac{1}{z' s}\Big) \, \left[
  G_{12}^i (z' s) - \Gamma_{10 , 21}^i (z' s) \right].
\end{align}

The story for the polarized gluon emissions in the second line of
\eq{e:Gevol3}, however, is significantly more complicated.  The reason
is that the transverse integration does not generate a logarithm of
the energy, so the whole kernel is not DLA.  (After integration,
$\ln\frac{1}{x_{21} \Lambda}$ becomes $\ln\frac{1}{x_{10} \Lambda}$
and not a logarithm of the energy.)  It would seem, then, that the
polarized emissions only generate one logarithm of energy from the
$z'$ integral and can be neglected compared to the DLA evolution of
the eikonal terms.

\begin{figure}[ht]
\begin{center}
\includegraphics[width= 0.5 \textwidth]{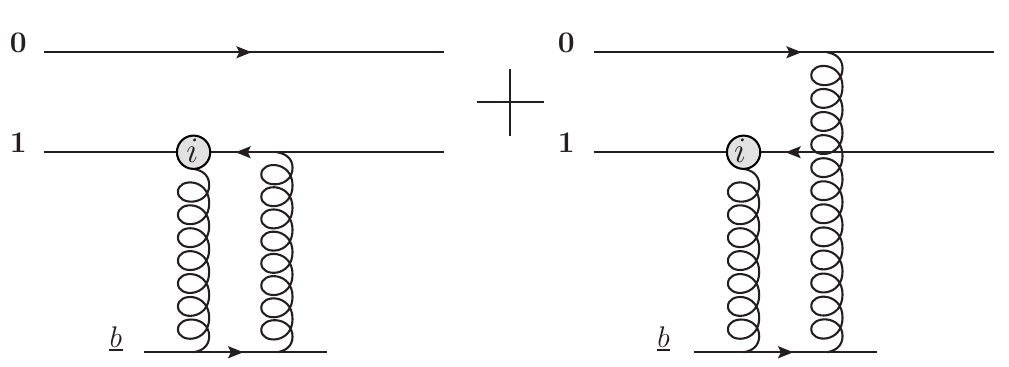} 
\caption{Diagrams contributing the initial conditions for $G^i$ and
  $\Gamma^i$ in Eq.~\eqref{Ginit}.}
\label{fig:Ginit}
\end{center}
\end{figure}
%

This, however, is not quite the case, because of the initial
conditions.  The initial conditions for the polarized dipole operator
$G_{10}^i$, taken in the quark target model at a fixed impact
parameter, can be obtained by computing the diagrams shown in
\fig{fig:Ginit}:
\begin{align}
  G^{i \, (0)}_{10} (z) = \Gamma^{i \, (0)}_{10, 21} (z) = -
  \frac{\alpha_s^2 C_F}{N_c} \epsilon^{ij} \frac{({\un x}_1 - {\un
      b})^j}{|{\un x}_1 - {\un b}|^2} \, \ln \frac{|{\un x}_1 - {\un
      b}|}{|{\un x}_0 - {\un b}|}.
\end{align} 
Integrating over the impact parameters yields
\begin{align}
  \label{Ginit}
  \int d^2 b_{10} \, G^{i \, (0)}_{10} (z s) = \int d^2 b_{10} \, \Gamma^{i \,
    (0)}_{10, 21} (z s) = - \frac{\alpha_s^2 C_F}{N_c} \, \pi \,
  \epsilon^{ij} \, x_{10}^j \, \ln \frac{1}{x_{10} \, \Lambda},
\end{align} 
which is independent of the energy.  By contrast, the dipoles
$\tr\left[ V_{\ul 0} V_{\ul 2}^{pol \, \dagger} \right]$ and
$\tr\left[ V_{\ul 2}^{pol} V_{\ul 1}^\dagger \right]$ in \eq{e:Gevol3}
are the ones which enter the evolution \eqref{e:oldevol1} of the quark
helicity distribution.  Their initial conditions are given by
Eq.~(13b) in \cite{Kovchegov:2016zex} for the impact-parameter
integrated case. Keeping only the gluon-exchange part of that
expression,
\begin{align}\label{init}
  \int d^2 b_{10} \, G^{(0)}_{10} (z s) = \int d^2 b_{10} \, \Gamma^{(0)}_{10 ,
    21} (z s) = - \frac{\alpha_s^2 C_F}{N_c} \pi \, \ln (z s \,
  x_{10}^2),
\end{align} 
we see that $\int d^2 b_{10} \, G_{10}^i$ in \eqref{Ginit} is suppressed by
a logarithm of energy compared to $\int d^2 b_{10} \, G_{10}$ in
\eqref{init}.

This implies that $G^i$ starts energy-independent and, after one step
of eikonal evolution, acquires two logarithms of energy.  On
the other hand, $G$ and $\Gamma$ can mix into $G^i$ through the second
line of \eq{e:Gevol3}, picking up one logarithm of energy from the
evolution.  But since $G$ and $\Gamma$ start off with one logarithm of
energy from the initial conditions, both of these two contributions
are of the same order.  Subsequent evolution in the eikonal $G^i
, \Gamma^i \rightarrow G^i , \Gamma^i$ channel and the prior evolution
\eqref{e:oldevol1} in the polarized $G , \Gamma \rightarrow G, \Gamma$
channel, are both double logarithmic.

Therefore, we conclude that we must keep all of \eq{e:Gevol3}, and we
are left with a transverse integral for the polarized emissions which
covers the entire plane. The resulting kernel in the second line of
\eq{e:Gevol3} is leading-logarithmic (LLA). This is similar to the
unpolarized BFKL/BK/JIMWLK evolution, which also has a LLA kernel,
without any logarithm of energy coming from the transverse coordinate
integral. In the unpolarized evolution case at LLA one does not need
to impose the lifetime ordering condition which would restrict the
transverse integrals (see \cite{Beuf:2014uia,Iancu:2015vea} for the
higher-order corrections though). The same is true here: the
transverse integral in the second line of \eq{e:Gevol3} is
unconstrained. 

This leads to a problem though: with an unconstrained integral the
second line of \eq{e:Gevol3} we cannot tell whether the dipole $21$
is smaller than the dipole $20$ ($x_{21} \ll x_{20}$) or vice versa
($x_{20} \ll x_{21}$) or both dipoles are large $x_{21} \sim x_{20}
\gg x_{10}$. This was not necessary for the LLA unpolarized dipole
evolution \cite{Mueller:1993rr,Mueller:1994jq,Mueller:1995gb}, since
there the subsequent evolution in all the daughter dipoles was
independent of other dipoles and their sizes. This is not the case for
our DLA helicity evolution \eqref{e:oldevol1}, where the subsequent
evolution in a given dipole can make it a ``neighbor dipole'' if the
adjacent dipole (produced in the same step of evolution) was smaller
in the transverse plane. 

\begin{figure}[ht]
\begin{center}
\includegraphics[width= \textwidth]{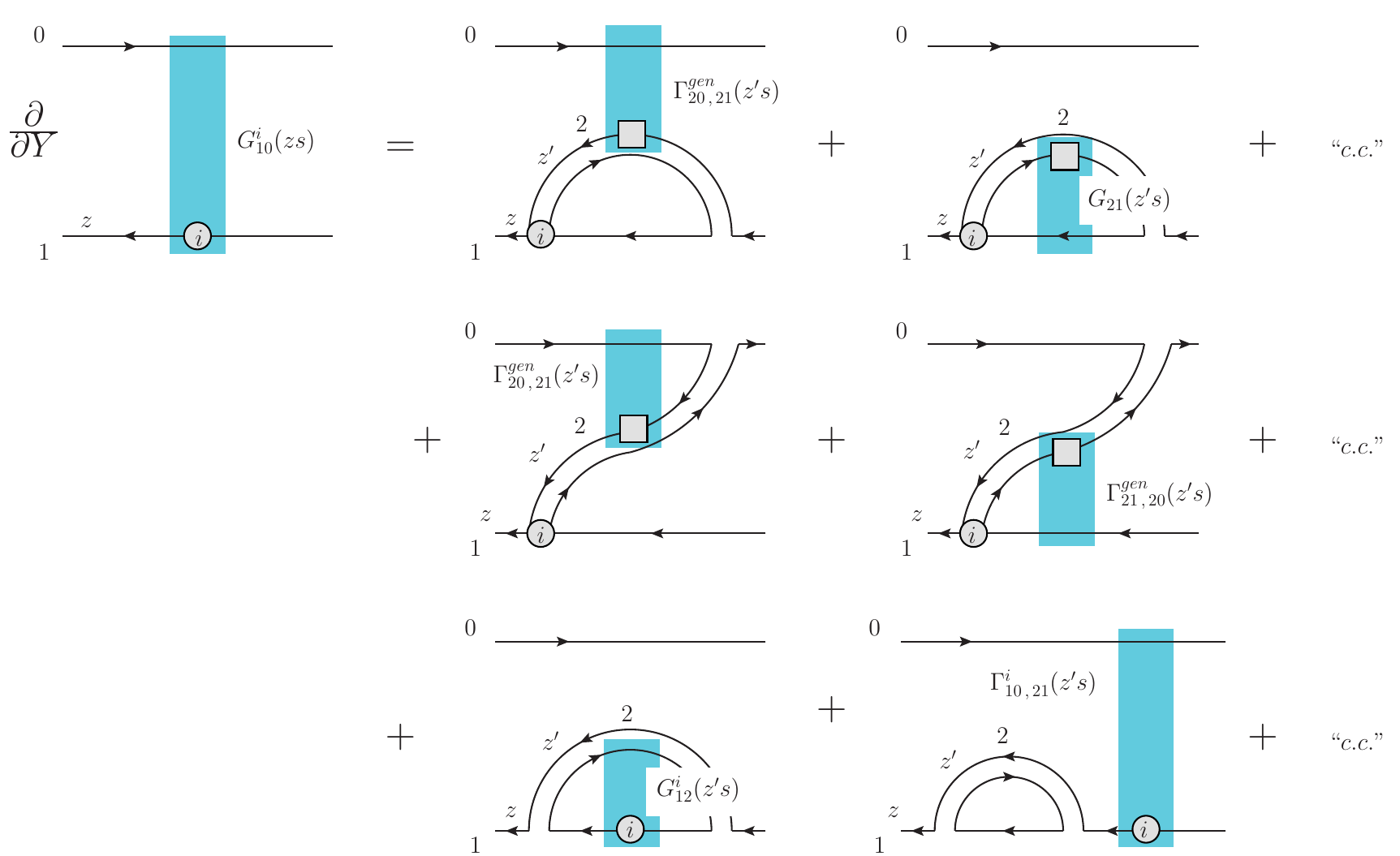} 
\caption{Linearized large-$N_c$ evolution of the new dipole function
  $G^i$ as written in \eq{M:evol4_1}.  Because there is no universal
  DLA parameter for the various terms, we have no a priori constraint
  on the relative sizes of $x_{20}$ and $x_{21}$, which makes
  enforcing lifetime ordering in these dipoles more subtle.  We must
  distinguish between the ladder emission of polarized gluons (top
  line), which are constrained by the lifetime of dipole $21$ only,
  and the non-ladder emission of polarized gluons (middle line), which
  are constrained by the lifetimes of both dipoles $20$ and $21$.  The
  ``+ c.c.'' stands for adding mirror-reflected diagrams as well as
  the true complex conjugates in which line $0$ becomes a polarized
  quark line.}
\label{f:MGievol}
\end{center}
\end{figure}
%

By our power counting, the subsequent evolution for the correlators in
the second line of \eq{e:Gevol3} should be DLA. Hence it should be
expressed in terms of the DLA amplitudes $G$ and $\Gamma$. Consider
specifically diagram V in \fig{fig:Gievol_v2}. When $x_{20} \ll
x_{21}$, the subsequent evolution in dipole $20$ is given by $G_{20}
(z' s)$. Conversely, when $x_{21} \ll x_{20}$, the subsequent
evolution in dipole $20$ is given by $\Gamma_{20,21} (z' s)$. With the
DLA accuracy of this subsequent evolution we can not distinguish, say,
$x_{21} < x_{20}$ from $x_{21} \ll x_{20}$. Therefore, to include both
the $x_{21} < x_{20}$ and $x_{21} > x_{20}$ regions of integration in
the second line of \eq{e:Gevol3} we define a new amplitude
\begin{align} 
  \label{Gen2}
  \Gamma^{gen}_{20,21} (z' s) = \theta (x_{20} - x_{21}) \,
  \Gamma_{20,21} (z' s) + \theta (x_{21} - x_{20}) \, G_{20} (z' s).
\end{align}
This amplitude $\Gamma^{gen}$ encompasses both regions of transverse
plane with the DLA accuracy, and is thus the proper amplitude to use
for diagrams IV, IV' in \fig{fig:Gievol_v2} when describing the
subsequent evolution in dipole $20$ and in diagrams V, V' when
describing the evolution in either of the daughter dipoles, $20$ or
$21$.

\begin{figure}[ht]
\begin{center}
\includegraphics[width= \textwidth]{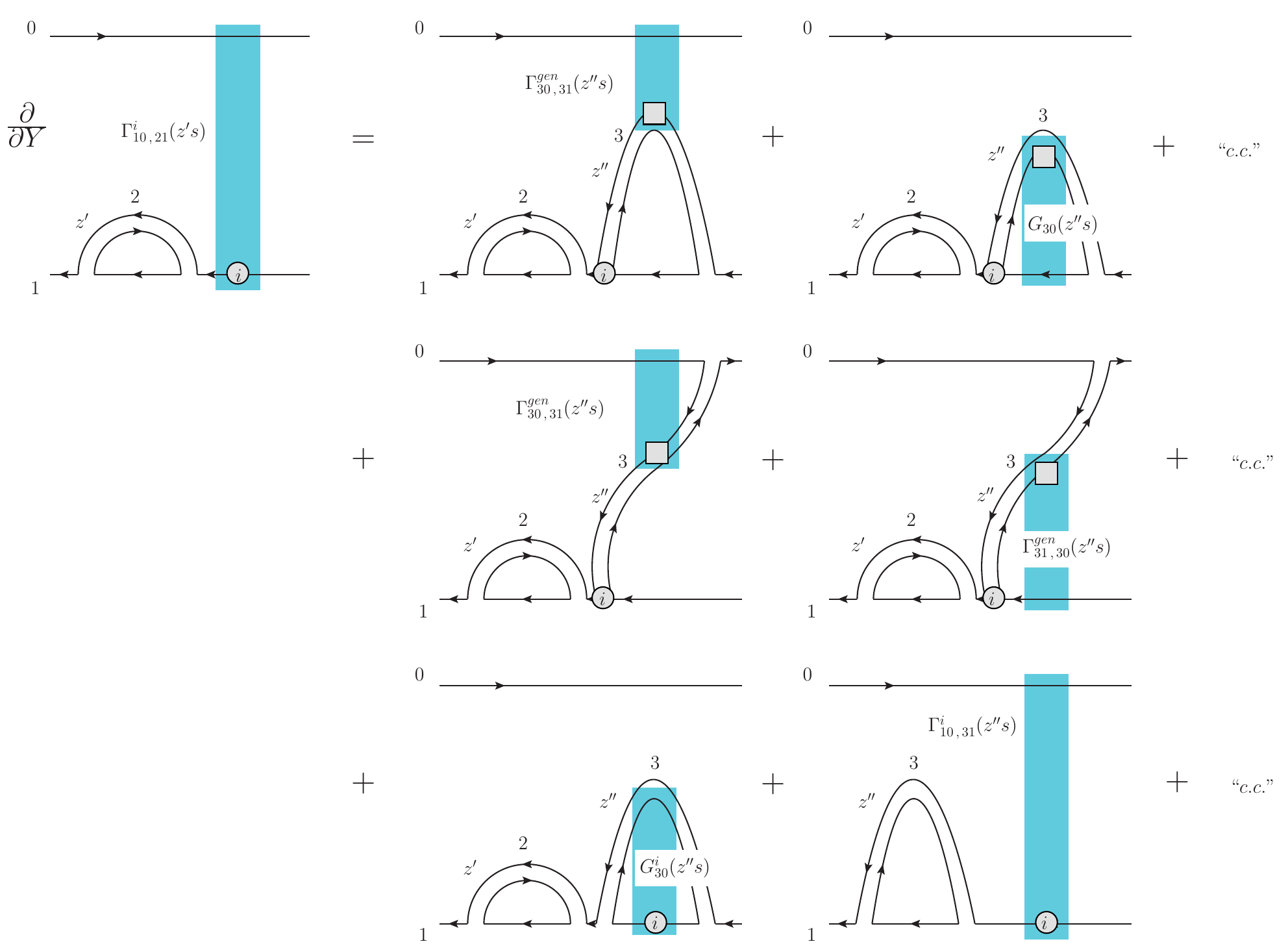} 
\caption{Linearized large-$N_c$ evolution of the new dipole function
  $\Gamma^i$ as written in \eq{M:evol4_2}.  Because there is no
  universal DLA parameter for the various terms, we have no a priori
  constraint on the relative sizes of $x_{30}$ and $x_{31}$, which
  makes enforcing lifetime ordering in these dipoles more subtle.  We
  must distinguish between the ladder emission of polarized gluons
  (top line), which are constrained by the lifetime of dipole $31$
  only, and the non-ladder emission of polarized gluons (middle line),
  which are constrained by the lifetimes of both dipoles $30$ and
  $31$.  The ``+ c.c.'' stands for adding mirror-reflected diagrams as
  well as the true complex conjugates in which line $1$ becomes a
  polarized quark line.}
\label{f:MGamievol}
\end{center}
\end{figure}
%

As a result of this analysis, we obtain the large-$N_c$ evolution
equations relevant for the dipole gluon helicity distribution,
\begin{subequations} \label{M:evol4}
\begin{align} 
  \label{M:evol4_1}
  G_{1 0}^{i} (z s) &= G_{1 0}^{i \, (0)} (z s) + \frac{\alpha_s
    N_c}{2\pi^2} \int\limits_{\frac{\Lambda^2}{s}}^z \frac{dz'}{z'}
  \int d^2 x_2 \, \ln\frac{1}{x_{21}^2 \Lambda^2} \: \frac{ \epsilon_T^{i
      j} \, (x_{21})_\bot^j }{ x_{21}^2 } \: \Big[ \Gamma_{20 \, , \,
    21}^{gen} (z' s) + G_{21} (z' s) \Big]
\notag \\ &
- \frac{\alpha_s N_c}{2\pi^2} \int\limits_{\frac{\Lambda^2}{s}}^z
\frac{dz'}{z'} \int d^2 x_2 \, \ln\frac{1}{x_{21}^2 \Lambda^2} \: \frac{
  \epsilon_T^{i j} \, (x_{20})_\bot^j }{ x_{20}^2 } \: \Big[
\Gamma_{20 \, , \, 21}^{gen} (z' s) + \Gamma_{21 \, , \, 20}^{gen} (z'
s) \Big]
\notag \\ &
+ \frac{\alpha_s N_c}{2\pi^2} \int\limits_{\frac{1}{x_{10}^2 s}}^z
\frac{dz'}{z'} \int \frac{d^2 x_2}{x_{21}^2} \: \theta\Big(x_{10}^2 -
x_{21}^2\Big) \, \theta\Big(x_{21}^2 - \frac{1}{z' s}\Big) \, \Big[
G_{12}^i (z' s) - \Gamma_{10 \, , \, 21}^i (z' s) \Big] ,
\\ \notag \\ \label{M:evol4_2}
\Gamma^i_{10, \, 21} (z' s) &= G_{10}^{i \, (0)} (z' s) +
\frac{\alpha_s N_c}{2\pi^2} \int\limits_{\frac{\Lambda^2}{s}}^{z'}
\frac{dz''}{z''} \int d^2 x_3 \, \ln\frac{1}{x_{31}^2 \Lambda^2} \: \frac{
  \epsilon_T^{i j} \, (x_{31})_\bot^j }{x_{31}^2} \: \Big[ \Gamma_{30
  \, , \, 31}^{gen} (z'' s) + G_{31} (z'' s) \Big]
\notag \\ &
- \frac{\alpha_s N_c}{2\pi^2} \int\limits_{\frac{\Lambda^2}{s}}^{z'}
\frac{dz''}{z''} \int d^2 x_3 \, \ln\frac{1}{x_{31}^2 \Lambda^2} \: \frac{
  \epsilon_T^{i j} \, (x_{30})_\bot^j }{x_{30}^2} \: \Big[ \Gamma_{30
  \, , \, 31}^{gen} (z'' s) + \Gamma_{31 \, , \, 30}^{gen} (z''
s)\Big]
\notag \\ &
+ \frac{\alpha_s N_c}{2\pi^2} \int\limits_{\frac{1}{x_{10}^2 s}}^{z'}
\frac{dz''}{z''} \int \frac{d^2 x_3}{x_{31}^2} \: \theta\Big(
\min\left[ x_{10}^2 \, , \, x_{21}^2 \tfrac{z'}{z''} \right] -
x_{31}^2 \Big) \: \theta\Big(x_{31}^2 - \frac{1}{z'' s} \Big) \, \Big[
G_{13}^i (z'' s) - \Gamma^i_{10 \, , \, 31} (z'' s) \Big] ,
\end{align}
\end{subequations}
which are illustrated in Figs.~\ref{f:MGievol} and
\ref{f:MGamievol}. The solution of these equations with the help of
\eq{M:dipdef} will give us the small-$x$ asymptotics of the dipole
gluon helicity TMD and, through this, of the gluon helicity PDF.

%
\section{Solution of the Evolution Equations for the Dipole Gluon
  Helicity}
\label{sec:solution}
%
%

%
\subsection{Structure of the Evolution Equations}
%

We will now proceed to simplify and solve the evolution equations
\eqref{M:evol4} for the polarized dipole amplitude $G_{10}^i$ at small
$x$.  First, it is convenient to convert from the vector quantity
$G_{10}^i (z s)$ to the scalar functions $G_1 (x_{10}^2 , z s)$ and
$G_2 (x_{10}^2 , z s)$ by integrating over impact parameters $\int d^2
b_{10} = \int d^2 b_{20} = \int d^2 b_{21}$ and using the
decomposition \eqref{decomp}. The same decomposition is applied to the
impact-parameter integral of $\Gamma^i_{10, \, 21} (z' s)$. From
\eq{M:Gdipcoord}, we see that the dipole gluon helicity distribution
couples to the $G_2$ function, which can be extracted using the
projection
\begin{align}
\label{proj}
  G_2 (x_{10}^2 , z s) = - \frac{(x_{10})_\bot^i \epsilon_T^{i
      j}}{x_{10}^2} \int d^2 b_{10} \: G_{10}^j (z s) .
\end{align}
In doing the impact parameter integral, the $G_{12}^i$ term from the
unpolarized eikonal evolution (third line of \eqref{M:evol4_1})
drops out due to the angular integration.  Similarly, the $G_{21}$
term in the polarized ladder evolution (first line of
\eqref{M:evol4_1}) appears to vanish due to the angular integral.
However, the radial integral in the kernel is potentially IR divergent
without this term, so we will keep this contribution for now.  After
performing the impact parameter integral of Eqs.~\eqref{M:evol4} along
with the projection \eqref{proj}, we obtain
\begin{subequations} \label{M:evol6}
  \begin{align} 
    \label{M:evol6_1}
    G_2 (x_{10}^2 , z s) &= G_2^{(0)} (x_{10}^2 , z s) +
    \frac{\alpha_s N_c}{2\pi^2} \int\limits_{\frac{\Lambda^2}{s}}^z
    \frac{dz'}{z'} \int d^2 x_2 \, \ln\frac{1}{x_{21}^2 \Lambda^2} \:
    \frac{\ul{x}_{10} \cdot \ul{x}_{21}}{x_{10}^2 \, x_{21}^2} \:
    \Big[ \Gamma_{gen} (x_{20}^2 , x_{21}^2 , z' s) + G(x_{21}^2 , z'
    s) \Big]
\notag \\ &
- \frac{\alpha_s N_c}{2\pi^2} \int\limits_{\frac{\Lambda^2}{s}}^z
\frac{dz'}{z'} \int d^2 x_2 \, \ln\frac{1}{x_{21}^2 \Lambda^2} \:
\frac{\ul{x}_{10} \cdot \ul{x}_{20}}{x_{10}^2 \, x_{20}^2} \: \Big[
\Gamma_{gen} (x_{20}^2 , x_{21}^2 , z' s) + \Gamma_{gen} (x_{21}^2 ,
x_{20}^2 , z' s)\Big]
\notag \\ &
- \frac{\alpha_s N_c}{2\pi} \int\limits_{\frac{1}{x_{10}^2 s}}^z
\frac{dz'}{z'} \int\limits_{\frac{1}{z' s}}^{x_{10}^2} \frac{d
  x_{21}^2}{x_{21}^2} \, \Gamma_2 (x_{10}^2 , x_{21}^2 , z' s),
\\ \notag \\ \label{M:evol6_2}
\Gamma_2 (x_{10}^2 , x_{21}^2 , z' s) &= G_2^{(0)} (x_{10}^2 , z' s) +
\frac{\alpha_s N_c}{2\pi^2} \int\limits_{\frac{\Lambda^2}{s}}^{z'}
\frac{dz''}{z''} \int d^2 x_3 \, \ln\frac{1}{x_{31}^2 \Lambda^2} \:
\frac{\ul{x}_{10} \cdot \ul{x}_{31}}{x_{10}^2 \, x_{31}^2} \: \Big[
\Gamma_{gen} (x_{30}^2 , x_{31}^2 , z'' s) + G(x_{31}^2 , z' s) \Big]
\notag \\ &
- \frac{\alpha_s N_c}{2\pi^2} \int\limits_{\frac{\Lambda^2}{s}}^{z'}
\frac{dz''}{z''} \int d^2 x_3 \, \ln\frac{1}{x_{31}^2 \Lambda^2} \:
\frac{\ul{x}_{10} \cdot \ul{x}_{30}}{x_{10}^2 \, x_{30}^2} \: \Big[
\Gamma_{gen} (x_{30}^2 , x_{31}^2 , z'' s) + \Gamma_{gen} (x_{31}^2 ,
x_{30}^2 , z'' s)\Big]
\notag \\ &
- \frac{\alpha_s N_c}{2\pi} \int\limits_{\frac{1}{x_{10}^2 s}}^{z'}
\frac{dz''}{z''} \int\limits_{\frac{1}{z'' s}}^{\min\left[ x_{10}^2 \,
    , \, x_{21}^2 \tfrac{z'}{z''} \right]} \frac{dx_{31}^2}{x_{31}^2}
\, \Gamma_2 (x_{10}^2 , x_{31}^2 , z'' s) .
\end{align}
\end{subequations}
We have defined an impact-parameter integrated amplitude
$\Gamma_{gen}$ by (cf. \eq{Gen2})
\begin{align} 
  \label{Gen}
  \Gamma_{gen} (x_{20}, x_{21}, z' s) = \theta (x_{20} - x_{21}) \,
  \Gamma (x_{20}, x_{21}, z' s) + \theta (x_{21} - x_{20}) \, G
  (x_{20}, z' s).
\end{align}
This function can be easily found using the analytic solution
\eqref{M:asymsol} for the asymptotics of $G$ and $\Gamma$ at high
energies.

The initial conditions for the scalar functions $G_2$ and $\Gamma_2$
in Eqs.~\eqref{M:evol6} follow from \eq{Ginit}:
\begin{align} 
  \label{Ginit2}
  G^{(0)}_2 (x_{10}^2 , z) = \Gamma^{(0)}_2 (x^2_{10}, x_{21}^2 , z')
  = - \frac{\alpha_s^2 C_F}{N_c} \, \pi \, \ln \frac{1}{x_{10} \,
    \Lambda}.
\end{align} 

It is useful to check that the transverse coordinate integral in the
LLA kernel of Eqs.~\eqref{M:evol6} (the first two lines of
\eqref{M:evol6_1} and \eqref{M:evol6_2}) is convergent. To see this,
let us use \eq{Gen} and \eq{M:asymsol} in Eqs.~\eqref{M:evol6}, to
check the behavior of the integrands in the $x_{21}^2 \gg x_{10}^2$
and $x_{21}^2 \ll x_{10}^2$ limits.  Although individual terms appear
to be logarithmically divergent in the IR, the sum of the terms scales
as
\begin{align}
  \int\limits^{\infty} \frac{dx_{21}^2}{(x_{21}^2)^{1.5-\alpha_h^q}}
  \ln\frac{1}{x_{21}^2 \Lambda^2} ,
\end{align}
which is convergent for $\alpha_h^q < \thalf$.  Noting from \eq{M:ahel}
that $\alpha_h^q \sim \sqrt{\alpha_s} \ll 1$, we conclude that this
integral is convergent in the IR for perturbative $\alpha_s$.  In the
UV, the terms converge as
\begin{align}
  \int\limits_0 dx_{21}^2 \, \ln\frac{1}{x_{21}^2 \Lambda^2} \,
  (x_{21}^2)^{c \: \alpha_h^q}
\end{align}
with $c$ a positive constant depending on the term.  We therefore
conclude that the transverse coordinate integral in
Eqs.~\eqref{M:evol6} is convergent in both the UV and IR limits.

The most intricate part of Eqs.~\eqref{M:evol6} is the treatment of
the non-logarithmic transverse integral; we want to evaluate it as
completely as possible within our DLA accuracy. Focusing on the
evolution of $G_2$ in \eq{M:evol6_1}, that integral is
\begin{align} 
  \label{e:Jdef}
  J \equiv & \frac{\alpha_s N_c}{2\pi^2}
  \int\limits_{\frac{\Lambda^2}{s}}^z \frac{dz'}{z'} \int d^2 x_2 \,
  \ln\frac{1}{x_{21}^2 \Lambda^2} \: \frac{\ul{x}_{10} \cdot
    \ul{x}_{21}}{x_{10}^2 \, x_{21}^2} \Big[ \Gamma_{gen} (x_{20}^2 ,
  x_{21}^2 , z' s) + G (x_{21}^2 , z' s) \Big]
\notag \\ & 
- \frac{\alpha_s N_c}{2\pi^2} \int\limits_{\frac{\Lambda^2}{s}}^z
\frac{dz'}{z'} \int d^2 x_2 \: \ln\frac{1}{x_{21}^2 \Lambda^2} \:
\frac{\ul{x}_{10} \cdot \ul{x}_{20}}{x_{10}^2 \, x_{20}^2} \: \Big[
\Gamma_{gen} (x_{20}^2 , x_{21}^2 , z' s) + \Gamma_{gen} (x_{21}^2 ,
x_{20}^2 , z' s) \Big].
\end{align}
Next we insert the expression \eqref{Gen} for $\Gamma_{gen}$ and the
asymptotic solutions \eqref{M:asymsol}, scaling out the various
power-counting parameters:
\begin{align} 
  \label{M:J1}
  J = \frac{\alpha_s N_c}{2\pi^2} \int\limits_{\frac{\Lambda^2}{s}}^z
  \frac{dz'}{z'} \, (z' s \, x_{10}^2)^{\alpha_h^q} \: G_{0} \:
  j(x_{10}^2)
%
%
  = \left( \frac{\alpha_s N_c}{2\pi^2} \, \frac{1}{ \alpha_h^q } G_{0}
  \right) \, j(x_{10}^2) \, (z s \, x_{10}^2)^{\alpha_h^q} ,
\end{align}
where
\begin{align} 
  \label{M:J2}
  j(x_{10}^2) &\equiv \frac{1}{G_{0}} \int d^2 x_2 \,
  \ln\frac{1}{x_{21}^2 \Lambda^2} \, (z' s \, x_{10}^2)^{-\alpha_h^q}
\notag \\ & \times
\Bigg\{ \frac{\ul{x}_{10}}{x_{10}^2} \cdot \left(
  \frac{\ul{x}_{21}}{{x}_{21}^2} - \frac{\ul{x}_{20}}{{x}_{20}^2}
\right) \Big[ \theta(x_{21}^2 - x_{20}^2) \, G(x_{20}^2 , z' s) +
\theta (x_{20}^2 - x_{21}^2) \, \Gamma(x_{20}^2 , x_{21}^2 , z' s)
\Big]
\notag \\ & \hspace{1cm} -
\left( \frac{ \ul{x}_{10} \cdot \ul{x}_{20} }{ x_{10}^2 \, x_{20}^2 }
\right) \Big[ \theta(x_{20}^2 - x_{21}^2) \, G(x_{21}^2 , z' s) +
\theta (x_{21}^2 - x_{20}^2) \, \Gamma(x_{21}^2 , x_{20}^2 , z' s)
\Big]
\notag \\ & \hspace{1cm} +
\left( \frac{ \ul{x}_{10} \cdot \ul{x}_{21} }{ x_{10}^2 \, x_{21}^2 }
\right) \: G(x_{21}^2 , z' s) \Bigg\}.
\end{align}
Using the expressions in \eqref{M:asymsol} we write
\begin{align}
j(x_{10}^2) & =
 \frac{1}{3} \int d^2 x_2 \, \ln\frac{1}{x_{21}^2 \Lambda^2} \, 
\notag \\ & \times
\Bigg\{ \frac{\ul{x}_{10}}{x_{10}^2} \cdot \left(
  \frac{\ul{x}_{21}}{{x}_{21}^2} - \frac{\ul{x}_{20}}{{x}_{20}^2}
\right) \left[ \theta (x_{21}^2 - x_{20}^2) \, \left(
    \frac{x_{20}^2}{x_{10}^2} \right)^{\alpha_h^q} + \:\: \theta
  (x_{20}^2 - x_{21}^2) \, \left( \frac{x_{21}^2}{x_{10}^2}
  \right)^{\alpha_h^q} \, \left( 4 \left( \frac{x_{20}^2}{x_{21}^2}
    \right)^{\frac{\alpha_h^q}{4}} - 3 \right) \right]
\notag \\ & \hspace{1cm}
- \left( \frac{\ul{x}_{10} \cdot \ul{x}_{20}}{x_{10}^2 \, x_{20}^2}
\right) \left[ \theta (x_{20}^2 - x_{21}^2) \, \left(
    \frac{x_{21}^2}{x_{10}^2} \right)^{\alpha_h^q} + \:\: \theta
  (x_{21}^2 - x_{20}^2) \, \left( \frac{x_{20}^2}{x_{10}^2}
  \right)^{\alpha_h^q} \, \left(4 \left( \frac{x_{21}^2}{x_{20}^2}
    \right)^{\frac {\alpha_h^q}{4}} - 3 \right) \right]
\notag \\ & \hspace{1cm} \label{J1}
+ \left( \frac{\ul{x}_{10} \cdot \ul{x}_{21}}{x_{10}^2 \, x_{21}^2}
\right) \, \left( \frac{x_{21}^2}{x_{10}^2} \right)^{\alpha_h^q}
\Bigg\}.
\end{align}

Consider the DLA power counting in \eq{M:J1}.  This step of evolution
contains an explicit factor of $\alpha_s$, together with
$\frac{1}{\alpha_h^q}$ and $G_{0}$.  From \eq{M:ahel} we see that
$\frac{1}{\alpha_h^q} \sim \frac{1}{\sqrt{\alpha_s}}$, and from
\eq{init}, we see that the scaling initial conditions $G_{0}$ contain
a relative logarithm of energy, which also scales as
$\frac{1}{\sqrt{\alpha_s}}$ in the DLA power counting ($\alpha_s
\ln^2\tfrac{s}{\Lambda^2} \sim 1$ such that $\ln\tfrac{s}{\Lambda^2}
\sim \tfrac{1}{\sqrt{\alpha_s}}$ ).  The factor in parentheses in
\eqref{M:J1} is therefore an $\ord{1}$ step of evolution in this
limit, and the energy dependence $(z s \, x_{10}^2)^{\alpha_h^q}$ is
also an $\ord{1}$ resummation.  Next we note that the quantity
$j(x_{10}^2)$ in \eq{M:J2} is independent of the energy $z' s$, is a
dimensionless function of $x_{10}$ and $\Lambda$, and converges in the
IR, such that the IR cutoff $\Lambda$ enters only in a single
logarithm in the integrand.  Therefore, the general form of
$j(x_{10}^2)$ can be written as
\begin{align}
j(x_{10}^2) = f_1 (\alpha_s) \: \ln\frac{1}{x_{10} \Lambda} + f_2 (\alpha_s) ,
\end{align}
where $f_1$ and $f_2$ are some functions only of $\as$ and contain no
additional logarithms of energy or of $x_{10}$.  The residual
$\alpha_s$ dependence in $f_1$ and $f_2$ is thus not enhanced by any
logarithms and only contributes to higher-order non-logarithmic
corrections.  In this spirit we therefore set $\as \to 0$ in
\eq{M:J2}, replacing both $G$ and $\Gamma$ from \eqref{M:asymsol} by
$\frac{1}{3} G_{0}$, obtaining
\begin{align}
  j(x_{10}^2) = \frac{2}{3} \int d^2 x_2 \, \ln\frac{1}{x_{21}^2
    \Lambda^2} \, \frac{\ul{x}_{10}}{x_{10}^2} \cdot \left(
    \frac{\ul{x}_{21}}{{x}_{21}^2} - \frac{\ul{x}_{20}}{{x}_{20}^2}
  \right) .
\end{align}
The integral now is at most log-divergent in $x_{21}$, and even that
divergence is zero after the angular integrations. Writing $d^2 x_2 =
x_{21} d x_{21} d \phi$ we can eliminate the first term in parentheses
after the angular averaging.
\footnote{It appears important to first
  choose the integration variables for the whole integral, and then
  integrate both terms in parenthesis using the same variables. If one
  simply discards the first term in parentheses, and writes $d^2 x_2 =
  x_{20} d x_{20} d \phi'$ for the second term, the result appears to
  be IR divergent again due to an illegal variable shift in one of two
  divergent terms of an overall convergent integral.}
Angular integration in the second term gives (see Eq.~(A.14) of
\cite{Kovchegov:2012mbw})
\begin{align}
  j(x_{10}^2) = - \frac{4 \pi}{3} \frac{1}{x_{10}^2}
  \int\limits_0^\infty d x_{21} \, x_{21} \, \ln\frac{1}{x_{21}^2
    \Lambda^2} \, \theta (x_{10} - x_{21})
%
%
= - \frac{4 \pi}{3} \ln\frac{1}{x_{10} \Lambda} - \frac{2 \pi}{3}.
\end{align}
Neglecting the constant compared to the logarithm and substituting our
result back into \eq{M:J1} we arrive at
\begin{align}
  \label{J2}
  J = - \left( \frac{\alpha_s N_c}{3 \pi} \frac{1}{\alpha_h^q} \,
    G_{0} \right) \left( z s \, x_{10}^2 \right)^{\alpha_h^q} \,
  \ln\frac{1}{x_{10}^2 \Lambda^2}.
\end{align}
Employing \eq{J2} in Eqs.~\eqref{M:evol6} to replace the terms
containing $\Gamma_{gen}$ and $G$ yields
\begin{subequations} \label{Y:evol6}
  \begin{align} 
    \label{Y:evol6_1}
    G_2 (x_{10}^2 , z s) &= G_2^{(0)} (x_{10}^2 , z s) - \left(
      \frac{\alpha_s N_c}{3 \pi} \frac{1}{\alpha_h^q} G_{0} \right)
    \left( z s \, x_{10}^2 \right)^{\alpha_h^q} \, \ln\frac{1}{x_{10}^2
      \Lambda^2}
\notag \\ &
- \frac{\alpha_s N_c}{2\pi} \int\limits_{\frac{1}{x_{10}^2 s}}^z
\frac{dz'}{z'} \int\limits_{\frac{1}{z' s}}^{x_{10}^2} \frac{d
  x_{21}^2}{x_{21}^2} \, \Gamma_2 (x_{10}^2 , x_{21}^2 , z' s) ,
\\ \notag \\ \label{Y:evol6_2}
\Gamma_2 (x_{10}^2 , x_{21}^2 , z' s) &= G_2^{(0)} (x_{10}^2 , z' s) -
\left( \frac{\alpha_s N_c}{3 \pi} \frac{1}{\alpha_h^q} G_{0} \right)
\left( z' s \, x_{10}^2 \right)^{\alpha_h^q} \, \ln\frac{1}{x_{10}^2
  \Lambda^2} \notag \\ &
- \frac{\alpha_s N_c}{2\pi} \int\limits_{\frac{1}{x_{10}^2 s}}^{z'}
\frac{dz''}{z''} \int\limits_{\frac{1}{z'' s}}^{\min\left[ x_{10}^2 \,
    , \, x_{21}^2 \tfrac{z'}{z''} \right]} \frac{dx_{31}^2}{x_{31}^2}
\, \Gamma_2 (x_{10}^2 , x_{31}^2 , z'' s) .
\end{align}
\end{subequations}
This leaves the simplified equations \eqref{Y:evol6} amenable to
analytic solution, which we will pursue next.

%
\subsection{High-Energy Asymptotics}
%

To begin, it is convenient to rescale the functions $G_2$ and
$\Gamma_2$ to eliminate the constants:
\begin{subequations} \label{M:scale}
\begin{align}
  G_2 &\equiv \left( - \frac{\alpha_s N_c}{3 \pi} \frac{1}{\alpha_h^q}
    G_{0} \, \ln\frac{1}{x_{10}^2 \Lambda^2} \right) \: {\bar G}_2 =
  \left( - \frac{G_{0}}{2\sqrt{3}} \, \sqrt{\frac{\alpha_s N_c}{2
        \pi}} \, \ln\frac{1}{x_{10}^2 \Lambda^2} \right) \: {\bar G}_2 ,
\\
\Gamma_2 &\equiv \left( - \frac{\alpha_s N_c}{3 \pi}
  \frac{1}{\alpha_h^q} G_{0} \, \ln\frac{1}{x_{10}^2 \Lambda^2} \right) \:
{\bar \Gamma}_2 = \left( - \frac{G_{0}}{2\sqrt{3}} \,
  \sqrt{\frac{\alpha_s N_c}{2 \pi}} \, \ln\frac{1}{x_{10}^2 \Lambda^2}
\right) \: {\bar \Gamma}_2,
\end{align}
\end{subequations}
which casts \eq{Y:evol6} into the form
\begin{subequations} \label{M:evol7}
\begin{align} 
  \label{M:evol7_1}
  \bar{G}_2 (x_{10}^2 , z s) &= \left( z s \,
    x_{10}^2 \right)^{\alpha_h^q} - \frac{\alpha_s N_c}{2\pi}
  \int\limits_{\frac{1}{x_{10}^2 s}}^z \frac{dz'}{z'}
  \int\limits_{\frac{1}{z' s}}^{x_{10}^2} \frac{d x_{21}^2}{x_{21}^2}
  \, \bar{\Gamma}_2 (x_{10}^2 , x_{21}^2 , z' s) ,
\\ \notag \\ \label{M:evol7_2}
\bar{\Gamma}_2 (x_{10}^2 , x_{21}^2 , z' s) &= 
\left( z' s \, x_{10}^2 \right)^{\alpha_h^q} - \frac{\alpha_s
  N_c}{2\pi} \int\limits_{\frac{1}{x_{10}^2 s}}^{z'} \frac{dz''}{z''}
\int\limits_{\frac{1}{z'' s}}^{\min\left[ x_{10}^2 \, , \, x_{21}^2
    \tfrac{z'}{z''} \right]} \frac{dx_{31}^2}{x_{31}^2} \,
\bar{\Gamma}_2 (x_{10}^2 , x_{31}^2 , z'' s) ,
\end{align}
\end{subequations}
where we have neglected the initial conditions for $G_2$ and
$\Gamma_2$ as small when compared to the $J$-term from \eq{J2}.
Introducing the logarithmic variables
\begin{subequations}
\label{e:eta_s}
\begin{align}
\eta &\equiv \sqrt{\frac{\alpha_s N_c}{2\pi}} \ln\frac{z s}{\Lambda^2} , \hspace{0.25cm}&
s_{10} &\equiv \sqrt{\frac{\alpha_s N_c}{2\pi}} \ln\frac{1}{x_{10}^2 \Lambda^2} , 
\\
\eta' &\equiv \sqrt{\frac{\alpha_s N_c}{2\pi}} \ln\frac{z' s}{\Lambda^2} , \hspace{0.25cm}&
s_{21} &\equiv \sqrt{\frac{\alpha_s N_c}{2\pi}} \ln\frac{1}{x_{21}^2 \Lambda^2} ,
\\
\eta'' &\equiv \sqrt{\frac{\alpha_s N_c}{2\pi}} \ln\frac{z'' s}{\Lambda^2} , \hspace{0.25cm}&
s_{31} &\equiv \sqrt{\frac{\alpha_s N_c}{2\pi}} \ln\frac{1}{x_{31}^2 \Lambda^2},
\end{align}
\end{subequations}
along with the scaling variables
\begin{subequations}\label{scal_var}
\begin{align}
\chi &\equiv \eta - s_{10} = \sqrt{\frac{\alpha_s N_c}{2\pi}} \ln (z s \, x_{10}^2), & &
\\
\zeta &\equiv \eta' - s_{10} = \sqrt{\frac{\alpha_s N_c}{2\pi}} \ln (z' s \, x_{10}^2)
, \hspace{0.25cm}&
\zeta' &\equiv \eta' - s_{21} = \sqrt{\frac{\alpha_s N_c}{2\pi}} \ln (z' s \, x_{21}^2),
\\
\xi &\equiv \eta'' - s_{10} = \sqrt{\frac{\alpha_s N_c}{2\pi}} \ln (z'' s \, x_{10}^2) 
, \hspace{0.25cm}&
\xi' &\equiv \eta'' - s_{31} = \sqrt{\frac{\alpha_s N_c}{2\pi}} \ln (z'' s \, x_{31}^2) ,
\end{align}
\end{subequations}
and the rescaled intercept as $\hat{\alpha}^q_h \equiv
\frac{4}{\sqrt{3}}$, we can rewrite Eqs.~\eqref{M:evol7} in the simple
form
\begin{subequations} \label{Y:evol7}
  \begin{align} 
    \label{Y:evol7_1}
    \bar{G}_2 (\chi) &= 
    e^{\hat{\alpha}^q_h \chi} - \int\limits_{0}^\chi d \zeta
    \int\limits_{0}^{\zeta} d \zeta' \, \bar{\Gamma}_2 (\zeta ,
    \zeta'),
\\ \label{Y:evol7_2}
\bar{\Gamma}_2 ( \zeta , \zeta') &= 
e^{\hat{\alpha}^q_h \zeta} - \int\limits_{0}^{\zeta'} d\xi
\int\limits_{0}^{\xi} d \xi' \, \bar{\Gamma}_2 ( \xi , \xi') -
\int\limits_{\zeta'}^\zeta d \xi \int\limits_{0}^{\zeta'} d \xi' \,
\bar{\Gamma}_2 (\xi , \xi').
\end{align}
\end{subequations}
Let us emphasize that, although we have expressed Eqs.~\eqref{Y:evol7}
in terms of scaling variables, we have not imposed a scaling form on
the functions, rather it resulted naturally from the form of the
equations.

Following the procedure used in \cite{Kovchegov:2017jxc} to obtain an
analytic solution for the quark helicity distribution, we first
differentiate Eqs.~\eqref{Y:evol7} to get
\begin{subequations} \label{Y:evol8}
\begin{align} 
  \label{Y:evol8_1}
  \frac{\partial}{\partial \chi} {\bar G}_2 (\chi) &=
 \hat{\alpha}^q_h\, e^{\hat{\alpha}^q_h \chi} - \int\limits_{0}^{\chi} d \zeta' \, {\bar
    \Gamma}_2 (\chi , \zeta'),
\\ \label{Y:evol8_2}
\frac{\partial}{\partial \zeta} {\bar \Gamma}_2 (\zeta, \zeta') &=
\hat{\alpha}^q_h\,e^{\hat{\alpha}^q_h \zeta} -
\int\limits_{0}^{\zeta'} d \xi' \, {\bar \Gamma}_2 (\zeta , \xi'),
\end{align}
\end{subequations}
with the boundary condition
\begin{align} 
  \label{e:Bdy} {\bar \Gamma}_2 (\zeta', \zeta') = {\bar G}_2 (\zeta')
  .
\end{align}
Next, we introduce the Laplace transforms
\begin{subequations} \label{e:Mellin}
\begin{align}
  & \hspace{-0.175cm} \! {\bar G}_2 (\chi) = \int \frac{d \omega}{2
    \pi i} \, e^{\omega \, \chi} \, {\bar G}_{2 \omega} ,\!  &
  \hspace{-0.15cm} &{\bar \Gamma}_2 (\zeta, \zeta') = \int \frac{d
    \omega}{2 \pi i} \, e^{\omega \, \zeta'} \, {\bar \Gamma}_{2
    \omega} (\zeta) ,
\\
& \hspace{-0.175cm} {\bar G}_{2 \omega} = \int\limits_0^\infty d\chi
\, e^{- \omega \chi} \, {\bar G}_2 (\chi),\!\!  & \hspace{-0.15cm} &
{\bar \Gamma}_{2 \omega} (\zeta) = \int\limits_0^\infty d\zeta' \,
e^{-\omega \zeta'} {\bar \Gamma}_2 (\zeta, \zeta'),
\end{align}
\end{subequations}
and start by focusing on \eq{Y:evol8_2}, obtaining
\begin{align}
  \frac{\partial}{\partial \zeta} {\bar \Gamma}_{2 \omega} (\zeta) =
  \frac{\hat{\alpha}^q_h}{\omega}\, e^{\hat{\alpha}^q_h \zeta} -
  \frac{1}{\omega}\, {\bar \Gamma}_{2 \omega} (\zeta).
\end{align}
This ODE is straightforward to solve, and the solution reads
\begin{align}
  {\bar \Gamma}_{2 \omega} (\zeta) = \frac{\hat{\alpha}^q_h}{1 + \hat{\alpha}^q_h \,
    \omega} \, e^{\hat{\alpha}^q_h \zeta} + \hat{\alpha}^q_h\,{C}_{\omega} \,
  e^{-\frac{\zeta}{\omega}}
\end{align}
with the integration ``constant'' $C_\omega$, such that
\begin{align} 
  {\bar \Gamma}_2 (\zeta, \zeta') & = \int \frac{d \omega}{2 \pi i} \,
  e^{\omega \, \zeta'} \, \left[ \frac{\hat{\alpha}^q_h}{1 + \hat{\alpha}^q_h \,
      \omega} \, e^{\hat{\alpha}^q_h \zeta} + \hat{\alpha}^q_h\,{C}_{\omega} \,
    e^{-\frac{\zeta}{\omega}} \right].
\end{align}
Collecting the pole at $\omega = - \frac{1}{\hat{\alpha}^q_h}$ and
using the boundary condition \eqref{e:Bdy} to obtain the corresponding
solution for $G$, we have
\begin{subequations} \label{M:soln1}
  \begin{align} 
    \label{G1} {\bar G}_2 (\chi) &= 
    e^{\left(\hat{\alpha}^q_h - \frac{1}{\hat{\alpha}^q_h}\right)
      \chi} + \int \frac{d \omega}{2 \pi i} \, \hat{\alpha}^q_h\,{C}_{\omega} \,
    e^{\left( \omega - \frac{1}{\omega}\right) \chi} ,
\\ \label{Gam1}
{\bar \Gamma}_2 (\zeta, \zeta') &= 
e^{\hat{\alpha}^q_h \zeta - \frac{1}{\hat{\alpha}^q_h} \zeta'} + \int
\frac{d \omega}{2 \pi i} \, \hat{\alpha}^q_h\,{C}_{\omega} \, e^{\omega \,
  \zeta'-\frac{\zeta}{\omega}} .
\end{align}
\end{subequations}

The integration constants $C_\omega$ can be constrained by
back-substituting the solution \eqref{M:soln1} into the differential
equations \eqref{Y:evol8}.  Plugging \eq{Gam1} into \eq{Y:evol8_2} we
arrive at the condition
\begin{align}
  \int \frac{d \omega}{2 \pi i} \, \frac{1}{\omega} C_{\omega} \,
  e^{-\frac{\zeta}{\omega}} =0,
\end{align}
and similarly, using \eq{G1} in \eq{Y:evol8_1}, we obtain
\begin{align}
  \label{Ceq}
  \int \frac{d \omega}{2 \pi i} \, \omega \, C_{\omega} \, e^{\left(
      \omega - \frac{1}{\omega} \right) \, \zeta} =
  \frac{1}{(\hat{\alpha}^q_h)^2} \, e^{\left(\hat{\alpha}^q_h -
      \frac{1}{\hat{\alpha}^q_h}\right) \zeta} .
\end{align}
This equation is hard to solve exactly, but it is straightforward to
match the large-$\zeta$ asymptotics.  In \eq{Ceq}, there is a pole at
$\omega =0$ in the exponent which can be shown to give a contribution
that asymptotes to zero as $\zeta \to \infty$ (see Appendix~\ref{A}
for the calculation). Hence, to make \eq{Ceq} be valid at $\zeta
\to \infty$, we simply need $C_\omega$ to contain a pole $\omega =
\hat{\alpha}^q_h$, with an appropriate choice of the coefficient:
\begin{align}\label{Csol}
  C_\omega = \frac{1}{(\hat{\alpha}_h^{q})^3} \frac{1}{\omega -
    \hat{\alpha}^q_h} .
\end{align}
We verify explicitly in Appendix~\ref{A} that \eq{Csol} solves
\eq{Ceq} in the large-$\zeta$ asymptotics and that the $\omega = 0$
pole is suppressed.

The asymptotic solution to Eqs.~\eqref{Y:evol7} is thus (using
$\hat{\alpha}^q_h = \frac{4}{\sqrt{3}}$)
\begin{subequations}\label{GGam_sol}
\begin{align}\label{Gsol_analytic1}
  \bar{G}_2 (\chi \gg 1) &= \left( 1 +
    \frac{1}{(\hat{\alpha}_h^{q})^2} \right) e^{\left(\hat{\alpha}^q_h
      - \frac{1}{\hat{\alpha}^q_h}\right) \chi} = \frac{19}{16} \,
  e^{\frac{13}{4\sqrt{3}} \chi},
\\
\bar{\Gamma}_2 (\zeta \gg 1 , \zeta' \gg 1) &=
e^{\hat{\alpha}^q_h \zeta -
    \frac{\zeta'}{\hat{\alpha}^q_h}} +
  \frac{1}{(\hat{\alpha}_h^{q})^2} \, e^{\hat{\alpha}^q_h \zeta' -
    \frac{\zeta}{\hat{\alpha}^q_h}} = 
e^{ \frac{4}{\sqrt{3}} \zeta - \frac{\sqrt{3}}{4} \zeta'} +
  \frac{3}{16} \, e^{ \frac{4}{\sqrt{3}} \zeta' - \frac{\sqrt{3}}{4}
    \zeta} . \label{Gamma_sol_analytic1}
\end{align}
\end{subequations}

\begin{figure}[ht]
\begin{center}
\includegraphics[width= 0.45 \textwidth]{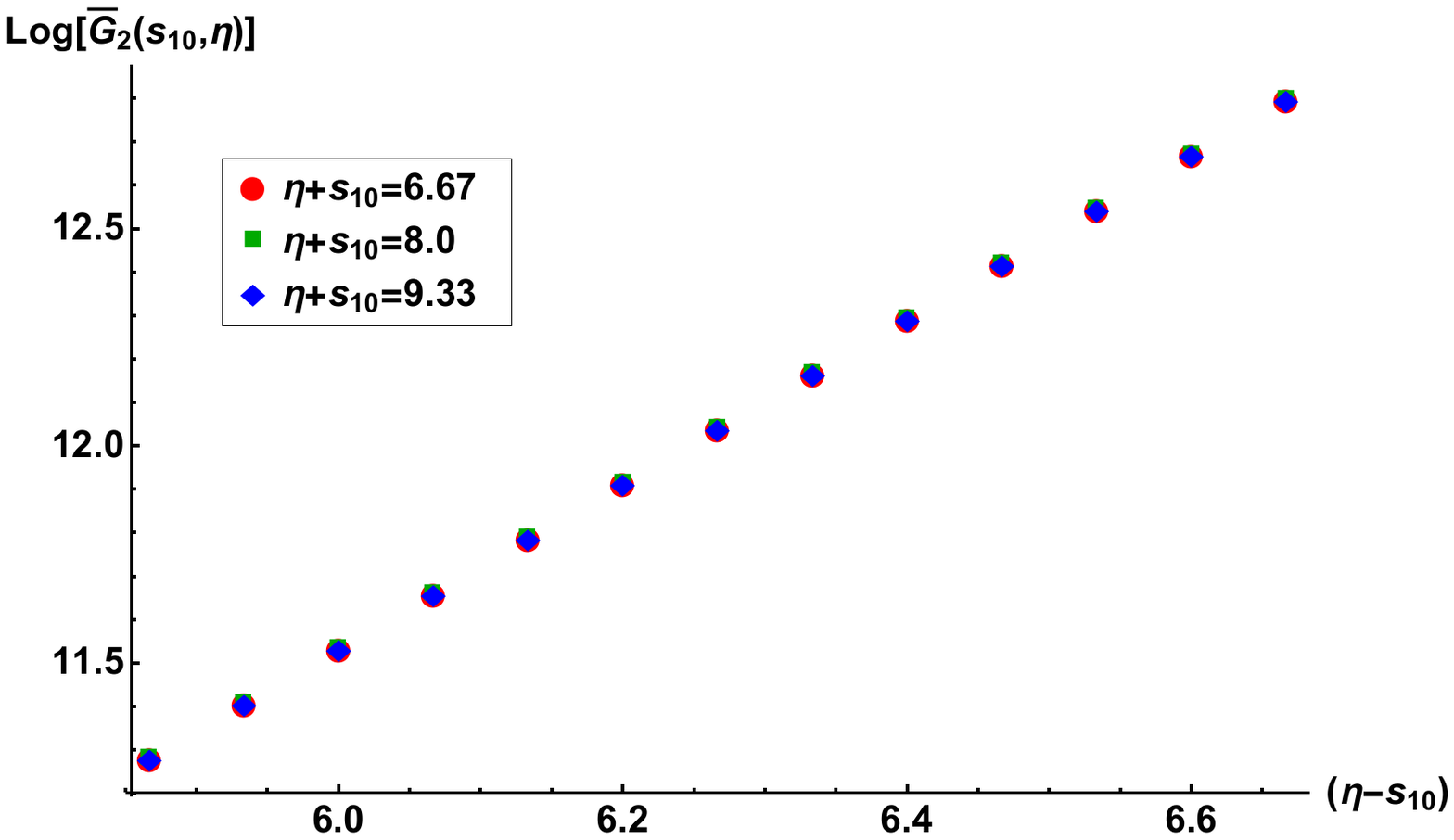} \hspace{1mm} \includegraphics[width= 0.45 \textwidth]{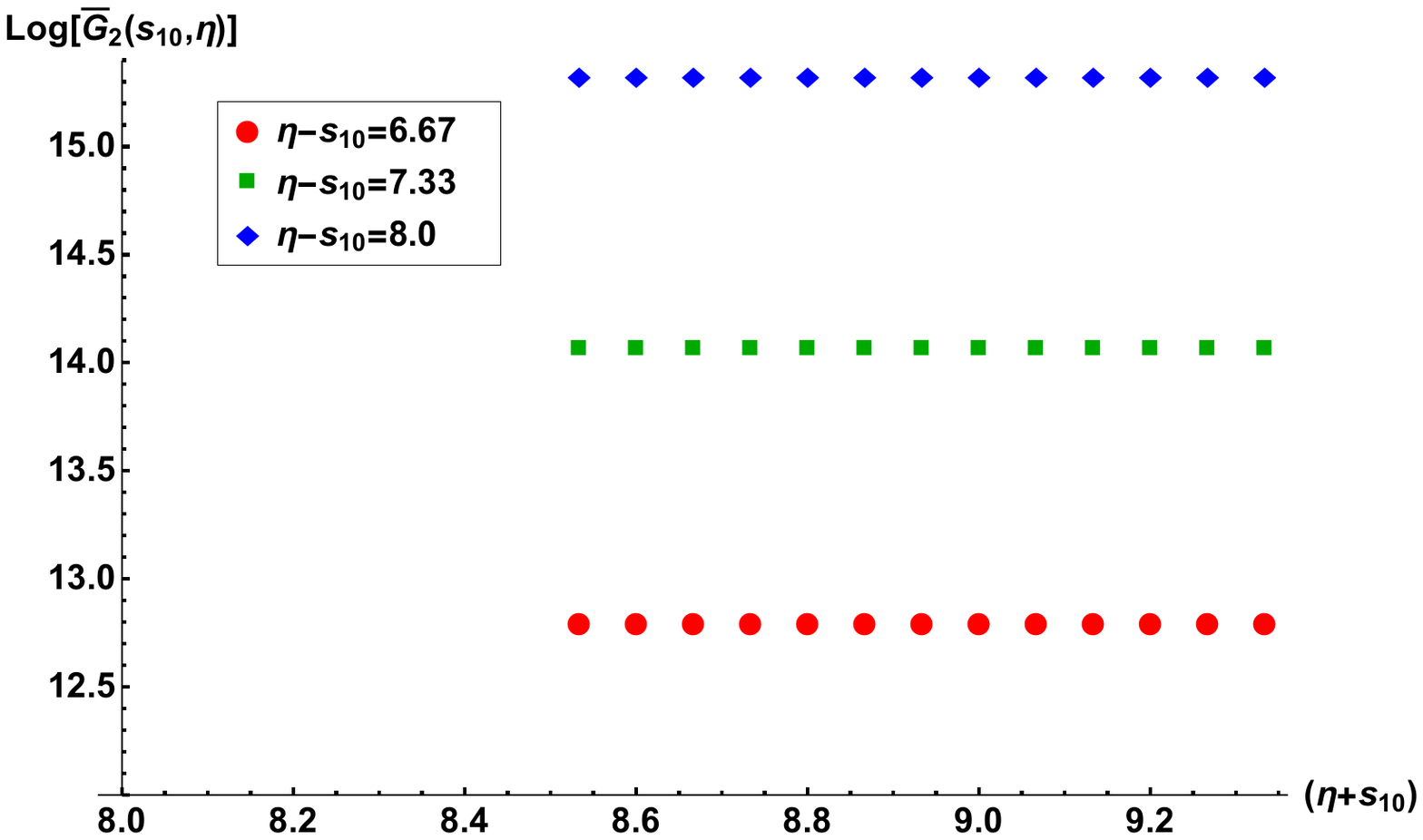}
\caption{Numerical solution of Eqs.~\eqref{Y:evol7} for $\ln {\bar
    G}_2$ plotted as a function of $\eta - s_{10}$ (for three
  different values of $\eta + s_{10}$) in the left panel and as a
  function of $\eta + s_{10}$ (for three different values of $\eta -
  s_{10}$) in the right panel. Both panels demonstrate that ${\bar
    G}_2$ is only a function of $\eta - s_{10}$, as expected from
  \eq{Gsol_analytic1}.}
\label{f:Gcheck}
\end{center}
\end{figure}

Our analytic solution can be cross-checked numerically. We did this by
solving Eqs.~\eqref{Y:evol7} on a discretized grid, exactly analogous
to what we did in Ref.~\cite{Kovchegov:2016weo}.  The resulting
numerical solution of ${\bar G}_2$ is shown in \fig{f:Gcheck} for a
grid spacing of 0.033.\footnote{This corresponds to using maximum
  $\eta$ and $s$ values (see Eqs.~(\ref{e:eta_s})) of 10 with a grid
  size of 300.}  These curves demonstrate the scaling behavior of
${\bar G}_2$ in agreement with our analytic result in
\eq{Gsol_analytic1}.  Moreover, from the slope of this curve we find
agreement with the exponent $13/(4\sqrt{3})$ of $\bar{G}_2$ to within
$1\%$.

%

\begin{figure}[ht]
\begin{center}
\includegraphics[width= 0.45 \textwidth]{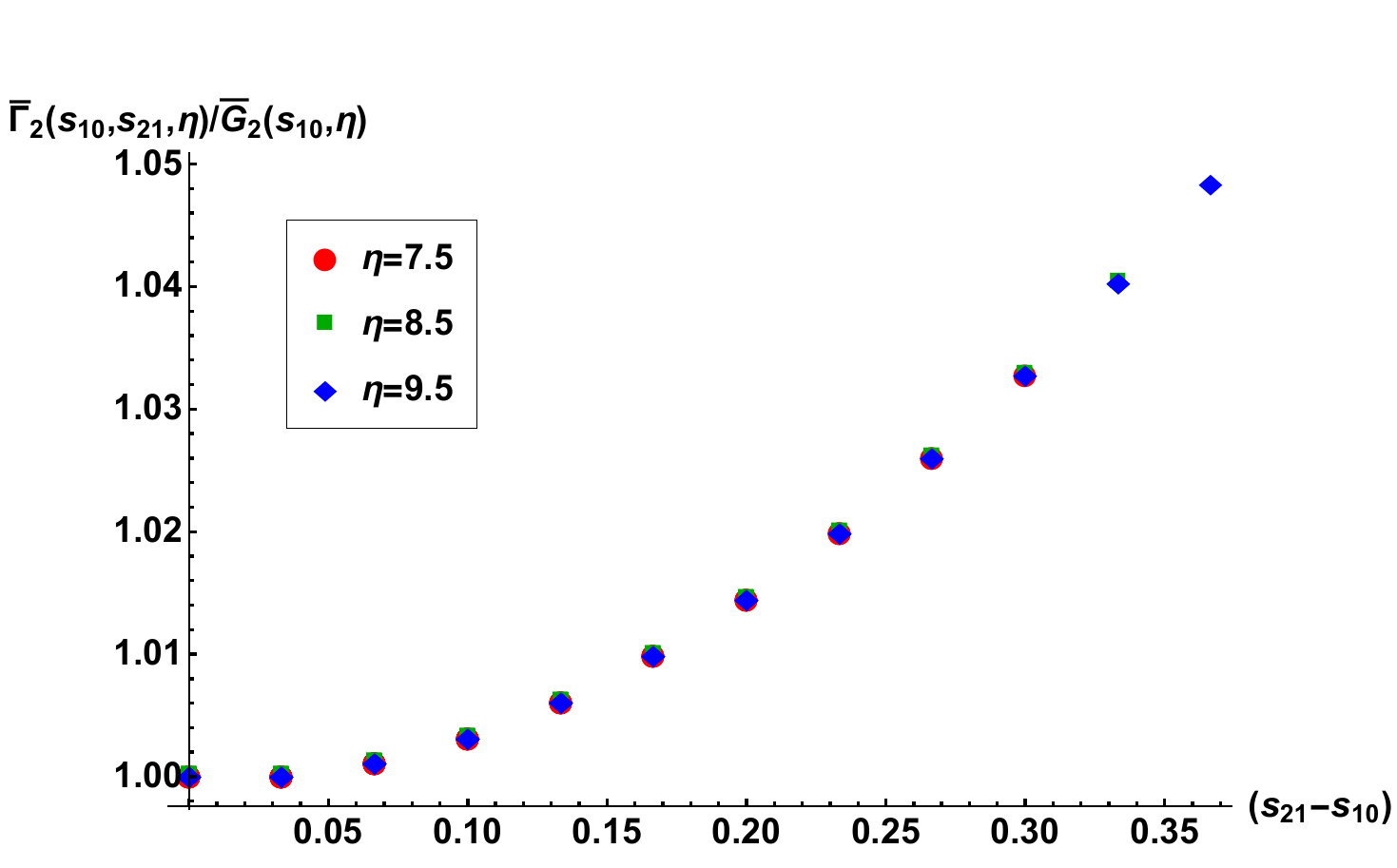} \hspace{1mm} \includegraphics[width= 0.45 \textwidth]{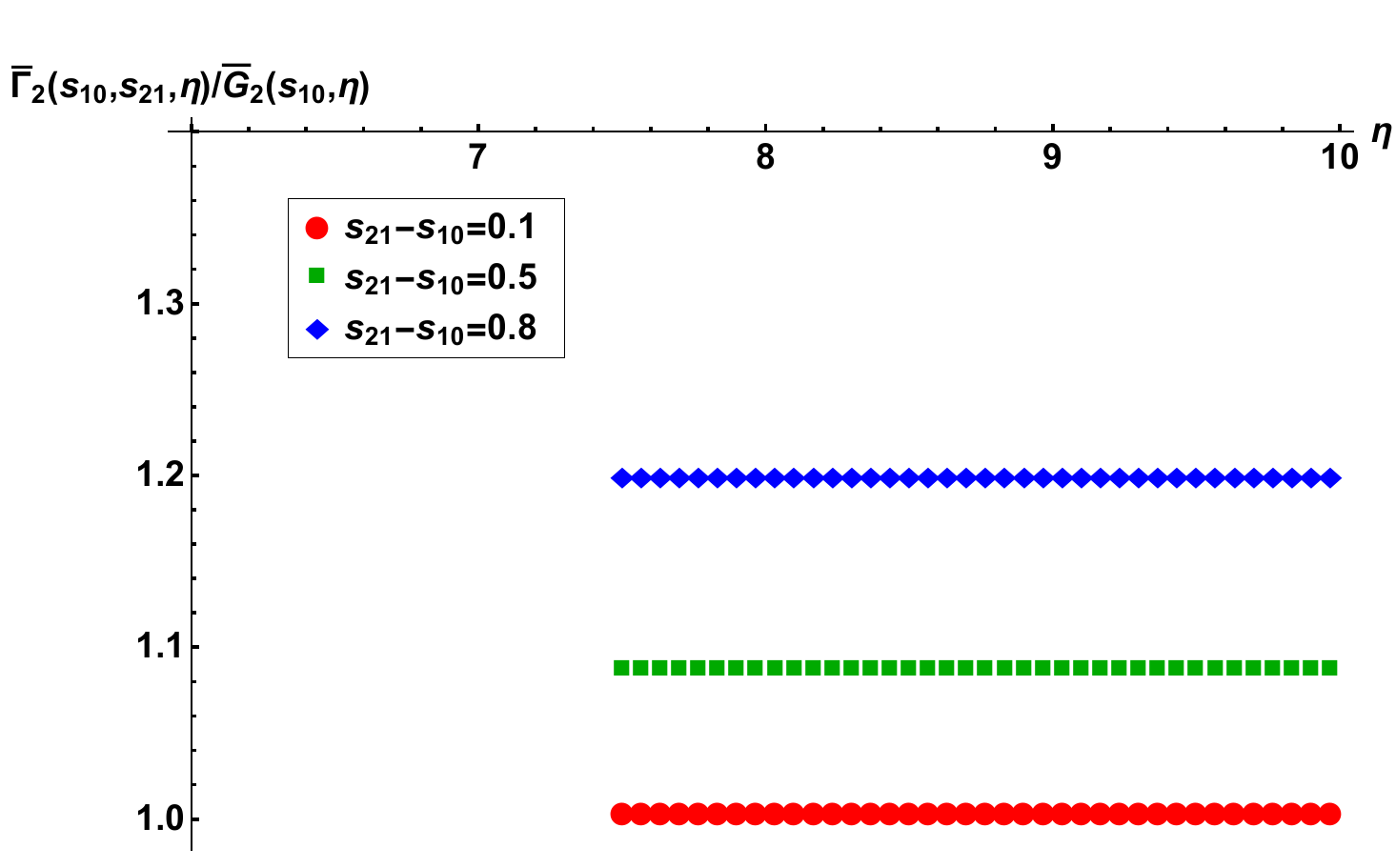}
\caption{Plot of the $\bar{\Gamma}_2/\bar{G}_2$ ratio given by the
  numerical solution of Eqs.~\eqref{Y:evol7} as a function of $s_{21}
  - s_{10}$ (for three different values of $\eta$) in the left panel
  and as a function of $\eta$ (for three different values of $s_{21} -
  s_{10}$) in the right panel. Both panels demonstrate that
  $\bar{\Gamma}_2/\bar{G}_2$ is only a function of $s_{21} - s_{10}$
  in agreement with \eq{eq:ratio}.}
\label{f:Gam_check}
\end{center}
\end{figure}

To cross-check our solution for ${\bar \Gamma}_2$ we take the ratio of
Eqs.~\eqref{Gamma_sol_analytic1} and \eqref{Gsol_analytic1} to obtain
\begin{align}
  \label{eq:ratio}
  \frac{\bar{\Gamma}_2 (\zeta, \zeta')}{\bar{G}_2 (\zeta)} =
  \frac{16}{19} \left[ e^{ \frac{\zeta-\zeta'}{\hat{\alpha}^q_h}} +
    \frac{1}{(\hat{\alpha}_h^{q})^2} \, e^{\hat{\alpha}^q_h (\zeta' -
      \zeta)} \right] = \frac{16}{19} \left[ e^{ \frac{\sqrt{3}}{4} \,
      ( s_{21} - s_{10})} + \frac{3}{16} \, e^{- \frac{4}{\sqrt{3}} \, (
      s_{21} - s_{10})} \right].
\end{align}
The ratio $\bar{\Gamma}_2/\bar{G}_2$ given by our numerical solution
is shown in \fig{f:Gam_check}. The plots demonstrate that the ratio
$\bar{\Gamma}_2/\bar{G}_2$ is only a function of $s_{21} - s_{10}$, in
agreement with our analytical result (\ref{eq:ratio}).  We likewise
were able to confirm in the physical region $s_{10} < s_{21} < \eta$
the functional form of (\ref{eq:ratio}), where we found agreement with
the exponent $\sqrt{3}/4$ to within $5\%$ and the coefficient $16/19$
to within $<0.5\%$.  Thus, we have numerically confirmed our analytic
solution for both $\bar{G}_2$ and $\bar{\Gamma}_2$.

%

%

Finally, converting Eqs.~\eqref{GGam_sol} back into the standard
variables by using Eqs.~\eqref{scal_var} and reinserting the scaling
factors from \eq{M:scale} gives us our final answer
\begin{subequations}
  \begin{align} 
    \label{G3}
    G_2 ( x_{10}^2 , z s) & \approx - \frac{19}{32 \sqrt{3} }
    \sqrt{\frac{\as \, N_c}{2 \pi}} \, G_{0} \: \ln\frac{1}{x_{10}^2
      \Lambda^2} \: (z s \, x_{10}^2)^{\frac{13}{4 \sqrt{3}}\sqrt{\frac{\alpha_s N_c} {2\pi}}} ,
\\ \label{Gam3}
\Gamma_2 (x_{10}^2 , x_{21}^2 , z' s) & \approx - \frac{1}{2 \sqrt{3}}
\sqrt{\frac{\as \, N_c}{2 \pi}} \, G_{0} \: \ln\frac{1}{x_{10}^2
  \Lambda^2} \left[ (z' s \, x_{10}^2)^{\frac{4}{\sqrt{3}}\sqrt{\frac{\alpha_s N_c} {2\pi}}} \, (z' s \,
  x_{21}^2)^{-\frac{\sqrt{3}}{4}\sqrt{\frac{\alpha_s N_c} {2\pi}}}\right. \nonumber\\
  &\hspace{4.75cm} \left.+\, \frac{3}{16} \, (z' s \,
  x_{21}^2)^{\frac{4}{\sqrt{3}}\sqrt{\frac{\alpha_s N_c} {2\pi}}} \, (z' s \,
  x_{10}^2)^{-\frac{\sqrt{3}}{4}\sqrt{\frac{\alpha_s N_c} {2\pi}}} \right] .
 \end{align}
\end{subequations}

The asymptotic solution \eqref{G3} for the polarized dipole amplitude
$G_2$ is the central result of this work.  Substituting the solution
\eq{G3} into \eq{M:Gdipcoord} yields the small-$x$ asymptotics of the
dipole gluon helicity distribution:
\begin{align}
  g_{1L}^{G \, dip} (x, k_T^2) \sim G_2 (x_{10}^2 , z s =
  \tfrac{Q^2}{x}) \sim \left( \frac{1}{x} \right)^{\alpha^G_h}
\end{align}
with the gluon helicity intercept
\begin{align} \label{e:Gint} \alpha^G_h = \frac{13}{4 \sqrt{3}} \,
  \sqrt{\frac{\as \, N_c}{2 \pi}} \approx 1.88 \, \sqrt{\frac{\as \,
      N_c}{2 \pi}} .
\end{align}
Strictly speaking, this intercept has been obtained by solving the
small-$x$ evolution equations \eqref{M:evol4} applicable to the dipole
gluon helicity distribution \eqref{M:Gdipcoord}.  The
Weizs\"{a}cker-Williams gluon helicity distribution \eqref{M:GWWcoord}
is defined by a different operator \eqref{eq:Gijdef} than the dipole
gluon helicity distribution \eqref{eq:Gidef}, and in general will have
different evolution equations than \eqref{M:evol4}.  While we leave
the derivation and solution of these evolution equations for future
work, we note that both the dipole and WW gluon helicity TMDs must
give the same gluon helicity PDF $\Delta G$ when integrated over all
$k_T$.  Integrating Eqs.~\eqref{M:Gdipcoord} and \eqref{M:GWWcoord}
over the transverse momentum to obtain the collinear gluon helicity
distribution $\Delta G$, we confirm that both distributions reduce to
a common operator, and that all three distributions possess the same
small-$x$ asymptotics:
\begin{align} 
  \label{e:coll}
  \Delta G (x, Q^2) &= \int d^2 k \, g_{1L}^{G \, WW} (x, k_T^2) =
  \int d^2 k \, g_{1L}^{G \, dip} (x, k_T^2)
\notag \\ &=
\frac{1}{\alpha_s \, 2\pi^2} \int d^2 x_0 \, \epsilon_T^{i j}
\left\langle \tr\left[ (V_{\ul 0}^{pol})_\bot^i \,
    \left(\frac{\partial}{\partial (x_0)_\bot^j} V_{\ul 0}^\dagger
    \right) \right] + \cc \right\rangle
\notag \\ &=
\frac{- 2 N_c}{\alpha_s \pi^2} \left[ \left( 1 + x_{10}^2
    \frac{\partial}{\partial x_{10}^2} \right) \: G_2 (x_{10}^2 , z s
  = \tfrac{Q^2}{x}) \right]_{x_{10}^2 = \tfrac{1}{Q^2}} .
\end{align}
We conclude that
\begin{align} \label{e:MAINRESULT}
  \Delta G (x, Q^2) \sim \left(\frac{1}{x}\right)^{\alpha_h^G} \sim
  \left(\frac{1}{x}\right)^{\frac{13}{4 \sqrt{3}} \, \sqrt{\frac{\as
        \, N_c}{2 \pi}}} \sim \left(\frac{1}{x}\right)^{1.88 \,
    \sqrt{\frac{\as \, N_c}{2 \pi}}}.
\end{align}
Thus, we see that the small-$x$ asymptotics of these three
distributions ($\Delta G$, $g_{1L}^{G\,dip}$, $g_{1L}^{G\,WW}$) -- and, indeed, {\it{all}} possible definitions of gluon
helicity TMDs -- are universal and governed by the gluon helicity
intercept \eqref{e:Gint}.

%
\section{Phenomenology of the Gluon Spin at Small {\bf \it  x}}
\label{sec:pheno}
%

In this section we give an estimate for the gluon spin $S_G$ in
(\ref{eq:net_spin}) based on our gluon helicity intercept
(\ref{e:Gint}).  The gluon spin has been a topic of intense
investigation, with only recent experiments showing that it can give a
more substantial fraction of the proton's spin than once
thought~\cite{Adamczyk:2014ozi,Adare:2015ozj}. Nevertheless, the
estimates of $S_G$ are still plagued by the lack of data below $x =
0.05$, which causes large uncertainties in this quantity (see, e.g.,
Ref.~\cite{Aschenauer:2015ata}), and is one of the main motivations
for the construction of an Electron-Ion Collider.  However, we
emphasize that once our theoretical calculations of the gluon (and
quark) helicity intercepts push beyond the current approximations and
include, e.g., large-$N_c\&N_f$, running coupling, and LLA
corrections, one could use these results in future extractions of the
already existing data to provide strong constraints on the small-$x$
behavior of the helicity PDFs, and, consequently, the quark and gluon
spin.  (We mention that helicity PDFs have been extracted by several
groups, e.g., DSSV~\cite{deFlorian:2009vb,deFlorian:2014yva},
JAM~\cite{Jimenez-Delgado:2013boa,Sato:2016tuz,Ethier:2017zbq},
LSS~\cite{Leader:2005ci,Leader:2010rb,Leader:2014uua},
NNPDF~\cite{Ball:2013lla,Nocera:2014gqa}.)

\begin{figure}[ht]
\centering \includegraphics[scale=0.5]{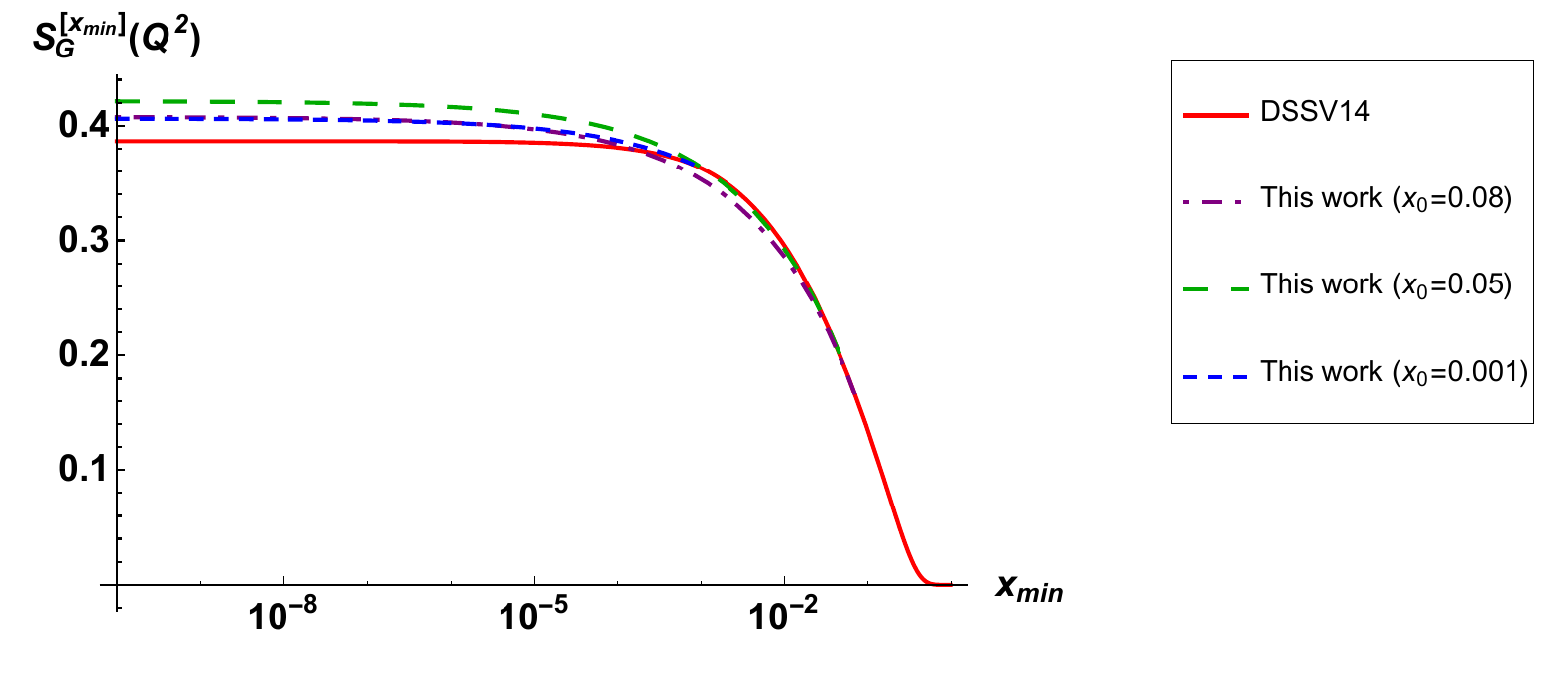} 
\caption{Plot of $S_G^{[x_{min}]}\!(Q^2)$ vs.~$x_{min}$ at
  $Q^2=10\,{\rm GeV}^2$.  The solid curve is from
  DSSV14~\cite{deFlorian:2014yva}.  The dot-dashed, long-dashed, and
  short-dashed curves are from various small-$x$ modifications of
  $\Delta G(x,Q^2)$ at $x_0=0.08,\,0.05,\,0.001$, respectively, using
  our gluon helicity intercept (see the text for
  details).} \label{f:run_spinG}
\end{figure}

In order to calculate $S_G$, we need input for the gluon helicity PDF
$\Delta G(x, Q^2)$, and we focus here on the fit from
DSSV14~\cite{deFlorian:2014yva}.  We proceed through a simple
approach, which we also employed in Ref.~\cite{Kovchegov:2016weo} for
an estimate of the quark spin based on (\ref{q_asym}), and leave a
more rigorous phenomenological study for future work.  First, we
attach a curve $\Delta\tilde{G}(x,Q^2) = N\,x^{-\alpha_h^G}$ (with
$\alpha_h^G$ given in (\ref{e:Gint})) to the DSSV14 result for $\Delta
G(x,Q^2)$ at a particular small-$x$ point $x_0$.  We fix the
normalization $N$ by requiring $\Delta\tilde{G}(x_0,Q^2) = \Delta
G(x_0,Q^2)$.  Then we calculate the truncated integral
\begin{align} 
  \label{e:run_intG}
  S_G^{[x_{min}]}\!(Q^2)\equiv \int_{x_{min}}^1 \!dx \,\Delta
  G(x,Q^2)
\end{align}
of the modified gluon helicity PDF
\begin{align} 
  \label{e:run_int_modG}
  \Delta G_{mod}(x,Q^2) \equiv & \ \theta(x-x_{0}) \, \Delta G(x,Q^2)
  + \theta(x_0-x) \, \Delta\tilde{G}(x,Q^2)
\end{align}
for different $x_0$ values. The results are shown in
Fig.~\ref{f:run_spinG} for $Q^2 = 10\,{\rm GeV^2}$ and
$\alpha_s\approx 0.25$, in which case $\alpha_h^G \approx 0.65$.
We see that the small-$x$ evolution of $\Delta G(x,Q^2)$ gives about a
$5\div 10 \%$ increase to the gluon spin, depending on where in $x$
the effects set in and on the parameterization of the gluon helicity
PDF at higher $x$.  Again we emphasize that the first principles
results of this work (along with that for the
quark~\cite{Kovchegov:2015pbl,Kovchegov:2016weo,Kovchegov:2017jxc})
can be included in future extractions of helicity PDFs, especially
once the present large-$N_c$ approximation is relaxed, which will
provide strong constraints on the small-$x$ behavior of the quark and
gluon spin. 


Saturation effects may also impact the amount of spin carried by
small-$x$ quarks and gluons. The small-$x$ asymptotics of $\Delta G$
found here and the small-$x$ asymptotics of $\Delta q$ found in
\cite{Kovchegov:2015pbl,Kovchegov:2016weo,Kovchegov:2017jxc} are such
that $x \Delta G \to 0$ and $x \Delta q \to 0$ as $x \to 0$. Hence the
helicity PDFs will not violate unitarity at small $x$. However, as one
can see from the helicity evolution equations including (LLA)
saturation effects, as derived in
\cite{Kovchegov:2015pbl,Kovchegov:2016zex}, saturation would
completely suppress the small-$x$ evolution of helicity PDFs, making
the effective $\alpha_h^q$ and $\alpha_h^G$ zero in the saturation
region (cf. \cite{Itakura:2003jp} for the flavor non-singlet
unpolarized quark distribution). Therefore, a very small amount of the
proton spin should reside in the saturation region. This observation
can become an important component of the future small-$x$ helicity PDF
phenomenology.

%
\section{Conclusions}
\label{sec:concl}
%
%

In this paper, we have shown that the dipole gluon helicity
distribution \eqref{M:dipdef} and the Weizs\"{a}cker-Williams gluon
helicity distribution \eqref{M:WWdef} at small $x$ are governed by
polarized dipole operators \eqref{eq:Gidef} and \eqref{eq:Gijdef},
respectively.  These operators are different from each other and from
the polarized dipole amplitude \eqref{eq:Gdef} which governs the quark
helicity distribution at small x.  For the case of the dipole gluon
helicity distribution, we have derived double-logarithmic small-$x$
evolution equations given by Eqs.~\eqref{M:evol4} in the large-$N_c$
limit.  These gluon helicity evolution equations mix with the
small-$x$ quark helicity evolution \eqref{M:asymsol}, but ultimately
result in a gluon helicity intercept \eqref{e:Gint} which is smaller
than the quark helicity intercept \eqref{M:ahel} by about $20\%$.  One
may speculate that the fact that $\alpha_h^G < \alpha_h^q$ is
partially responsible for the difficulty in experimentally detecting a
non-zero signal for $\Delta G$ at small-$x$.

The difference between the quark and gluon helicity intercepts
mathematically results from the fact that the small-$x$ evolution for
quark and gluon helicity is given by a coupled set of equations,
Eqs.~\eqref{e:oldevol1} and \eqref{M:evol4}. This is similar to the
Dokshitzer--Gribov--Lipatov--Altarelli--Parisi (DGLAP) evolution
equations \cite{Dokshitzer:1977sg,Gribov:1972ri,Altarelli:1977zs}
which mix the evolution of the (flavor-singlet) quark and gluon
distributions. Due to this mixing, the $Q^2$ dependence of quark and gluon
PDFs is different from each other. The unpolarized small-$x$ evolution
is different in this respect: at LLA the BFKL evolution is entirely
gluon-driven. The quark distribution is obtained from this evolution by
having a gluon at the end of BFKL ladder emit a $q\bar q$ pair. This
results in $x$-dependence of the (flavor-singlet) unpolarized quark
distribution at small $x$ being practically the same as that for the
gluons. In this paper we observed that for helicity TMDs and PDFs the
small-$x$ evolution mixes the contributions of quarks and gluons, resulting in
a different $x$-dependence of quark and gluon helicity PDFs.
This is indeed different from the $x$-dependence of unpolarized quark and
gluon PDFs resulting from BFKL evolution.

On a technical level, this reduction of the gluon helicity as compared to
the quark helicity can be attributed to the fact that the dipole gluon
helicity evolution receives contributions from the radiation of
virtual unpolarized gluons, but not real unpolarized gluons (the
bottom two diagrams of Fig.~\ref{f:MGievol}).  The physical reason for
this stems from the definition \eqref{eq:TMDdef} of what gluon
helicity really means: a circular flow of the gluon field-strength.
Maintaining this circular orientation during the small-$x$ evolution
requires that the angular correlations between the fields be
preserved, but in the DLA limit, the radiation of unpolarized gluons
is isotropic.  The resulting angular decorrelation causes the real
gluon emission term to drop out from the gluon helicity evolution
equations \eqref{M:evol6}, leaving only the virtual emissions.
Consequently, this leads to a depletion of the gluon helicity compared
to the quark helicity: the uncorrelated radiation of soft gluons
causes the gluon distribution to ``forget'' about polarized
interactions which take place later in the cascade.  Only cascades
which develop without such uncorrelated radiation contribute to the
gluon helicity.

The fact that gluon helicity, which relies upon the circular
transverse structure of the fields, is capable of decorrelating can
also be seen in the structure of the polarized Wilson lines.  The
polarized Wilson line \eqref{M:Vpol2} relevant for the gluon helicity
couples to a total derivative: the curl operator applied to the entire
Wilson line.  This is in contrast to the polarized Wilson line
\eqref{eq:Wpol2} relevant for the quark helicity, which couples to a
local derivative: the curl operator applied to a single point in the
polarized Wilson line.  This operator structure suggests that a
polarized interaction at any point in the cascade is sufficient to
contribute to the quark helicity, while only those polarized
interactions which preserve the angular correlations can contribute to
the gluon helicity.  Presumably, this fundamental difference between
the nature of quark and gluon helicity can be attributed to the fact
that the quark helicity \eqref{e:qTMD} is defined as a matrix element
of the axial vector current.  Until such accuracy that the evolution
becomes sensitive to the axial anomaly, the axial vector current which
defines the quark helicity is conserved during the evolution; a
coupling to the axial vector current anywhere in the evolution is
guaranteed to propagate back to contribute to the quark helicity
distribution.

We also note that the asymptotic solution \eqref{e:MAINRESULT} is an
important input to the proton spin puzzle and a first principles
prediction to be tested against phenomenological extractions.  The
total gluon polarization $S_G$ is far less constrained by experiments
than the quark polarization $S_q$, so this theoretical guidance on how
to extrapolate from data at finite $x$ down to $x \rightarrow 0$ can
provide a useful estimate of $S_G$.  In Sec.~\ref{sec:pheno} we gave
such an estimate of this quantity in a simple approach and found it
could increase the current DSSV extrapolation by $5\div 10\%$.  We
stress again that the results for the small $x$ behavior of the gluon
(and quark) from this work should be included in future helicity PDF
fits.

Additionally, a recent paper \cite{Hatta:2016aoc} has provided a
gauge-invariant definition of the gluon orbital angular momentum
operator in terms of Wilson lines at small $x$.  Deriving and solving
similar small-$x$ evolution equations for such an operator could
provide yet another piece of the proton spin decomposition at small
$x$.

In closing, we must emphasize a note of caution about the precise
values of our quark and gluon helicity intercepts: these numerical
values are the result of a leading-order DLA resummation at large
$N_c$, and they may receive significant corrections at higher orders
in $\alpha_s$, at finite $N_c$, and at $N_f \neq 0$ .  The
single-logarithmic corrections, which can include the effects of
parton saturation and multiple scattering, may be particularly
important.  Our calculation is also performed at fixed coupling at
this accuracy; to precisely set the scale of $\alpha_s$, a
higher-order calculation is needed.  Indeed, we know from the
unpolarized sector that running coupling corrections
\cite{Balitsky:2006wa,Gardi:2006rp,Kovchegov:2006vj,Kovchegov:2006wf}
play an essential role in slowing down the small-$x$ evolution
\cite{Albacete:2004gw,Albacete:2007yr} and bringing the theory in line
with experiment
\cite{Albacete:2009fh,Albacete:2010sy,Kuokkanen:2011je}.  As such,
while much work remains to be done in the intervening years, the
growing pool of spin-related operators whose small-$x$ asymptotics
have been calculated represents an important step in developing the
theoretical framework needed for a future Electron-Ion Collider.


%
\section*{Acknowledgments}
%

The authors are grateful to Ian Balitsky for his interest in their
work and for a number of helpful questions and suggestions.  We also
thank R. Sassot and W. Vogelsang for providing us with the parameters
and Fortran code of the DSSV14 fit.  This material is based upon work
supported by the U.S. Department of Energy, Office of Science, Office
of Nuclear Physics under Award Number DE-SC0004286 (YK), within the
framework of the TMD Topical Collaboration (DP), and DOE Contract
No. DE-AC52-06NA25396 (MS).  MS received additional support from the
U.S. Department of Energy, Office of Science under the DOE Early
Career Program. \\


%
\appendix
\section{A Cross-Check}
%

\renewcommand{\theequation}{A\arabic{equation}}
  \setcounter{equation}{0}
\label{A}

Substituting \eq{Csol} into the left-hand side of \eq{Ceq} we get
\begin{align}
  & \int \frac{d \omega}{2 \pi i} \, \omega \,    C_{\omega}  \,  e^{\left( \omega - \frac{1}{\omega} \right) \, \zeta} = \frac{3 \sqrt{3}}{64} \int \frac{d \omega}{2 \pi i} \, \omega \,    \frac{e^{\left( \omega - \frac{1}{\omega} \right) \, \zeta} }{ \omega - \frac{4}{\sqrt{3}}}  =  \frac{3}{16} \, e^{\frac{13}{4 \sqrt{3}} \, \zeta} + \frac{3 \sqrt{3}}{64}  \sum_{n=2}^\infty \, \frac{(-\zeta)^n}{n!} \, \frac{1}{(n-2)!} \,  \frac{d^{n-2}}{d \omega^{n-2}} \, \left( \frac{e^{\omega \, \zeta} }{\omega - \frac{4}{\sqrt{3}} } \right) \Bigg|_{\omega =0} \notag \\
  & = \frac{3}{16} \, e^{\frac{13}{4 \sqrt{3}}\, \zeta} + \frac{3
    \sqrt{3}}{64} \sum_{n=2}^\infty \, \frac{(-\zeta)^n}{n!} \,
  \frac{1}{(n-2)!} \, \sum_{m=0}^{n-2} \left(
\begin{array}{c}
n-2 \\
m
\end{array}
\right) \, \zeta^{n-2-m} \, (-1)^m \, m! \, \left( - \frac{\sqrt{3}}{4} \right)^{m+1} \notag \\
& = \frac{3}{16} \, e^{\frac{13}{4 \sqrt{3}}\, \zeta} + \frac{3 \sqrt{3}}{64} \sum_{n=2}^\infty \, \sum_{m=0}^{n-2}  \, \frac{(-\zeta^2)^{n-1}}{n! \, (n-2-m)!} \, \left( \frac{\sqrt{3}}{4 \, \zeta} \right)^m = \Bigg| k = n-2-m  \Bigg|   \\
& = \frac{3}{16} \, e^{\frac{13}{4 \sqrt{3}}\, \zeta} + \frac{3 \sqrt{3}}{64}  \sum_{n=2}^\infty \, \sum_{k=0}^{n-2}  \, \frac{(-\zeta^2)^{n-1}}{n! \, k!} \, \left( \frac{\sqrt{3}}{4 \, \zeta} \right)^{n-2-k}  = \frac{3}{16} \, e^{\frac{13}{4 \sqrt{3}}\, \zeta} + \frac{3 \sqrt{3}}{64} \sum_{k=0}^\infty \, \sum_{n=k+2}^{\infty}  \, \frac{(-\zeta^2)^{n-1}}{n! \, k!} \, \left( \frac{\sqrt{3}}{4 \, \zeta} \right)^{n-2-k}  \notag \\
& = \Bigg| {l=n-k-2} \Bigg| = \frac{3}{16} \, e^{\frac{13}{4
    \sqrt{3}}\, \zeta} + \frac{3 \sqrt{3}}{64} \sum_{k=0}^\infty \,
\sum_{l=0}^{\infty} \, \frac{(-\zeta^2)^{l+k+1}}{(l+k+2)! \, k!} \,
\left( \frac{\sqrt{3}}{4 \, \zeta} \right)^{l} = \frac{3}{16} \,
e^{\frac{13}{4 \sqrt{3}}\, \zeta} - \frac{3 \sqrt{3}}{64}
\sum_{l=0}^{\infty} \, \left( - \frac{\sqrt{3}}{4} \right)^{l} \,
J_{l+2} (2 \zeta). \notag
\end{align}

While the remaining sum cannot be cast in a form of a single function,
we can deduce its large-$\zeta$ asymptotics:
\begin{align}
  \sum_{l=0}^{\infty} \, \left( - \frac{\sqrt{3}}{4} \right)^{l} \,
  J_{l+2} (2 \zeta) \Bigg|_{\zeta \to \infty} \longrightarrow
  \sum_{l=0}^{\infty} \, \left( - \frac{\sqrt{3}}{4} \right)^{l} \,
  \sqrt{\frac{1}{\pi \, \zeta}} \, \cos \left( 2 \zeta - \frac{\pi \,
      l}{2} - \frac{5 \pi}{4} \right) \sim \frac{1}{\sqrt{\zeta}}
  \longrightarrow 0.
\end{align}
We conclude that 
\begin{align}
  \frac{3 \sqrt{3}}{64} \int \frac{d \omega}{2 \pi i} \, \omega \,
  \frac{e^{\left( \omega - \frac{1}{\omega} \right) \, \zeta} }{
    \omega - \frac{4}{\sqrt{3}}} = \frac{3}{16} \, e^{\frac{13}{4
      \sqrt{3}}\, \zeta} + {\cal O} \left( \frac{1}{\sqrt{\zeta}}
  \right)
\end{align}
and, hence, \eq{Csol} solves \eq{Ceq} in the large-$\zeta$
asymptotics.



\begin{thebibliography}{10}

\bibitem{Accardi:2012qut}
A.~Accardi, J.~Albacete, M.~Anselmino, N.~Armesto, E.~Aschenauer et~al.,
  \emph{{Electron Ion Collider: The Next QCD Frontier - Understanding the glue
  that binds us all}},  \href{https://arxiv.org/abs/1212.1701}{{\ttfamily
  1212.1701}}.

\bibitem{Aschenauer:2013woa}
E.~C. Aschenauer et~al., \emph{{The RHIC Spin Program: Achievements and Future
  Opportunities}},  \href{https://arxiv.org/abs/1304.0079}{{\ttfamily
  1304.0079}}.

\bibitem{Aschenauer:2015eha}
E.-C. Aschenauer et~al., \emph{{The RHIC SPIN Program: Achievements and Future
  Opportunities}},  \href{https://arxiv.org/abs/1501.01220}{{\ttfamily
  1501.01220}}.

\bibitem{Aschenauer:2016our}
E.-C. Aschenauer et~al., \emph{{The RHIC Cold QCD Plan for 2017 to 2023: A
  Portal to the EIC}},  \href{https://arxiv.org/abs/1602.03922}{{\ttfamily
  1602.03922}}.

\bibitem{Jaffe:1989jz}
R.~L. Jaffe and A.~Manohar, \emph{{The G(1) Problem: Fact and Fantasy on the
  Spin of the Proton}},
  \href{https://doi.org/10.1016/0550-3213(90)90506-9}{\emph{Nucl. Phys.}
  {\bfseries B337} (1990) 509--546}.

\bibitem{Ji:1996ek}
X.-D. Ji, \emph{{Gauge-Invariant Decomposition of Nucleon Spin}},
  \href{https://doi.org/10.1103/PhysRevLett.78.610}{\emph{Phys. Rev. Lett.}
  {\bfseries 78} (1997) 610--613},
  [\href{https://arxiv.org/abs/hep-ph/9603249}{{\ttfamily hep-ph/9603249}}].

\bibitem{Ji:2012sj}
X.~Ji, X.~Xiong and F.~Yuan, \emph{{Proton Spin Structure from Measurable
  Parton Distributions}},
  \href{https://doi.org/10.1103/PhysRevLett.109.152005}{\emph{Phys. Rev. Lett.}
  {\bfseries 109} (2012) 152005},
  [\href{https://arxiv.org/abs/1202.2843}{{\ttfamily 1202.2843}}].

\bibitem{Leader:2013jra}
E.~Leader and C.~Lorce, \emph{{The angular momentum controversy: What’s it
  all about and does it matter?}},
  \href{https://doi.org/10.1016/j.physrep.2014.02.010}{\emph{Phys.Rept.}
  {\bfseries 541} (2014) 163--248},
  [\href{https://arxiv.org/abs/1309.4235}{{\ttfamily 1309.4235}}].

\bibitem{Kovchegov:2015pbl}
Y.~V. Kovchegov, D.~Pitonyak and M.~D. Sievert, \emph{{Helicity Evolution at
  Small-x}}, \href{https://doi.org/10.1007/JHEP01(2016)072}{\emph{JHEP}
  {\bfseries 01} (2016) 072},
  [\href{https://arxiv.org/abs/1511.06737}{{\ttfamily 1511.06737}}].

\bibitem{Kovchegov:2015zha}
Y.~V. Kovchegov and M.~D. Sievert, \emph{{Calculating TMDs of a Large Nucleus:
  Quasi-Classical Approximation and Quantum Evolution}},
  \href{https://doi.org/10.1016/j.nuclphysb.2015.12.008}{\emph{Nucl. Phys.}
  {\bfseries B903} (2016) 164--203},
  [\href{https://arxiv.org/abs/1505.01176}{{\ttfamily 1505.01176}}].

\bibitem{Balitsky:1995ub}
I.~Balitsky, \emph{{Operator expansion for high-energy scattering}},
  \href{https://doi.org/10.1016/0550-3213(95)00638-9}{\emph{Nucl. Phys.}
  {\bfseries B463} (1996) 99--160},
  [\href{https://arxiv.org/abs/hep-ph/9509348}{{\ttfamily hep-ph/9509348}}].

\bibitem{Balitsky:1998ya}
I.~Balitsky, \emph{Factorization and high-energy effective action},
  {\emph{Phys. Rev.} {\bfseries D60} (1999) 014020},
  [\href{https://arxiv.org/abs/hep-ph/9812311}{{\ttfamily hep-ph/9812311}}].

\bibitem{Kovchegov:1999yj}
Y.~V. Kovchegov, \emph{Small-x {$F_2$} structure function of a nucleus
  including multiple pomeron exchanges}, {\emph{Phys. Rev.} {\bfseries D60}
  (1999) 034008}, [\href{https://arxiv.org/abs/hep-ph/9901281}{{\ttfamily
  hep-ph/9901281}}].

\bibitem{Kovchegov:1999ua}
Y.~V. Kovchegov, \emph{Unitarization of the {BFKL} pomeron on a nucleus},
  {\emph{Phys. Rev.} {\bfseries D61} (2000) 074018},
  [\href{https://arxiv.org/abs/hep-ph/9905214}{{\ttfamily hep-ph/9905214}}].

\bibitem{Jalilian-Marian:1997dw}
J.~Jalilian-Marian, A.~Kovner and H.~Weigert, \emph{The {Wilson}
  renormalization group for low x physics: Gluon evolution at finite parton
  density}, {\emph{Phys. Rev.} {\bfseries D59} (1998) 014015},
  [\href{https://arxiv.org/abs/hep-ph/9709432}{{\ttfamily hep-ph/9709432}}].

\bibitem{Jalilian-Marian:1997gr}
J.~Jalilian-Marian, A.~Kovner, A.~Leonidov and H.~Weigert, \emph{The {Wilson}
  renormalization group for low x physics: Towards the high density regime},
  {\emph{Phys. Rev.} {\bfseries D59} (1998) 014014},
  [\href{https://arxiv.org/abs/hep-ph/9706377}{{\ttfamily hep-ph/9706377}}].

\bibitem{Iancu:2001ad}
E.~Iancu, A.~Leonidov and L.~D. McLerran, \emph{{The renormalization group
  equation for the color glass condensate}},
  \href{https://doi.org/10.1016/S0370-2693(01)00524-X}{\emph{Phys. Lett.}
  {\bfseries B510} (2001) 133--144}.

\bibitem{Iancu:2000hn}
E.~Iancu, A.~Leonidov and L.~D. McLerran, \emph{Nonlinear gluon evolution in
  the color glass condensate. {I}}, {\emph{Nucl. Phys.} {\bfseries A692} (2001)
  583--645}, [\href{https://arxiv.org/abs/hep-ph/0011241}{{\ttfamily
  hep-ph/0011241}}].

\bibitem{Kuraev:1977fs}
E.~A. Kuraev, L.~N. Lipatov and V.~S. Fadin, \emph{{The Pomeranchuk
  singlularity in non-Abelian gauge theories}}, {\emph{Sov. Phys. JETP}
  {\bfseries 45} (1977) 199--204}.

\bibitem{Balitsky:1978ic}
I.~Balitsky and L.~Lipatov, \emph{{The Pomeranchuk Singularity in Quantum
  Chromodynamics}}, {\emph{Sov.J.Nucl.Phys.} {\bfseries 28} (1978) 822--829}.

\bibitem{Kirschner:1983di}
R.~Kirschner and L.~Lipatov, \emph{{Double Logarithmic Asymptotics and Regge
  Singularities of Quark Amplitudes with Flavor Exchange}},
  \href{https://doi.org/10.1016/0550-3213(83)90178-5}{\emph{Nucl.Phys.}
  {\bfseries B213} (1983) 122--148}.

\bibitem{Kirschner:1985cb}
R.~Kirschner, \emph{{Regge Asymptotics of Scattering Amplitudes in the
  Logarithmic Approximation of {QCD}}},
  \href{https://doi.org/10.1007/BF01559604}{\emph{Z. Phys.} {\bfseries C31}
  (1986) 135}.

\bibitem{Kirschner:1994vc}
R.~Kirschner, \emph{{Regge asymptotics of scattering with flavor exchange in
  QCD}}, \href{https://doi.org/10.1007/BF01624588}{\emph{Z.Phys.} {\bfseries
  C67} (1995) 459--466},
  [\href{https://arxiv.org/abs/hep-th/9404158}{{\ttfamily hep-th/9404158}}].

\bibitem{Kirschner:1994rq}
R.~Kirschner, \emph{{Reggeon interactions in perturbative QCD}},
  \href{https://doi.org/10.1007/BF01556138}{\emph{Z.Phys.} {\bfseries C65}
  (1995) 505--510}, [\href{https://arxiv.org/abs/hep-th/9407085}{{\ttfamily
  hep-th/9407085}}].

\bibitem{Griffiths:1999dj}
S.~Griffiths and D.~Ross, \emph{{Studying the perturbative Reggeon}},
  \href{https://doi.org/10.1007/s100529900240}{\emph{Eur.Phys.J.} {\bfseries
  C12} (2000) 277--286},
  [\href{https://arxiv.org/abs/hep-ph/9906550}{{\ttfamily hep-ph/9906550}}].

\bibitem{Itakura:2003jp}
K.~Itakura, Y.~V. Kovchegov, L.~McLerran and D.~Teaney, \emph{{Baryon stopping
  and valence quark distribution at small x}},
  \href{https://doi.org/10.1016/j.nuclphysa.2003.10.016}{\emph{Nucl. Phys.}
  {\bfseries A730} (2004) 160--190},
  [\href{https://arxiv.org/abs/hep-ph/0305332}{{\ttfamily hep-ph/0305332}}].

\bibitem{Bartels:2003dj}
J.~Bartels and M.~Lublinsky, \emph{{Quark anti-quark exchange in gamma* gamma*
  scattering}},
  \href{https://doi.org/10.1088/1126-6708/2003/09/076}{\emph{JHEP} {\bfseries
  0309} (2003) 076}, [\href{https://arxiv.org/abs/hep-ph/0308181}{{\ttfamily
  hep-ph/0308181}}].

\bibitem{Bartels:1995iu}
J.~Bartels, B.~Ermolaev and M.~Ryskin, \emph{{Nonsinglet contributions to the
  structure function g1 at small x}}, {\emph{Z.Phys.} {\bfseries C70} (1996)
  273--280}, [\href{https://arxiv.org/abs/hep-ph/9507271}{{\ttfamily
  hep-ph/9507271}}].

\bibitem{Bartels:1996wc}
J.~Bartels, B.~Ermolaev and M.~Ryskin, \emph{{Flavor singlet contribution to
  the structure function G(1) at small x}},
  \href{https://doi.org/10.1007/s002880050285}{\emph{Z.Phys.} {\bfseries C72}
  (1996) 627--635}, [\href{https://arxiv.org/abs/hep-ph/9603204}{{\ttfamily
  hep-ph/9603204}}].

\bibitem{Ermolaev:1999jx}
B.~I. Ermolaev, M.~Greco and S.~I. Troian, \emph{{QCD running coupling effects
  for the nonsinglet structure function at small $x$}},
  \href{https://doi.org/10.1016/S0550-3213(99)00812-3}{\emph{Nucl. Phys.}
  {\bfseries B571} (2000) 137--150},
  [\href{https://arxiv.org/abs/hep-ph/9906276}{{\ttfamily hep-ph/9906276}}].

\bibitem{Ermolaev:2000sg}
B.~I. Ermolaev, M.~Greco and S.~I. Troyan, \emph{{Intercepts of the nonsinglet
  structure functions}},
  \href{https://doi.org/10.1016/S0550-3213(00)00647-7}{\emph{Nucl. Phys.}
  {\bfseries B594} (2001) 71--88},
  [\href{https://arxiv.org/abs/hep-ph/0009037}{{\ttfamily hep-ph/0009037}}].

\bibitem{Ermolaev:2003zx}
B.~I. Ermolaev, M.~Greco and S.~I. Troyan, \emph{{Running coupling effects for
  the singlet structure function $g_1$ at small $x$}},
  \href{https://doi.org/10.1016/j.physletb.2003.11.016}{\emph{Phys. Lett.}
  {\bfseries B579} (2004) 321--330},
  [\href{https://arxiv.org/abs/hep-ph/0307128}{{\ttfamily hep-ph/0307128}}].

\bibitem{Ermolaev:2009cq}
B.~I. Ermolaev, M.~Greco and S.~I. Troyan, \emph{{Overview of the spin
  structure function $g_1$ at arbitrary $x$ and $Q^2$}},
  \href{https://doi.org/10.1393/ncr/i2010-10052-3}{\emph{Riv. Nuovo Cim.}
  {\bfseries 33} (2010) 57--122},
  [\href{https://arxiv.org/abs/0905.2841}{{\ttfamily 0905.2841}}].

\bibitem{Mueller:1993rr}
A.~H. Mueller, \emph{{Soft gluons in the infinite momentum wave function and
  the BFKL pomeron}},
  \href{https://doi.org/10.1016/0550-3213(94)90116-3}{\emph{Nucl. Phys.}
  {\bfseries B415} (1994) 373--385}.

\bibitem{Mueller:1994jq}
A.~H. Mueller and B.~Patel, \emph{Single and double {BFKL} pomeron exchange and
  a dipole picture of high-energy hard processes}, {\emph{Nucl. Phys.}
  {\bfseries B425} (1994) 471--488},
  [\href{https://arxiv.org/abs/hep-ph/9403256}{{\ttfamily hep-ph/9403256}}].

\bibitem{Mueller:1994gb}
A.~H. Mueller, \emph{{Unitarity and the BFKL pomeron}},
  \href{https://doi.org/10.1016/0550-3213(94)00480-3}{\emph{Nucl. Phys.}
  {\bfseries B437} (1995) 107--126},
  [\href{https://arxiv.org/abs/hep-ph/9408245}{{\ttfamily hep-ph/9408245}}].

\bibitem{Kovchegov:2016zex}
Y.~V. Kovchegov, D.~Pitonyak and M.~D. Sievert, \emph{{Helicity Evolution at
  Small $x$: Flavor Singlet and Non-Singlet Observables}},
  \href{https://doi.org/10.1103/PhysRevD.95.014033}{\emph{Phys. Rev.}
  {\bfseries D95} (2017) 014033},
  [\href{https://arxiv.org/abs/1610.06197}{{\ttfamily 1610.06197}}].

\bibitem{Kovchegov:2016weo}
Y.~V. Kovchegov, D.~Pitonyak and M.~D. Sievert, \emph{{Small-$x$ asymptotics of
  the quark helicity distribution}},
  \href{https://doi.org/10.1103/PhysRevLett.118.052001}{\emph{Phys. Rev. Lett.}
  {\bfseries 118} (2017) 052001},
  [\href{https://arxiv.org/abs/1610.06188}{{\ttfamily 1610.06188}}].

\bibitem{Kovchegov:2017jxc}
Y.~V. Kovchegov, D.~Pitonyak and M.~D. Sievert, \emph{{Small-$x$ Asymptotics of
  the Quark Helicity Distribution: Analytic Results}},
  \href{https://arxiv.org/abs/1703.05809}{{\ttfamily 1703.05809}}.

\bibitem{Dominguez:2011wm}
F.~Dominguez, C.~Marquet, B.-W. Xiao and F.~Yuan, \emph{{Universality of
  Unintegrated Gluon Distributions at small x}},
  \href{https://doi.org/10.1103/PhysRevD.83.105005}{\emph{Phys.Rev.} {\bfseries
  D83} (2011) 105005}, [\href{https://arxiv.org/abs/1101.0715}{{\ttfamily
  1101.0715}}].

\bibitem{Metz:2011wb}
A.~Metz and J.~Zhou, \emph{{Distribution of linearly polarized gluons inside a
  large nucleus}},
  \href{https://doi.org/10.1103/PhysRevD.84.051503}{\emph{Phys.Rev.} {\bfseries
  D84} (2011) 051503}, [\href{https://arxiv.org/abs/1105.1991}{{\ttfamily
  1105.1991}}].

\bibitem{Dominguez:2011gc}
F.~Dominguez, A.~Mueller, S.~Munier and B.-W. Xiao, \emph{{On the small-x
  evolution of the color quadrupole and the Weizs\'acker-Williams gluon
  distribution}},
  \href{https://doi.org/10.1016/j.physletb.2011.09.104}{\emph{Phys.Lett.}
  {\bfseries B705} (2011) 106--111},
  [\href{https://arxiv.org/abs/1108.1752}{{\ttfamily 1108.1752}}].

\bibitem{deFlorian:2014yva}
D.~de~Florian, R.~Sassot, M.~Stratmann and W.~Vogelsang, \emph{{Evidence for
  polarization of gluons in the proton}},
  \href{https://doi.org/10.1103/PhysRevLett.113.012001}{\emph{Phys. Rev. Lett.}
  {\bfseries 113} (2014) 012001},
  [\href{https://arxiv.org/abs/1404.4293}{{\ttfamily 1404.4293}}].

\bibitem{Mulders:1995dh}
P.~J. Mulders and R.~D. Tangerman, \emph{{The Complete tree level result up to
  order 1/Q for polarized deep inelastic leptoproduction}},
  \href{https://doi.org/10.1016/0550-3213(95)00632-X}{\emph{Nucl. Phys.}
  {\bfseries B461} (1996) 197--237},
  [\href{https://arxiv.org/abs/hep-ph/9510301}{{\ttfamily hep-ph/9510301}}].

\bibitem{Beuf:2014uia}
G.~Beuf, \emph{{Improving the kinematics for low-$x$ QCD evolution equations in
  coordinate space}},
  \href{https://doi.org/10.1103/PhysRevD.89.074039}{\emph{Phys. Rev.}
  {\bfseries D89} (2014) 074039},
  [\href{https://arxiv.org/abs/1401.0313}{{\ttfamily 1401.0313}}].

\bibitem{Mueller:1989st}
A.~H. Mueller, \emph{{Small x Behavior and Parton Saturation: A QCD Model}},
  {\emph{Nucl. Phys.} {\bfseries B335} (1990) 115}.

\bibitem{Kovchegov:2012mbw}
Y.~V. Kovchegov and E.~Levin, \emph{{Quantum chromodynamics at high energy}},
  vol.~33.
\newblock Cambridge University Press, 2012.

\bibitem{Mulders:2000sh}
P.~J. Mulders and J.~Rodrigues, \emph{{Transverse momentum dependence in gluon
  distribution and fragmentation functions}},
  \href{https://doi.org/10.1103/PhysRevD.63.094021}{\emph{Phys. Rev.}
  {\bfseries D63} (2001) 094021},
  [\href{https://arxiv.org/abs/hep-ph/0009343}{{\ttfamily hep-ph/0009343}}].

\bibitem{Collins:1981uw}
J.~C. Collins and D.~E. Soper, \emph{{Parton Distribution and Decay
  Functions}},
  \href{https://doi.org/10.1016/0550-3213(82)90021-9}{\emph{Nucl.Phys.}
  {\bfseries B194} (1982) 445}.

\bibitem{Ji:2005nu}
X.-d. Ji, J.-P. Ma and F.~Yuan, \emph{{Transverse-momentum-dependent gluon
  distributions and semi-inclusive processes at hadron colliders}},
  \href{https://doi.org/10.1088/1126-6708/2005/07/020}{\emph{JHEP} {\bfseries
  07} (2005) 020}, [\href{https://arxiv.org/abs/hep-ph/0503015}{{\ttfamily
  hep-ph/0503015}}].

\bibitem{Bomhof:2006dp}
C.~J. Bomhof, P.~J. Mulders and F.~Pijlman, \emph{{The Construction of
  gauge-links in arbitrary hard processes}},
  \href{https://doi.org/10.1140/epjc/s2006-02554-2}{\emph{Eur. Phys. J.}
  {\bfseries C47} (2006) 147--162},
  [\href{https://arxiv.org/abs/hep-ph/0601171}{{\ttfamily hep-ph/0601171}}].

\bibitem{Sievert:2014psa}
M.~D. Sievert, \emph{{Transverse Spin and Classical Gluon Fields: Combining Two
  Perspectives on Hadronic Structure}},
  \href{https://arxiv.org/abs/1407.4047}{{\ttfamily 1407.4047}}.

\bibitem{Hatta:2016aoc}
Y.~Hatta, Y.~Nakagawa, F.~Yuan, Y.~Zhao and B.~Xiao, \emph{{Gluon orbital
  angular momentum at small-$x$}},
  \href{https://doi.org/10.1103/PhysRevD.95.114032}{\emph{Phys. Rev.}
  {\bfseries D95} (2017) 114032},
  [\href{https://arxiv.org/abs/1612.02445}{{\ttfamily 1612.02445}}].

\bibitem{Chirilli:2015fza}
G.~A. Chirilli, Y.~V. Kovchegov and D.~E. Wertepny, \emph{{Regularization of
  the Light-Cone Gauge Gluon Propagator Singularities Using Sub-Gauge
  Conditions}}, \href{https://doi.org/10.1007/JHEP12(2015)138}{\emph{JHEP}
  {\bfseries 12} (2015) 138},
  [\href{https://arxiv.org/abs/1508.07962}{{\ttfamily 1508.07962}}].

\bibitem{Kovchegov:1996ty}
Y.~V. Kovchegov, \emph{Non-abelian {Weizs\"{a}cker-Williams} field and a two-
  dimensional effective color charge density for a very large nucleus},
  {\emph{Phys. Rev.} {\bfseries D54} (1996) 5463--5469},
  [\href{https://arxiv.org/abs/hep-ph/9605446}{{\ttfamily hep-ph/9605446}}].

\bibitem{Kovchegov:1997pc}
Y.~V. Kovchegov, \emph{{Quantum structure of the non-Abelian
  Weizs\"{a}cker-Williams field for a very large nucleus}}, {\emph{Phys. Rev.}
  {\bfseries D55} (1997) 5445--5455},
  [\href{https://arxiv.org/abs/hep-ph/9701229}{{\ttfamily hep-ph/9701229}}].

\bibitem{Balitsky:2015qba}
I.~Balitsky and A.~Tarasov, \emph{{Rapidity evolution of gluon TMD from low to
  moderate x}}, \href{https://doi.org/10.1007/JHEP10(2015)017}{\emph{JHEP}
  {\bfseries 10} (2015) 017},
  [\href{https://arxiv.org/abs/1505.02151}{{\ttfamily 1505.02151}}].

\bibitem{Mueller:2012bn}
A.~Mueller and S.~Munier, \emph{{$p_{\perp}$-broadening and production
  processes versus dipole/quadrupole amplitudes at next-to-leading order}},
  \href{https://doi.org/10.1016/j.nuclphysa.2012.08.005}{\emph{Nucl.Phys.}
  {\bfseries A893} (2012) 43--86},
  [\href{https://arxiv.org/abs/1206.1333}{{\ttfamily 1206.1333}}].

\bibitem{Kovchegov:2018znm}
Y.~V. Kovchegov and M.~D. Sievert, \emph{{Small-$x$ Helicity Evolution: an
  Operator Treatment}},  \href{https://arxiv.org/abs/1808.09010}{{\ttfamily
  1808.09010}}.

\bibitem{Iancu:2015vea}
E.~Iancu, J.~D. Madrigal, A.~H. Mueller, G.~Soyez and D.~N. Triantafyllopoulos,
  \emph{{Resumming double logarithms in the QCD evolution of color dipoles}},
  \href{https://doi.org/10.1016/j.physletb.2015.03.068}{\emph{Phys. Lett.}
  {\bfseries B744} (2015) 293--302},
  [\href{https://arxiv.org/abs/1502.05642}{{\ttfamily 1502.05642}}].

\bibitem{Mueller:1995gb}
A.~H. Mueller, \emph{Unitarity and the {BFKL} pomeron}, {\emph{Nucl. Phys.}
  {\bfseries B437} (1995) 107--126},
  [\href{https://arxiv.org/abs/hep-ph/9408245}{{\ttfamily hep-ph/9408245}}].

\bibitem{Adamczyk:2014ozi}
{\scshape STAR} collaboration, L.~Adamczyk et~al., \emph{{Precision Measurement
  of the Longitudinal Double-spin Asymmetry for Inclusive Jet Production in
  Polarized Proton Collisions at $\sqrt{s}=200$ GeV}},
  \href{https://doi.org/10.1103/PhysRevLett.115.092002}{\emph{Phys. Rev. Lett.}
  {\bfseries 115} (2015) 092002},
  [\href{https://arxiv.org/abs/1405.5134}{{\ttfamily 1405.5134}}].

\bibitem{Adare:2015ozj}
{\scshape PHENIX} collaboration, A.~Adare et~al., \emph{{Inclusive cross
  section and double-helicity asymmetry for $\pi^{0}$ production at midrapidity
  in $p$$+$$p$ collisions at $\sqrt{s}=510$ GeV}},
  \href{https://doi.org/10.1103/PhysRevD.93.011501}{\emph{Phys. Rev.}
  {\bfseries D93} (2016) 011501},
  [\href{https://arxiv.org/abs/1510.02317}{{\ttfamily 1510.02317}}].

\bibitem{Aschenauer:2015ata}
E.~C. Aschenauer, R.~Sassot and M.~Stratmann, \emph{{Unveiling the Proton Spin
  Decomposition at a Future Electron-Ion Collider}},
  \href{https://doi.org/10.1103/PhysRevD.92.094030}{\emph{Phys. Rev.}
  {\bfseries D92} (2015) 094030},
  [\href{https://arxiv.org/abs/1509.06489}{{\ttfamily 1509.06489}}].

\bibitem{deFlorian:2009vb}
D.~de~Florian, R.~Sassot, M.~Stratmann and W.~Vogelsang, \emph{{Extraction of
  Spin-Dependent Parton Densities and Their Uncertainties}},
  \href{https://doi.org/10.1103/PhysRevD.80.034030}{\emph{Phys. Rev.}
  {\bfseries D80} (2009) 034030},
  [\href{https://arxiv.org/abs/0904.3821}{{\ttfamily 0904.3821}}].

\bibitem{Jimenez-Delgado:2013boa}
P.~Jimenez-Delgado, A.~Accardi and W.~Melnitchouk, \emph{{Impact of hadronic
  and nuclear corrections on global analysis of spin-dependent parton
  distributions}},
  \href{https://doi.org/10.1103/PhysRevD.89.034025}{\emph{Phys. Rev.}
  {\bfseries D89} (2014) 034025},
  [\href{https://arxiv.org/abs/1310.3734}{{\ttfamily 1310.3734}}].

\bibitem{Sato:2016tuz}
{\scshape Jefferson Lab Angular Momentum} collaboration, N.~Sato,
  W.~Melnitchouk, S.~E. Kuhn, J.~J. Ethier and A.~Accardi, \emph{{Iterative
  Monte Carlo analysis of spin-dependent parton distributions}},
  \href{https://doi.org/10.1103/PhysRevD.93.074005}{\emph{Phys. Rev.}
  {\bfseries D93} (2016) 074005},
  [\href{https://arxiv.org/abs/1601.07782}{{\ttfamily 1601.07782}}].

\bibitem{Ethier:2017zbq}
J.~J. Ethier, N.~Sato and W.~Melnitchouk, \emph{{First simultaneous extraction
  of spin-dependent parton distributions and fragmentation functions from a
  global QCD analysis}},  \href{https://arxiv.org/abs/1705.05889}{{\ttfamily
  1705.05889}}.

\bibitem{Leader:2005ci}
E.~Leader, A.~V. Sidorov and D.~B. Stamenov, \emph{{Longitudinal polarized
  parton densities updated}},
  \href{https://doi.org/10.1103/PhysRevD.73.034023}{\emph{Phys. Rev.}
  {\bfseries D73} (2006) 034023},
  [\href{https://arxiv.org/abs/hep-ph/0512114}{{\ttfamily hep-ph/0512114}}].

\bibitem{Leader:2010rb}
E.~Leader, A.~V. Sidorov and D.~B. Stamenov, \emph{{Determination of Polarized
  PDFs from a QCD Analysis of Inclusive and Semi-inclusive Deep Inelastic
  Scattering Data}},
  \href{https://doi.org/10.1103/PhysRevD.82.114018}{\emph{Phys. Rev.}
  {\bfseries D82} (2010) 114018},
  [\href{https://arxiv.org/abs/1010.0574}{{\ttfamily 1010.0574}}].

\bibitem{Leader:2014uua}
E.~Leader, A.~V. Sidorov and D.~B. Stamenov, \emph{{New analysis concerning the
  strange quark polarization puzzle}},
  \href{https://doi.org/10.1103/PhysRevD.91.054017}{\emph{Phys. Rev.}
  {\bfseries D91} (2015) 054017},
  [\href{https://arxiv.org/abs/1410.1657}{{\ttfamily 1410.1657}}].

\bibitem{Ball:2013lla}
{\scshape NNPDF} collaboration, R.~D. Ball, S.~Forte, A.~Guffanti, E.~R.
  Nocera, G.~Ridolfi and J.~Rojo, \emph{{Unbiased determination of polarized
  parton distributions and their uncertainties}},
  \href{https://doi.org/10.1016/j.nuclphysb.2013.05.007}{\emph{Nucl. Phys.}
  {\bfseries B874} (2013) 36--84},
  [\href{https://arxiv.org/abs/1303.7236}{{\ttfamily 1303.7236}}].

\bibitem{Nocera:2014gqa}
{\scshape NNPDF} collaboration, E.~R. Nocera, R.~D. Ball, S.~Forte, G.~Ridolfi
  and J.~Rojo, \emph{{A first unbiased global determination of polarized PDFs
  and their uncertainties}},
  \href{https://doi.org/10.1016/j.nuclphysb.2014.08.008}{\emph{Nucl. Phys.}
  {\bfseries B887} (2014) 276--308},
  [\href{https://arxiv.org/abs/1406.5539}{{\ttfamily 1406.5539}}].

\bibitem{Dokshitzer:1977sg}
Y.~L. Dokshitzer, \emph{{Calculation of the Structure Functions for Deep
  Inelastic Scattering and $e^+ e^-$ Annihilation by Perturbation Theory in
  Quantum Chromodynamics}}, {\emph{Sov. Phys. JETP} {\bfseries 46} (1977)
  641--653}.

\bibitem{Gribov:1972ri}
V.~N. Gribov and L.~N. Lipatov, \emph{{Deep inelastic e p scattering in
  perturbation theory}}, {\emph{Sov. J. Nucl. Phys.} {\bfseries 15} (1972)
  438--450}.

\bibitem{Altarelli:1977zs}
G.~Altarelli and G.~Parisi, \emph{{Asymptotic Freedom in Parton Language}},
  \href{https://doi.org/10.1016/0550-3213(77)90384-4}{\emph{Nucl. Phys.}
  {\bfseries B126} (1977) 298}.

\bibitem{Balitsky:2006wa}
I.~I. Balitsky, \emph{{Quark Contribution to the Small-$x$ Evolution of Color
  Dipole}}, {\emph{Phys. Rev. D} {\bfseries 75} (2007) 014001},
  [\href{https://arxiv.org/abs/hep-ph/0609105}{{\ttfamily hep-ph/0609105}}].

\bibitem{Gardi:2006rp}
E.~Gardi, J.~Kuokkanen, K.~Rummukainen and H.~Weigert, \emph{Running coupling
  and power corrections in nonlinear evolution at the high-energy limit},
  {\emph{Nucl. Phys.} {\bfseries A784} (2007) 282--340},
  [\href{https://arxiv.org/abs/hep-ph/0609087}{{\ttfamily hep-ph/0609087}}].

\bibitem{Kovchegov:2006vj}
Y.~Kovchegov and H.~Weigert, \emph{{Triumvirate of Running Couplings in
  Small-$x$ Evolution}}, {\emph{Nucl. Phys. {\bf A}} {\bfseries 784} (2007)
  188--226}, [\href{https://arxiv.org/abs/hep-ph/0609090}{{\ttfamily
  hep-ph/0609090}}].

\bibitem{Kovchegov:2006wf}
Y.~V. Kovchegov and H.~Weigert, \emph{{Quark loop contribution to BFKL
  evolution: Running coupling and leading-N(f) NLO intercept}},
  \href{https://doi.org/10.1016/j.nuclphysa.2007.03.008}{\emph{Nucl. Phys.}
  {\bfseries A789} (2007) 260--284},
  [\href{https://arxiv.org/abs/hep-ph/0612071}{{\ttfamily hep-ph/0612071}}].

\bibitem{Albacete:2004gw}
J.~L. Albacete, N.~Armesto, J.~G. Milhano, C.~A. Salgado and U.~A. Wiedemann,
  \emph{{Numerical analysis of the Balitsky-Kovchegov equation with running
  coupling: Dependence of the saturation scale on nuclear size and rapidity}},
  {\emph{Phys. Rev.} {\bfseries D71} (2005) 014003},
  [\href{https://arxiv.org/abs/hep-ph/0408216}{{\ttfamily hep-ph/0408216}}].

\bibitem{Albacete:2007yr}
J.~L. Albacete and Y.~V. Kovchegov, \emph{Solving high energy evolution
  equation including running coupling corrections}, {\emph{Phys. Rev.}
  {\bfseries D75} (2007) 125021}, [\href{https://arxiv.org/abs/arXiv:0704.0612
  [hep-ph]}{{\ttfamily arXiv:0704.0612 [hep-ph]}}].

\bibitem{Albacete:2009fh}
J.~L. Albacete, N.~Armesto, J.~G. Milhano and C.~A. Salgado, \emph{{Non-linear
  QCD meets data: A global analysis of lepton- proton scattering with running
  coupling BK evolution}},
  \href{https://doi.org/10.1103/PhysRevD.80.034031}{\emph{Phys. Rev.}
  {\bfseries D80} (2009) 034031},
  [\href{https://arxiv.org/abs/0902.1112}{{\ttfamily 0902.1112}}].

\bibitem{Albacete:2010sy}
J.~L. Albacete, N.~Armesto, J.~G. Milhano, P.~Quiroga-Arias and C.~A. Salgado,
  \emph{{AAMQS: A non-linear QCD analysis of new HERA data at small-x including
  heavy quarks}},
  \href{https://doi.org/10.1140/epjc/s10052-011-1705-3}{\emph{Eur. Phys. J.}
  {\bfseries C71} (2011) 1705},
  [\href{https://arxiv.org/abs/1012.4408}{{\ttfamily 1012.4408}}].

\bibitem{Kuokkanen:2011je}
J.~Kuokkanen, K.~Rummukainen and H.~Weigert, \emph{{HERA-Data in the Light of
  Small x Evolution with State of the Art NLO Input}},
  \href{https://doi.org/10.1016/j.nuclphysa.2011.10.006}{\emph{Nucl. Phys.}
  {\bfseries A875} (2012) 29--93},
  [\href{https://arxiv.org/abs/1108.1867}{{\ttfamily 1108.1867}}].

\end{thebibliography}

\providecommand{\href}[2]{#2}\begingroup\raggedright\endgroup

\end{document}